\documentclass[reprint,prl,preprintnumbers,superscriptaddress,longbibliography]{revtex4-2}
\usepackage{orcidlink}
\usepackage{graphicx}
\usepackage{dcolumn}
\usepackage{bm}
\usepackage{textgreek}
\usepackage{placeins}
\usepackage{braket}
\usepackage{physics}
\usepackage{overpic}
\usepackage{amssymb}
\usepackage{amsmath}
\usepackage{subfigure}
\usepackage{qcircuit}
\usepackage{tikz}
\usepackage{adjustbox}
\usepackage{longtable}
\usepackage{soul}

\newcommand{\SU}[1]{{\rm SU}({#1})}

\begin{document}

\title{Quantum Simulation of SU(3) Lattice Yang Mills Theory at Leading Order in Large N}

\author{Anthony N. Ciavarella \,\orcidlink{0000-0003-3918-4110}}
\email{anciavarella@lbl.gov}
\affiliation{Physics Division, Lawrence Berkeley National Laboratory, Berkeley, California 94720, USA}
\author{Christian W. Bauer \,\orcidlink{0000-0001-9820-5810}}
\email{cwbauer@lbl.gov}
\affiliation{Physics Division, Lawrence Berkeley National Laboratory, Berkeley, California 94720, USA}
\affiliation{Department of Physics, University of California, Berkeley, Berkeley, CA 94720}

\date{\today}

\begin{abstract}
Quantum simulations of the dynamics of QCD have been limited by the complexities of mapping the continuous gauge fields onto quantum computers.
By parametrizing the gauge invariant Hilbert space in terms of plaquette degrees of freedom, we show how the Hilbert space and interactions can be expanded in inverse powers of $N_c$.
At leading order in this expansion, the Hamiltonian simplifies dramatically, both in the required size of the Hilbert space as well as the type of interactions involved.
Adding a truncation of the resulting Hilbert space in terms of local energy states we give explicit constructions that allow simple representations of SU(3) gauge fields on qubits and qutrits. 
This formulation allows a simulation of the real time dynamics of a SU(3) lattice gauge theory on a $5\times5$ and $8\times8$ lattice on {\tt ibm\_torino} with a CNOT depth of 113.
\end{abstract}

\maketitle

The real time dynamics of strongly coupled quantum field theories such as quantum chromodynamics (QCD) are relevant to many processes in high energy physics. 
These include phenomena such as hadronization, jet fragmentation, and the behavior of matter under extreme conditions such as in the early universe. 
The numerical study of QCD on a lattice using Monte-Carlo (MC) integration has enabled precision non-perturbative calculations of a number of observables~\cite{wilson1974confinement,borsanyi2015ab,karsch2003hadron,tiburzi2017double,beane2015ab,savage2017proton}. 
However, for many observables such as the QCD shear viscosity and inelastic scattering amplitudes, Monte-Carlo integration is limited due to a sign problem~\cite{moore2020shear}.
Hamiltonian lattice QCD formulations promise to circumvent the sign problem, but are still exponentially difficult to simulate on classical computers.
Research in Hamiltonian formulations has gained importance recently due to advances in the development of quantum computers based on a number of different platforms, such as superconducting qubits, trapped ions, and neutral atoms~\cite{huang2020superconducting,bianchetti2010control,wang2022high,wallraff2004strong,chiorescu2004coherent,bruzewicz2019trapped,Henriet_2020,Browaeys_2020,Barredo_2020,bluvstein2022quantum,bluvstein2023logical}. It is anticipated that simulations performed on quantum computers will be able to directly probe real-time dynamics with polynomially scaling computational costs~\cite{feynman1981simulating,Bauer_2023,humble2022snowmass,humble2022snowmass2,beck2023quantum,dimeglio2023quantum,nielsen2001quantum}.
The continuous gauge fields need to be digitized to map them onto a quantum computer's discrete degrees of freedom. 
Common basis choices for the  
Hilbert space of a LGT correspond to choosing on each link group elements (magnetic basis)~\cite{alexandru2019gluon,lamm2019general,ji2020gluon,alexandru2022spectrum,alam2022primitive,Gustafson_2022,zache2023fermionqudit,gonzalez2022hardware}, group representations (electric basis)~\cite{byrnes2006simulating,zohar2012simulating,zohar2013quantum,zohar2013cold,zohar2015formulation,zohar2015quantum,zohar2022quantum,raychowdhury2020solving,raychowdhury2020loop,kadam2023loop,shaw2020quantum,ciavarella2022conceptual,ciavarella2022preparation,klco20202,ciavarella2021trailhead,kavaki2024square,rahman2022real,Atas_2021,Rahman:2022rlg,paulson2021simulating,halimeh2023spin,meurice2021theoretical,davoudi2020towards,davoudi2021search,davoudi2021search,belyansky2023highenergy,berenstein2023integrable,rigobello2023hadrons,kane2024nearlyoptimal,hariprakash2023strategies,su2024coldatom}, or a mixture of the two~\cite{grabowska2023overcoming,bauer2023efficient,kane2022efficient,bauer2023new}, and digitizations can be obtained in each of the choices.
These formal developments have been used to perform a number of quantum simulations on existing hardware including simulations of the Schwinger model, $1+1D$ QCD, SU(2) and SU(3) gauge theories on small lattices and some discrete groups~\cite{martinez2016real,klco20202,Rahman:2022rlg,rahman2022real,ciavarella2021trailhead,ciavarella2022preparation,alam2022primitive,Illa:2022jqb,Gustafson_2022,Atas_2021,farrell2023preparations,farrell2023preparations2,Atas:2022dqm,yang2020observation,Zhou_2022,Su_2023,zhang2023observation,mildenberger2022probing,ciavarella2023quantum,farrell2023scalable,farrell2024quantum,charles2023simulating,kavaki2024square,mueller2022quantum}. 
However, most quantum simulations of lattice gauge theories have been restricted to either small systems or one dimensional systems. 
Going beyond (1+1)D systems is limited by the complexity of implementing plaquette operators which are not present in one spatial dimension.

In this work we will use an electric basis, in which states are labeled by the representation of the gauge group at each link and gauge invariance can be implemented using local constraints that implement Gauss's law at each lattice site.
The electric basis can be digitized by truncating the allowed representations at each link, which amounts to limiting the local energy allowed. 
This can be done in a way that respects gauge invariance, and gauge invariance can be used to integrate out some unphysical states at the cost of a slight increase in the non-locality of the Hamiltonian~\cite{ciavarella2021trailhead}.

In this work we add an expansion in the number of colors $N_c$ to the electric basis formulation. 
It is known that such a $1/N_c$ expansion leads to simplifications in perturbative QCD (for a review, see~\cite{LUCINI201393,Manohar:1998xv}), and is a crucial ingredient in many calculational frameworks of QCD, most notably the parton shower approximation~\cite{Sjostrand:2006za,Bahr:2008pv}.
Combining a large $N_c$ expansion with the lattice formulation of QCD will enable lattice calculations to reproduce the results of these frameworks and determine $1/N_c$ corrections to them.
While the physical value of $N_c = 3$ is not particularly large, such expansions have been shown to be very successful phenomenologically~\cite{tHooft:1973alw,Sjostrand:2006za,Bahr:2008pv,PICH_2002,KAPLAN1996244}. Additionally, the large $N_c$ limit of QCD has been shown to be connected to models of quantum gravity through the AdS/CFT correspondence~\cite{maldacena1999large}.

The large $N_c$ limit can be understood as a classical limit~\cite{Witten1980,yaffe1982large} and by expanding in $1/N_c$ more non-classical features of the theory will be included in the quantum simulation. Note that the classical limit has a degree of freedom for each possible loop on the lattice which limits its applicability to simulating dynamics on classical computers~\cite{yaffe1982large,JEVICKI1983169,JEVICKI1984299}.

The Kogut Susskind Hamiltonian describing pure SU(3) LGT is given by
\begin{equation}
    \hat{H} = \frac{g^2}{2}\sum_{l \in \text{links}}  \hat{E}_{l}^2 -\frac{1}{2g^2} \sum_{p \in \text{plaquettes}} \left(\Box_p + \Box^\dagger_p\right) \,,
    \label{eq:LQCDFull}
\end{equation}
where $g$ is the strong coupling constant, $\hat{E}_{l}^2 =  \hat{E}_{l}^c  \hat{E}_{l}^c$ with $\hat{E}_{l}^c$ the SU(3) chromo-electric field on link $l$ and $\Box_p$ is the trace over color indices of the product of parallel transporters on plaquette $p$~\cite{kogut1975hamiltonian,kogut1979introduction,banks1977strong,jones1979lattice}. 
In the electric basis, the Hilbert space on each link is spanned by states $\ket{R,a,b}$ where $R$ is an irreducible representation of SU(3) and $a$ and $b$ label states in the representation $R$ acting from the left and right.

The SU(3) representation at each link on a point-split lattice can be labeled by the two quantum numbers $p$ and $q$, due to SU(3) being a rank two group.
A gauge invariant representation requires representations at each vertex to combine into a singlet.
This is most easily accomplished using point-split vertices and requiring that the quantum numbers at each 3-point vertex add to zero. 
This has previously been used in formulations of q-deformed lattice gauge theories~\cite{zache2023quantum,hayata2023q} and is very similar to the approach taken in Loop String Hadron formulations~\cite{raychowdhury2020solving,raychowdhury2020loop,davoudi2023general,kadam2023loop}.

As explained in Appendix A, an alternative labeling of a gauge invariant Hilbert state is obtained by specifying oriented closed loops, denoted by $L$, and the way the arrows at each link having more than one loop pass through are combined, denoted by $a$.
A basis state can therefore be written as $\ket{\{L_i, a_\ell\}}$. The only physically relevant states have nonzero overlap with those obtained by acting on the electric vacuum state $\ket{0}$ with 
an operator $\hat O_{\{P_p, \bar P_p\}}$ containing $P_p$ ($\bar P_p$) powers of the plaquette operator $\Box_p$ ($\Box^\dagger_p$) at each plaquette $p$. While other gauge invariant states do exist, they are in different topological sectors and do not need to be represented on the quantum computer as different topological sectors do not interact. For example, a state with a loop of electric flux winding across the entire lattice has a non-trivial winding number and is not coupled to the electric vacuum.
The operator $\hat O_{\{P_p, \bar P_p\}}$ allows us to define the state
\begin{align}
    \ket{\{P_p, \bar P_p\}} = \hat O_{\{P_p, \bar P_p\}} \ket{0}
    \,.
\end{align}

As shown in Appendix B, the overlap between $\ket{\{L_i, a_\ell\}}$ and $\ket{{\{P_p, \bar P_p\}}}$
at leading order in large $N_c$ is given by
\begin{align}
    \braket{\{L_i, a_\ell\}}{\{P_p, \bar P_p\}} \propto \prod_i N_c^{1 - m_i}
    \,,
\end{align}
where $m_i$ counts the total number of plaquettes encircled by each loop $L_i$.
Therefore, the only overlap that survives in the large $N_c$ limit is the one with states $\ket{\{L_i, a_\ell\}}$ for which each loop encircles exactly one plaquette, such that all $m_i = 1$.
This leads to the final result that in the large $N_c$ limit each state can be specified by the number of single-plaquette loops in the positive and negative direction at each plaquette and $a$ at each link traversed by multiple of these loops..
These states are orthonormal to each other, such that the Hilbert space is spanned by the basis
\begin{align}
    {\cal H} = {\rm span}\left\{ \ket{\{a_\ell, n_p, \bar n_p\}} \right\}
    \,.
\end{align}
Due to the suppression of larger loops, the dimension of the Hilbert space in the large $N_c$ limit is dramatically reduced.
Another important simplification is that in this formulation no virtual point-splitting is required. 

So far, our discussion has not used any truncation of the Hilbert space.
However, a truncation is necessary to map the theory onto the finite dimensional Hilbert space of a quantum computer.
As already discussed, a standard way of truncating the Hilbert space is to limit the energy stored in each link of the lattice, which in turn limits the Hilbert space to those states for which the Casimir at each link is below a certain value. 
This truncation preserves all symmetries of the Hamiltonian, most importantly gauge invariance, which guarantees the truncated theory either to have a lattice spacing that freezes out or goes to a theory with the correct gauge symmetry and matter content in the continuum limit~\cite{levin2005string}. 
Simulations of truncated $1+1D$ theories have demonstrated the freezing out of the lattice spacing~\cite{banuls2017efficient,haase2021resource,bruckmann20193,zache2022toward,qubitboson,araz2023toward}, although some infrared properties of the theory can still be recovered~\cite{liu2023phases}.
This truncation amounts to limiting the total value of $p+q$ at each link.
The simplest non-trivial truncation is to require $p+q \leq 1$ which only includes the fundamental and anti-fundamental representations. 
Working to leading order in large $N_c$, this allows at most one loop excitation at each plaquette.
Furthermore, a plaquette can only be excited if all adjacent plaquettes are in the ground state. 
The only allowed values for $\{n_p, \bar n_p\}$ are then $\{0,0\}$, $\{0,1\}$ or $\{1,0\}$, and no specification of $a_\ell$ is necessary.

The Hilbert space at this truncation can therefore be described by assigning a qutrit to each plaquette in the lattice. 
The states of the qutrit will be labelled by $\ket{0}$, $\ket{\circlearrowleft}$, and $\ket{\circlearrowright}$.
Physical states are subject to the constraint that neighboring plaquettes are not simultaneously excited. 
For example, in a two plaquette system, the states $\ket{\circlearrowleft}\ket{0}$ and $\ket{\circlearrowright}\ket{0}$ are physical while $\ket{\circlearrowright}\ket{\circlearrowright}$ and $\ket{\circlearrowleft} \ket{\circlearrowright}$ are not, since it would give rise to the common link having $p+q > 1$.
Similar constructions have been used to study SU(2) lattice gauge theory in the electric basis on plaquette chains and a hexagonal lattice~\cite{rahman2022real,muller2023simple,ebner2024eigenstate,yao20232,yao2023testing,ebner2024entanglement,turro2024classical}.
However, note that the basis given here can work in higher spatial dimensions and with periodic boundary conditions as there is no potential double counting of states at this truncation unlike previous work on the hexagonal lattice.

If one works to leading order in $N_c=3$ and in $2+1D$, the electric field operator for a link $\ell$ lying on plaquettes $p$ and $p'$ at this truncation can be written as
\begin{align}
    \hat{E_\ell}^2 = & \frac{4}{3}\left[ \ket{\circlearrowleft}_{p}\bra{\circlearrowleft}_{p}+ \ket{\circlearrowleft}_{p'}\bra{\circlearrowleft}_{p'}+ \left(\ket{\circlearrowleft} \leftrightarrow  \ket{\circlearrowright}\right)\right] \,,
\end{align}
where we have used the full expression of the Casimir of the fundamental representation $C_f = (N_c^2 - 1) / (2 N_c) = 4/3$.
The plaquette operator at position $p$ is given by
\begin{align}
    \hat{\Box}_{p} = &  \hat{P}_{0,p+\hat x} \hat{P}_{0,p-\hat x} \hat{P}_{0,p+\hat y} \hat{P}_{0,p-\hat y} \nonumber \\
    & \times \left(\ket{\circlearrowleft}_p \bra{0}_p + \ket{\circlearrowright}_p \bra{\circlearrowleft}_p + \ket{0}_p \bra{\circlearrowright}_p\right) \,,
\end{align}
where $\hat{P}_{0,p} = \ket{0}_p \bra{0}_p$ and $p\pm\hat x$ ($p\pm\hat y$) denotes the plaquette one position away in the $x$ ($y$) direction. 

This Hamiltonian has a charge conjugation (C) symmetry that causes states with the anti-symmetric combination $\frac{1}{\sqrt{2}}\left(\ket{\circlearrowleft} - \ket{\circlearrowright}\right)$ anywhere on the lattice to decouple from the rest of the Hilbert space. 
This decoupling can be seen by repeated applications of the plaquette operators to the electric vacuum. Explicitly, we have
\begin{align}
    \hat{\Box}_{p} & \ket{0} = \ket{\circlearrowleft} + \ket{\circlearrowright} \nonumber \\
    \hat{\Box}_{p} & \frac{1}{\sqrt{2}}\left(\ket{\circlearrowleft} + \ket{\circlearrowright}\right) = \sqrt{2} \ket{0} + \frac{1}{\sqrt{2}}\left(\ket{\circlearrowleft} + \ket{\circlearrowright} \right)  \,,
\end{align}
so the state $\frac{1}{\sqrt{2}}\left(\ket{\circlearrowleft} - \ket{\circlearrowright}\right)$ is never coupled to the rest of the Hilbert space.
One can therefore perform separate simulations for the C even and odd sector.
By assigning $\ket{1} = \frac{1}{\sqrt{2}}\left(\ket{\circlearrowleft} \pm \ket{\circlearrowright}\right)$, the C (anti)symmetric subspace can be described by assigning a qubit to each plaquette instead of a qutrit. 
As already mentioned, physical states have the constraint that neighboring qubits cannot both be in the $\ket{1}$ state. 
With this encoding, the Hamiltonian for the C even sector is given by
\begin{align}
    \hat{H} =& \sum_p \left(\frac{8}{3}g^2 - \frac{1}{2g^2}\right) \hat{P}_{1,p} \nonumber \\
    & - \frac{1}{g^2\sqrt{2}}\hat{P}_{0,p+\hat x} \hat{P}_{0,p-\hat x} \hat{P}_{0,p+\hat y} \hat{P}_{0,p-\hat y} \hat{X}_p \,,
    \label{eq:Ham3PXP}
\end{align}
where $\hat{P}_{1,p} = \ket{1}_p \bra{1}_p$ and $\hat{X}_p$ is the Pauli X operator acting on the qubit at plaquette $p$.
It is interesting to note that the plaquette operator at this truncation is a $PXP$ term. 
$PXP$ models have previously been studied as an effective Hamiltonian describing the low energy subspace of Rydberg atom arrays which can be described by Ising models~\cite{ebadi2021quantum,semeghini2021probing,Omran_2019}.
It has been studied in the context of thermalization where the presence of scar states has been demonstrated~\cite{nandkishore2015many,choi2019emergent,Moudgalya_2022,chandran2023quantum,surace2020lattice}. 
Eq.~\eqref{eq:Ham3PXP} can be described as a limit of an Ising model with fields in the $\hat{x}$ and $\hat{z}$ directions and that in this regime the Ising model has been shown to demonstrate confinement~\cite{kormos2017real,james2019nonthermal,robinson2019signatures}. 
This suggests that it may be possible to connect the presence of confinement in the Ising model to the physics of large $N_c$ Yang Mills.

There are three representations with $C \sim N_c$, namely $p = 2$, $q = 2$ and $p = q = 1$, and a next truncation in the large $N_c$ limit should include all three of those states. 
However, taking into account subleading $N_c$ corrections, the representation with the second smallest Casimir at large $N_c$ is the anti-symmetric combination of two fundamental representations with $C = N_c - 1-\frac{2}{N_c}$. Note that in ${\rm SU}(3)$, this is just the $\Bar{\mathbf{3}}$ representation.
Changing the truncation to include this representation allows neighboring plaquettes to be excited and includes states with vertices that have three incoming or outgoing $\mathbf{3}$ representations. 
As shown in more detail in the supplemental material, this truncation still fixes the representation on each link by the number of loops on the neighboring plaquettes, and each plaquette can still only be in the three possible states $\ket{0}$, $\ket{\circlearrowleft}$ and $\ket{\circlearrowright}$. 
A pair of neighboring plaquettes can only be in one of the following states $\ket{0}\ket{0}$, $\ket{0}\ket{\circlearrowleft}$, $\ket{\circlearrowleft}\ket{0}$, $\ket{0}\ket{\circlearrowright}$, $\ket{\circlearrowright}\ket{0}$, $\ket{\circlearrowleft}\ket{\circlearrowright}$ or $\ket{\circlearrowright}\ket{\circlearrowleft}$.
The Hilbert space is therefore still spanned by a qutrit at each plaquette. 
 
At this truncation, the electric field operator on a link shared between plaquettes $p$ and $p'$ is given by
\begin{align}
    \hat{E}^2 = & \frac{4}{3}\left[ \ket{\circlearrowleft}_{p}\bra{\circlearrowleft}_{p}
    + \ket{\circlearrowleft}_{p'}\bra{\circlearrowleft}_{p'}+  \left(\ket{\circlearrowleft} \leftrightarrow  \ket{\circlearrowright}\right)
    \right] \nonumber \\
     & - \frac{4}{3}\left[ \ket{\circlearrowleft}_{p}\bra{\circlearrowleft}_{p} \ket{\circlearrowright}_{p'}\bra{\circlearrowright}_{p'} + \left(\ket{\circlearrowleft} \leftrightarrow  \ket{\circlearrowright}\right)\right]
    \,.
\end{align}
The plaquette operator is given by
\begin{align}
    \hat{\Box}_p = & \sum_{c_k,s_i,s_f} \mathcal{M}_{c_1,c_2,c_3,c_4}^{s_i,s_f} \nonumber \\
    & \times \hat{P}_{c_1,p-\hat x} \hat{P}_{c_2,p+\hat x} \hat{P}_{c_3,p-\hat y} \hat{P}_{c_4,p+\hat y} \nonumber \\
    &\times\left(\ket{s_f} \bra{s_i} + \ket{s_i} \bra{s_f} \right) \, .
\end{align}
Using results in the supplemental material, it can be seen that $\mathcal{M}_{0,0,0,0}^{s_i,s_f}=1$ for all $s_i$ and $s_f$, and when the controls are not all $0$, $\mathcal{M}_{c_1,c_2,c_3,c_4}^{s_i,s_f}=3^{-n_e/2}$ for transitions between allowed physical states where $n_e$ is the number of excited neighboring plaquettes.

The Hamiltonians obtained at the two truncations discussed above have only negative off-diagonal elements. 
This means that their static properties can be studied using Monte Carlo techniques without a sign problem~\cite{bravyi2007complexity,suzuki2013transverse}. This can be helpful for quantum simulation as classical MC calculations can be used to generate ensembles of states that when averaged produce a thermal distribution~\cite{lamm2018simulation,harmalkar2020quantum,gustafson2021toward,blunt2014density,saroni2023reconstructing}. 
These states can be initialized on quantum computers which would allow for studying dynamics of this theory at finite temperature without a sign problem. 
Additionally, many of the variables used in scale setting in traditional lattice QCD are defined in terms of Euclidean correlation functions which are difficult to access on quantum computers~\cite{sommer2014scale,clemente2022strategies,paulson2021simulating,ciavarella2022preparationO3}. 
For these truncated Hamiltonians, classical MC calculations can be used to compute these variables to set the scale for a simulation with the same Hamiltonian on a quantum computer~\cite{clemente2022strategies}.

\begin{figure}
    \centering
    \includegraphics[width=8.6cm]{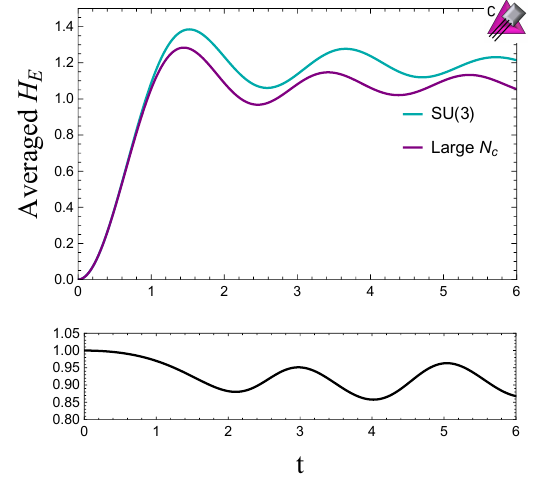}
    \caption{Calculation of $\frac{1}{T}\int_0^T dt \bra{\psi(t)} \hat{H}_E \ket{ \psi(t)}$ on a $4\times1$ lattice with periodic boundary conditions and $g=1$. The blue line shows the simulation for a SU(3) lattice gauge theory truncated at $p+q\leq1$, using the formalism introduced in Ref~\cite{ciavarella2021trailhead}. The purple line shows the time evolution computed with the large $N_c$ truncated Hamiltonian in Eq.~\eqref{eq:Ham3PXP}. The black line underneath shows the ratio of the large $N_c$ electric energy to the SU(3) electric energy.}
    \label{fig:SU3LargeNComparison}
\end{figure}

To probe the effects of working to leading order in large $N_c$, the electric vacuum was evolved in time on a $4\times1$ lattice with PBC. Fig.~\ref{fig:SU3LargeNComparison} shows the evolution of $\frac{1}{T} \int_0^T dt \bra{\psi(t)} \hat{H}_E \ket{ \psi(t)}$ as a function of $T$ for a SU(3) LGT truncated at $p+q\leq1$. At long times, this observable is expected to equilibrate to a thermal value determined by the inital state's energy. As Fig.~\ref{fig:SU3LargeNComparison} shows, the relative error from the large $N_c$ expansion is roughly $20\%$ which should be expected from expanding in $\frac{1}{N_c}$ with $N_c=3$.

As an example of how the formalism introduced in this work can be used for quantum simulation, the Hamiltonian in Eq.~\eqref{eq:Ham3PXP} was simulated on IBM's 133 qubit superconducting quantum computer {\tt ibm\_torino}~\cite{aleksandrowicz2019qiskit,ibmTorino}. 
Due to the connectivity of the hardware, open boundary conditions were used. 
Time evolution was implemented using Trotterized time evolution operators. 
Errors in the calculation were suppressed using $XX$ dynamical decoupling sequences and Pauli twirling~\cite{viola1999dynamical,urbanek2021mitigating,Rahman:2022rlg,rahman2022real}. 
Errors in the gates were mitigated using operator decoherence renormalization and CNOT noise extrapolations~\cite{urbanek2021mitigating,Rahman:2022rlg,rahman2022real,farrell2023scalable,asaduzzaman2024model,hidalgo2023quantum,kiss2024quantum,he2020zero,pascuzzi2022computationally}. 
Readout errors were mitigated using twirled readout error extinction (T-REX)~\cite{trexmit}. 
Since the number of Trotter steps that can be run on quantum hardware is limited, we utilize multiple time step sizes $\Delta t$ to obtain results at more $t$ values.
Increasing $\Delta t$ will increase the size of time discretization errors in the simulation, but this can be mitigated by choosing $\Delta t$ such that late time slices are sampled by multiple values of $\Delta t$, which then allows an extrapolation to small $\Delta t$. 
The details of the implementation of these techniques is described in the supplemental material.

\begin{figure}
    \centering
    \includegraphics[width=8.7cm]{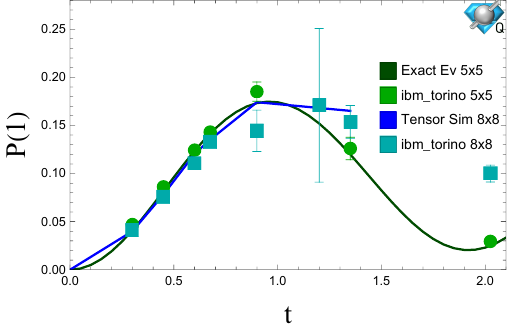}
    \caption{Average probability of a plaquette being excited from the electric vacuum as a function of time on a $4\times4$ and $7\times7$ plaquette lattice with open boundary conditions and $g=1$. The dark green points are an exact classical simulation. The dark blue points were obtained by tensor network simulations of up to two Trotter steps. The light blue and green points are the error mitigated results from {\tt ibm\_torino}.}
    \label{fig:torino_sim}
\end{figure}

A $5\times5$ lattice with $g=1$ was simulated using a set of 39 qubits on {\tt ibm\_torino}. 
Due to open boundary conditions, this lattice has $4\times4$ plaquettes. 
16 of the qubits were used to represent the Hilbert space of the theory and the remaining qubits were used to enable efficient communication between them. 
The system was initialized in the electric vacuum and the probability of a qubit being excited averaged over the lattice is shown in Fig.~\ref{fig:torino_sim}~\footnote{The icons in the corners of plots indicate if classical or quantum compute resources were used to perform the calculation~\cite{klco2020minimally} and are available at {\tt https://iqus.uw.edu/resources/icons/} }, showing results both from classical simulations of this relatively small system and from runs on quantum hardware using {\tt ibm\_torino}.
We observe good agreement between the classical simulation and the results from {\tt ibm\_torino}. 
Note that this observable is proportional to the electric energy in the system, and previous work has used the evolution of the electric energy as a probe of thermalization times in SU(2) lattice gauge theory~\cite{hayata2021thermalization}. 

Having validated the quantum circuits for the $5\times5$ lattice, an $8\times8$ lattice with open boundary conditions was simulated on {\tt ibm\_torino} as well. 
This requires 49 qubits to represent the state of the system and the remaining qubits are used to enable communication between them. 
The average probability of a plaquette being excited is shown in Fig.~\ref{fig:torino_sim}.
Simulating the time evolution for a system of this size is beyond the reach of brute force state vector simulation. 
Vacuum properties of large one dimensional systems can be simulated efficiently using tensor networks~\cite{white1992density,white1993density,verstraete2004matrix,haegeman2011time,haegeman2016unifying}, however performing real time evolution requires resources that grow with evolution time. Scaling tensor network calculations to multiple spatial dimensions is practically challenging~\cite{pang2020efficient}. 
The dark blue points in Fig.~\ref{fig:torino_sim} show tensor network simulations of up to two Trotter steps of the circuits that were implemented on {\tt ibm\_torino} using {\tt cuQuantum}~\cite{cuQuantum} on a single NVIDIA A100 GPU. 
Two Trotter steps took roughly one minute to run, however 3 Trotter steps did not finish running within 20 hours. 
For this reason, there is no extrapolation to $\Delta t = 0$ in the classical simulation or classical data for 3 Trotter steps. 
Due to the lack of the $\Delta t = 0$ extrapolation in the classical simulation and validation of the quantum circuits on a smaller lattice, it is expected that data from {\tt ibm\_torino} is a more accurate simulation of the dynamics of the system than the tensor network calculations. 
Note that further optimization of the classical simulation is likely to reduce the runtime, however this system is still in the regime where classical simulation is expected to be difficult. 
For reference, running and processing all of the quantum circuits for a single time step took roughly 7 minutes.

\FloatBarrier
In this work, a large $N_c$ expansion was combined with electric basis truncations of the Kogut-Susskind Hamiltonian. 
This led to significant simplifications of the Hamiltonian and enabled a quantum simulation of SU(3) lattice gauge theory in multiple spatial dimensions. 
It is expected that this formalism can be extended to $3$ spatial dimensions and to include matter. 
Going to subleading order in $1/N_c$ and to larger truncations should also be possible systematically.
The simplifications from truncating at some order in $1/N_c$ and success of large $N_c$ expansions may allow for near term simulations of phenomenologically relevant phenomena such as inelastic scattering, jet fragmentation or thermalization. 
Additionally, the connection of the large $N_c$ limit of ${\rm SU}(N_c)$ gauge theories to quantum gravity may allow quantum simulations of these truncations to give insights into some models of quantum gravity.

\begin{acknowledgements}
We would like to acknowledge helpful conversations with Ivan Burbano, Irian D'Andrea, Jesse Stryker, and Michael Kreshchuk. We would like to thank Martin Savage, Marc Illa, and Roland Farrell for many conversations related to quantum simulation. We would like to thank Aneesh Manohar for helpful discussions about large $N_c$ expansions. We would also like to acknowledge helpful conversations with Jad Halimeh about the emergence of $PXP$ models from certain limits of gauge theories. 
This material is based 
upon work supported by the U.S. Department of Energy, Office of Science, National Quantum Information 
Science Research Centers, Quantum Systems Accelerator. Additional support is acknowledged from  the U.S. Department of Energy (DOE), Office of Science under contract DE-AC02-05CH11231, partially through Quantum Information Science Enabled Discovery (QuantISED) for High Energy Physics (KA2401032).
This research used resources of the Oak Ridge Leadership Computing Facility (OLCF), which is a DOE Office of Science User Facility supported under Contract DE-AC05-00OR22725.
We acknowledge the use of IBM Quantum services for this work. The views expressed are those of the authors, and do not reflect the official policy or position of IBM or the IBM Quantum team. This research used resources of the National Energy Research Scientific Computing Center, which is supported by the Office of Science of the U.S. Department of Energy under Contract No. DE-AC02-05CH11231.
\end{acknowledgements}

\newpage
\textit{Appendix A: Graphical representation of Basis States.}

To explain the large $N_c$ counting employed in this work, it will be useful to develop some graphical notation for physical states on a lattice.
The Hilbert space describing a single link in a lattice gauge theory is spanned by electric basis states of the form $\ket{R,a,b}$ where $R$ is a representation of the gauge group, and $a$ and $b$ are indices that label states in the representation $R$ when acted from the left and right. 
Physical states are subject to a constraint from Gauss's law which requires that the sum of representations on each vertex of the lattice forms a singlet. 
On a lattice where each vertex is connected to at most three links, gauge invariant states can be specified by the representation $R$ on each link and a specification on each vertex of how the links add to form a singlet. 

\begin{figure}
    \includegraphics[width=8.6cm]{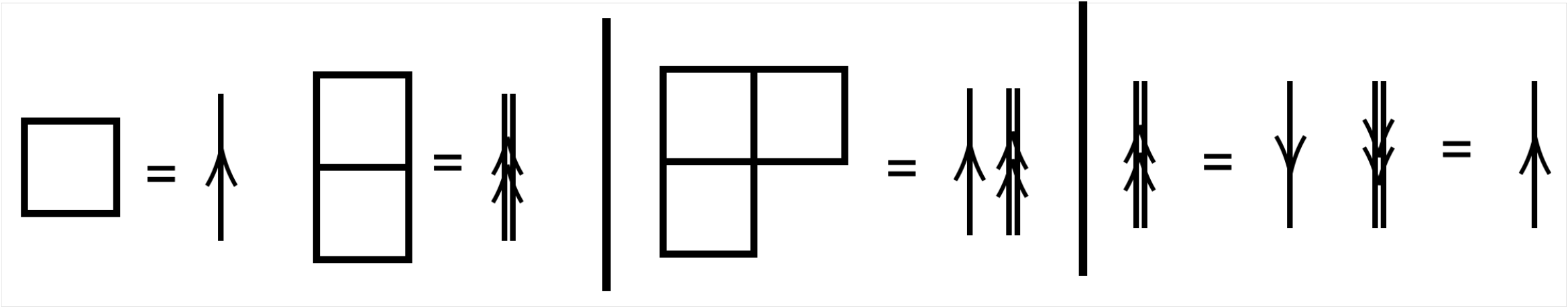}
    \caption{Graphical representations of Young diagrams in terms of arrows.}
    \label{fig:arrowRep}
\end{figure}

For \SU{N} gauge groups a representation $R$ can be labeled by a Young diagram, which can be specified through the number of columns with $1, 2, \ldots, N-1$ boxes. 
For \SU{3} only two numbers are required, and the labels are often chosen as $(p, q)$, with $p$ labeling the number of columns with a single box, and $q$ labeling the number of columns with two boxes.
It will be useful to obtain a representation of Young diagrams in terms of lines with arrows, as illustrated in Fig.~\ref{fig:arrowRep}. 
One can see that fundamental and anti-fundamental representations can be represented either by lines with a single arrow in one direction, or by lines with a double arrow in the opposite direction. 
More complex representations can be built by combining such lines together.

For lattices where vertices connect to more links, such as a square lattice in 2D or 3D, not all states that can be labeled by the representation above are linearly independent, leading to an ambiguity in labeling the basis states.
This is due to the so-called Mandelstam constraints, which relate contractions of representation indices across a vertex.
A point splitting procedure can be performed to split each vertex into three link vertices connected by virtual links, which lifts this ambiguity.
In this point-split lattice, the gauge invariant states can be specified with the same assignment of labels used on a trivalent lattice.

\begin{figure}
    \includegraphics[width=6.6cm]{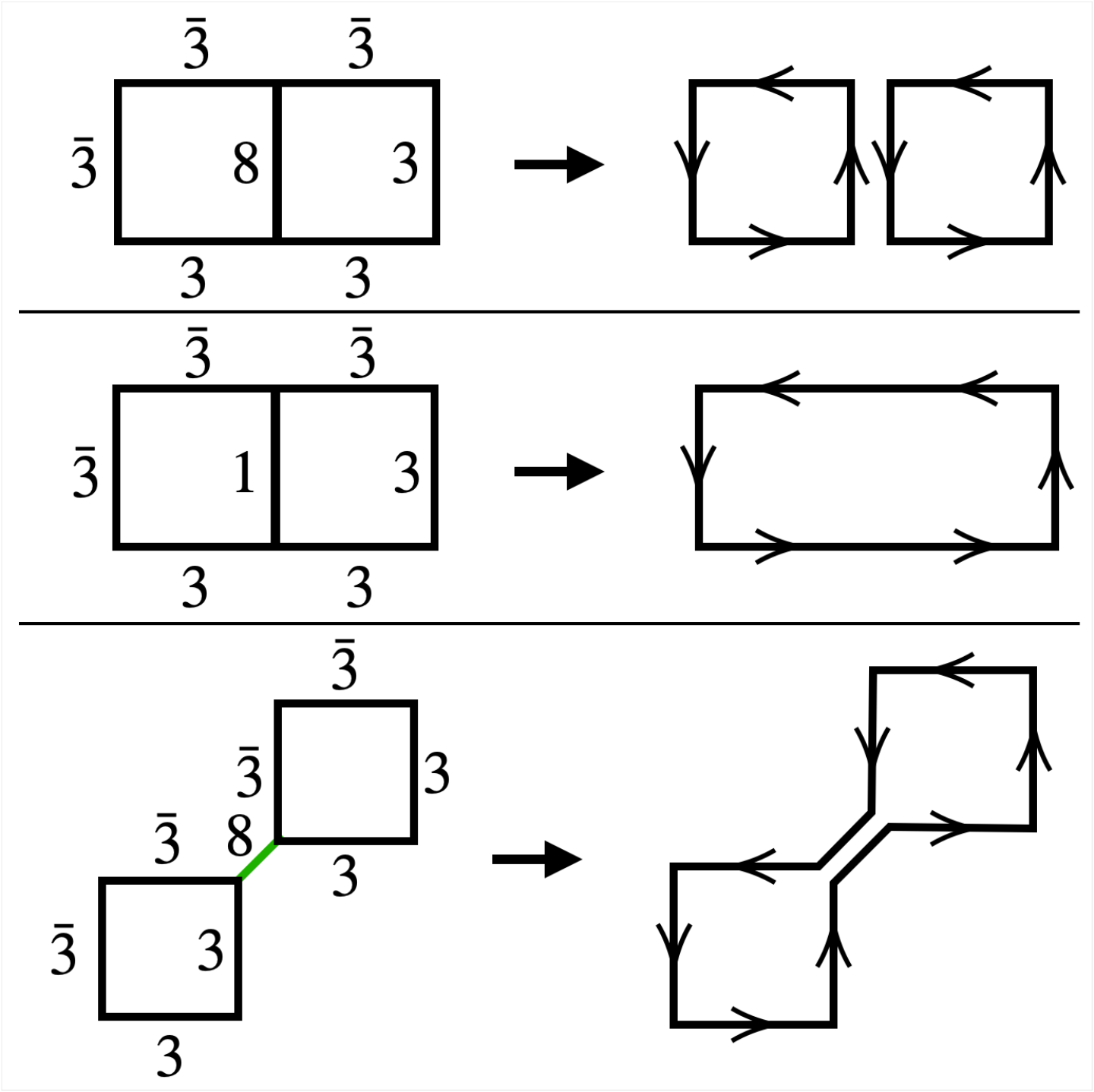}
    \caption{Graphical representations of basis states on a point-split lattice.}
    \label{fig:repRelation}
\end{figure}

There is an equivalent labeling of the states of the physical Hilbert space that will prove useful, using the arrow representation introduced above.
This is illustrated in Fig.~\ref{fig:repRelation}, and will be called a ``loop representation''\footnote{Note that this presentation only works as presented in the simplest topological sector of the allowed states. There are other states that allow for additional overall winding numbers, which will not be considered in this work. One can easily generalize the loop representation to also include winding loops.}.
In this representation each state is labeled by a set of loops $L_i$, together with a specification $a_\ell$, which denotes the way the arrows at each link $\ell$ having more than one loop pass through being combined. 
Each loop needs to specify which plaquettes are encircled and in which order, while $a_\ell$ contains the information on how to combine lines of multiple loops into single or double arrows.
Note that it might seem that there is an ambiguity in the choice of single arrows in one or double arrows in the other direction. 
This ambiguity is fixed by choosing representations with $p=0$ or $q=0$ to have only single arrows, and demanding that the number of arrows entering and leaving a vertex is conserved.
Due to the point splitting, closed loops have have the property that their lines can not cross each other, so they can not form knots or be twisted. 
A loop representation is therefore spanned by the states $\ket{\{L_i, a_\ell\}}.$
Note that this loop representation is simply a graphical representation of the states with definite representation at each link.
In particular, lines do not necessarily represent the tensor indices of a given representation, and lines being connected does not necessarily imply tensor indices being contracted.
\begin{figure*}
    \centering    \includegraphics[width=.9\textwidth]{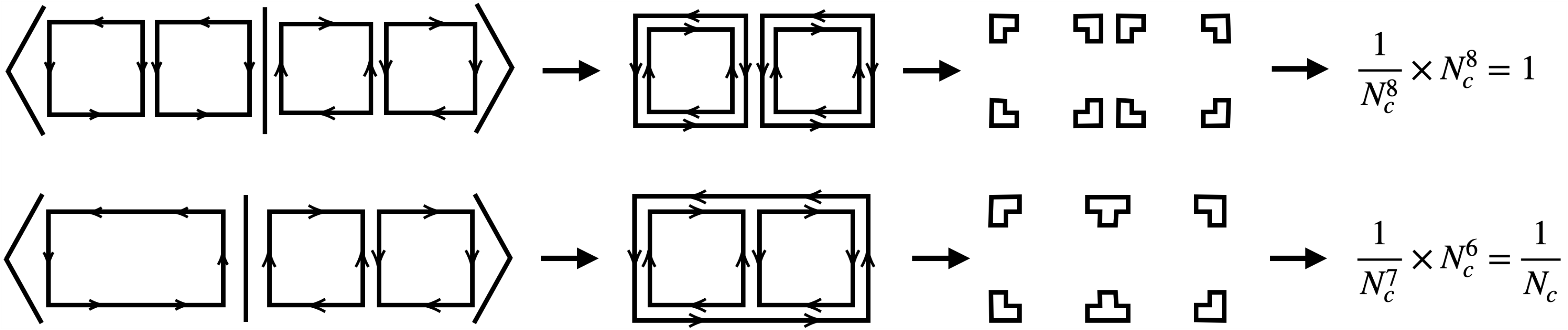}
    \caption{Graphical method to obtain the scaling of the overlap matrix $\braket{\{L_i\}}{\{P_p, \bar P_p\}}$. The top example contains two loops, each encircling a single plaquette $m_1 = m_2 = 1$, while the bottom example has a single loop encircling 2 loops $m_1 = 1$. This gives for the top example $q_1 = q_2 = 1+3\times 1 = 4$ and $v_1 = v_2 = 2 + 2\times 1 = 4$, giving the final scaling $N_c^0$. For the bottom example we have $q_1 = 1 + 2 \times 3 = 7$ and $v_1 = 2 + 2\times 2 = 6$, giving the final scaling $1/N_c$. }
    \label{fig:NScale}
\end{figure*}
Before moving on, we want to make it clear that the loop representation as given is likely not a computationally efficient representation of the Hilbert space, since loops are necessarily non-local objects which can in general span an arbitrary number of plaquettes. Its usefulness will come from applying the $1/N_c$ expansion.

The vacuum state in the interacting theory can be generated adiabatically from the vacuum state of the free electric theory (the vacuum at $g=\infty$) by acting with the operators of the interacting Hamiltonian, which are $\hat{\Box}$ operators at the different plaquettes or $\hat E_i^2$ operators at the different links. 
Excited states in the simpliest topological sector can be obtained by further applications of electric energy or plaquette operators. 
One can therefore classify all states in this sector by the minimum number of plaquette operators and its conjugate that are required to reach it from the vacuum
\begin{align}
	\ket{\{P_p, \bar P_p\}}\equiv \prod_p \hat{\Box}_p^{P_p} \hat{\Box}_p^{\dagger \bar P_p} \ket{0} 
	\,.
\end{align}
This state will be a linear combination of several electric basis states and one can write
\begin{align}
\label{eq:state_expansion}
    \ket{\{P_p, \bar P_p\}} = \sum_{\{L_i, a_\ell\}} \braket{\{L_i, a_\ell\}}{\{P_p, \bar P_p\}} \, \ket{L_i, a_\ell}
    \,.
\end{align}

\textit{Appendix B: Large $N_c$ Counting of States.}
The large $N_c$ scaling of a state, $\ket{\{L_i, a_\ell\}}$ is determined by the large $N_c$ expansion of $\braket{\{L_i, a_\ell\}}{\{P_p, \bar P_p\}}$ for the minimal choice of $P_p$ and $\bar P_p$ to obtain a nonzero overlap. 
Defining $\ket{\{L_i\}} = \prod_i U_{L_i} \ket{0}$ where $U_{L_i}$ is a product of parallel transporters along the loop $L_i$ and using that the overlap $\braket{\{L_i, a_\ell\}}{\{L_i\}}$ is ${\cal O}(1)$ in the $N_c$ scaling, the $N_c$ scaling is determined by the overlap $\braket{\{L_i\}}{\{P_p, \bar P_p\}}$.
This overlap can be evaluated in the magnetic basis through inserting $\mathbf{1} = \prod_{\text{links l}} \int dU_l \ket{U_l} \bra{U}_l$. 
To evaluate the large $N_c$ scaling of these integrals, the identity 
\begin{align}
    & \int dU \prod_{n=1}^q U_{i_n j_n} U^{*}_{i'_n j'_n} = \nonumber \\
    & \frac{1}{N_c^q} \sum_{\text{permutations k}} \prod_{n=1}^q \delta_{i_n i'_{k_n}} \delta_{j_n j'_{k_n}} + \mathcal{O}\left(\frac{1}{N_c^{q+1}}\right) \,,
    \label{eq:Uint}
\end{align}
will be used~\cite{weingarten1978asymptotic}. 
The large $N_c$ scaling will be determined by the permutation of indices contraction that gives the largest factors of $N_c$. 
A diagrammatic method of evaluating the large $N_c$ scaling is shown in Fig.~\ref{fig:NScale}. 

First, the plaquette operators being applied are placed over loops in the final state. 
To determine the powers of $N_c$ that come from contracting the Kronecker $\delta$s, one can erase the middle of each link in the diagram and connect the lines from the same vertex.
This leaves a set of $v$ closed loops involving one vertex each, and each of these closed loops contributes a factor of $N_c$ in the numerator. 
Each loop $L_i$ therefore contributes a factor $N_c^{v_i - q_i}$ and the the total $N_c$ scaling is given by
\begin{align}
    N_c^{v - q}\,, \qquad  q\equiv \sum_i q_i\,, \quad v\equiv \sum_i v_i
\end{align}
to the final overlap.
Since each $U_{ij}$ in Eq.~\eqref{eq:Uint} corresponds to a line in the figure, one immediately finds that $q = n_l/2$, where $n_l$ is the total number of lines on each link in the diagram. 
Denoting by $m_i$ the number of plaquettes encircled by each loop $L_i$, one needs $m_i$ plaquette operators for each loop. 
The total number of lines is then given by $n_l = 2 + 6m_i$, and the total number of closed loops $n_v$ is given by $2 + 2 m_i$ for each loop in the basis. Thus one finds
\begin{align}
    q_i = 1 + 3 m_i \,, \qquad v_i = 2 + 2m_i
    \,.
\end{align}
Putting this together, one finds that each loop contributes a factor of $N_c^{1 - m_i}$
to the overall scaling of the overlap, such that
\begin{align}
    \braket{\{L_i, a_\ell\}}{\{P_p, \bar P_p\}} \propto \prod_i N_c^{1 - m_i}
    \,,
\end{align}

This implies that the states that can be reached to leading order in $1/N_c$ are those that only involve loops $L_i$ with $m_i = 1$.
Therefore, the only overlap that survives in the large $N_c$ limit is the one with states $\ket{\{L_i, a_\ell\}}$ for which each loop encircles exactly one plaquette.
At order $1/N_c$, states with loops extending over two plaquettes will be present. Basis constructions similar to those in the main text can be used to represent these states on a quantum computer.

\newpage

\bibliography{ref}

\begin{thebibliography}{158}%
\makeatletter
\providecommand \@ifxundefined [1]{%
 \@ifx{#1\undefined}
}%
\providecommand \@ifnum [1]{%
 \ifnum #1\expandafter \@firstoftwo
 \else \expandafter \@secondoftwo
 \fi
}%
\providecommand \@ifx [1]{%
 \ifx #1\expandafter \@firstoftwo
 \else \expandafter \@secondoftwo
 \fi
}%
\providecommand \natexlab [1]{#1}%
\providecommand \enquote  [1]{``#1''}%
\providecommand \bibnamefont  [1]{#1}%
\providecommand \bibfnamefont [1]{#1}%
\providecommand \citenamefont [1]{#1}%
\providecommand \href@noop [0]{\@secondoftwo}%
\providecommand \href [0]{\begingroup \@sanitize@url \@href}%
\providecommand \@href[1]{\@@startlink{#1}\@@href}%
\providecommand \@@href[1]{\endgroup#1\@@endlink}%
\providecommand \@sanitize@url [0]{\catcode `\\12\catcode `\$12\catcode `\&12\catcode `\#12\catcode `\^12\catcode `\_12\catcode `\%12\relax}%
\providecommand \@@startlink[1]{}%
\providecommand \@@endlink[0]{}%
\providecommand \url  [0]{\begingroup\@sanitize@url \@url }%
\providecommand \@url [1]{\endgroup\@href {#1}{\urlprefix }}%
\providecommand \urlprefix  [0]{URL }%
\providecommand \Eprint [0]{\href }%
\providecommand \doibase [0]{https://doi.org/}%
\providecommand \selectlanguage [0]{\@gobble}%
\providecommand \bibinfo  [0]{\@secondoftwo}%
\providecommand \bibfield  [0]{\@secondoftwo}%
\providecommand \translation [1]{[#1]}%
\providecommand \BibitemOpen [0]{}%
\providecommand \bibitemStop [0]{}%
\providecommand \bibitemNoStop [0]{.\EOS\space}%
\providecommand \EOS [0]{\spacefactor3000\relax}%
\providecommand \BibitemShut  [1]{\csname bibitem#1\endcsname}%
\let\auto@bib@innerbib\@empty
\bibitem [{\citenamefont {Wilson}(1974)}]{wilson1974confinement}%
  \BibitemOpen
  \bibfield  {author} {\bibinfo {author} {\bibfnamefont {K.~G.}\ \bibnamefont {Wilson}},\ }\bibfield  {title} {\bibinfo {title} {Confinement of quarks},\ }\href {https://doi.org/10.1103/PhysRevD.10.2445} {\bibfield  {journal} {\bibinfo  {journal} {Phys. Rev. D}\ }\textbf {\bibinfo {volume} {10}},\ \bibinfo {pages} {2445} (\bibinfo {year} {1974})}\BibitemShut {NoStop}%
\bibitem [{\citenamefont {Borsanyi}\ \emph {et~al.}(2015)\citenamefont {Borsanyi}, \citenamefont {Durr}, \citenamefont {Fodor}, \citenamefont {Hoelbling}, \citenamefont {Katz}, \citenamefont {Krieg}, \citenamefont {Lellouch}, \citenamefont {Lippert}, \citenamefont {Portelli}, \citenamefont {Szabo},\ and\ \citenamefont {Toth}}]{borsanyi2015ab}%
  \BibitemOpen
  \bibfield  {author} {\bibinfo {author} {\bibfnamefont {S.}~\bibnamefont {Borsanyi}}, \bibinfo {author} {\bibfnamefont {S.}~\bibnamefont {Durr}}, \bibinfo {author} {\bibfnamefont {Z.}~\bibnamefont {Fodor}}, \bibinfo {author} {\bibfnamefont {C.}~\bibnamefont {Hoelbling}}, \bibinfo {author} {\bibfnamefont {S.~D.}\ \bibnamefont {Katz}}, \bibinfo {author} {\bibfnamefont {S.}~\bibnamefont {Krieg}}, \bibinfo {author} {\bibfnamefont {L.}~\bibnamefont {Lellouch}}, \bibinfo {author} {\bibfnamefont {T.}~\bibnamefont {Lippert}}, \bibinfo {author} {\bibfnamefont {A.}~\bibnamefont {Portelli}}, \bibinfo {author} {\bibfnamefont {K.~K.}\ \bibnamefont {Szabo}},\ and\ \bibinfo {author} {\bibfnamefont {B.~C.}\ \bibnamefont {Toth}},\ }\bibfield  {title} {\bibinfo {title} {Ab initio calculation of the neutron-proton mass difference},\ }\href {https://doi.org/10.1126/science.1257050} {\bibfield  {journal} {\bibinfo  {journal} {Science}\ }\textbf {\bibinfo {volume} {347}},\ \bibinfo {pages} {1452} (\bibinfo {year} {2015})},\
  \Eprint {https://arxiv.org/abs/https://www.science.org/doi/pdf/10.1126/science.1257050} {https://www.science.org/doi/pdf/10.1126/science.1257050} \BibitemShut {NoStop}%
\bibitem [{\citenamefont {Karsch}\ \emph {et~al.}(2003)\citenamefont {Karsch}, \citenamefont {Redlich},\ and\ \citenamefont {Tawfik}}]{karsch2003hadron}%
  \BibitemOpen
  \bibfield  {author} {\bibinfo {author} {\bibfnamefont {F.}~\bibnamefont {Karsch}}, \bibinfo {author} {\bibfnamefont {K.}~\bibnamefont {Redlich}},\ and\ \bibinfo {author} {\bibfnamefont {A.}~\bibnamefont {Tawfik}},\ }\bibfield  {title} {\bibinfo {title} {Hadron resonance mass spectrum and lattice {QCD} thermodynamics},\ }\href {https://doi.org/10.1140/epjc/s2003-01228-y} {\bibfield  {journal} {\bibinfo  {journal} {The European Physical Journal C}\ }\textbf {\bibinfo {volume} {29}},\ \bibinfo {pages} {549} (\bibinfo {year} {2003})}\BibitemShut {NoStop}%
\bibitem [{\citenamefont {Tiburzi}\ \emph {et~al.}(2017)\citenamefont {Tiburzi}, \citenamefont {Wagman}, \citenamefont {Winter}, \citenamefont {Chang}, \citenamefont {Davoudi}, \citenamefont {Detmold}, \citenamefont {Orginos}, \citenamefont {Savage},\ and\ \citenamefont {Shanahan}}]{tiburzi2017double}%
  \BibitemOpen
  \bibfield  {author} {\bibinfo {author} {\bibfnamefont {B.~C.}\ \bibnamefont {Tiburzi}}, \bibinfo {author} {\bibfnamefont {M.~L.}\ \bibnamefont {Wagman}}, \bibinfo {author} {\bibfnamefont {F.}~\bibnamefont {Winter}}, \bibinfo {author} {\bibfnamefont {E.}~\bibnamefont {Chang}}, \bibinfo {author} {\bibfnamefont {Z.}~\bibnamefont {Davoudi}}, \bibinfo {author} {\bibfnamefont {W.}~\bibnamefont {Detmold}}, \bibinfo {author} {\bibfnamefont {K.}~\bibnamefont {Orginos}}, \bibinfo {author} {\bibfnamefont {M.~J.}\ \bibnamefont {Savage}},\ and\ \bibinfo {author} {\bibfnamefont {P.~E.}\ \bibnamefont {Shanahan}} (\bibinfo {collaboration} {NPLQCD Collaboration}),\ }\bibfield  {title} {\bibinfo {title} {Double-$\ensuremath{\beta}$ decay matrix elements from lattice quantum chromodynamics},\ }\href {https://doi.org/10.1103/PhysRevD.96.054505} {\bibfield  {journal} {\bibinfo  {journal} {Phys. Rev. D}\ }\textbf {\bibinfo {volume} {96}},\ \bibinfo {pages} {054505} (\bibinfo {year} {2017})}\BibitemShut {NoStop}%
\bibitem [{\citenamefont {Beane}\ \emph {et~al.}(2015)\citenamefont {Beane}, \citenamefont {Chang}, \citenamefont {Detmold}, \citenamefont {Orginos}, \citenamefont {Parre\~no}, \citenamefont {Savage},\ and\ \citenamefont {Tiburzi}}]{beane2015ab}%
  \BibitemOpen
  \bibfield  {author} {\bibinfo {author} {\bibfnamefont {S.~R.}\ \bibnamefont {Beane}}, \bibinfo {author} {\bibfnamefont {E.}~\bibnamefont {Chang}}, \bibinfo {author} {\bibfnamefont {W.}~\bibnamefont {Detmold}}, \bibinfo {author} {\bibfnamefont {K.}~\bibnamefont {Orginos}}, \bibinfo {author} {\bibfnamefont {A.}~\bibnamefont {Parre\~no}}, \bibinfo {author} {\bibfnamefont {M.~J.}\ \bibnamefont {Savage}},\ and\ \bibinfo {author} {\bibfnamefont {B.~C.}\ \bibnamefont {Tiburzi}} (\bibinfo {collaboration} {NPLQCD Collaboration}),\ }\bibfield  {title} {\bibinfo {title} {Ab initio calculation of the $np\ensuremath{\rightarrow}d\ensuremath{\gamma}$ radiative capture process},\ }\href {https://doi.org/10.1103/PhysRevLett.115.132001} {\bibfield  {journal} {\bibinfo  {journal} {Phys. Rev. Lett.}\ }\textbf {\bibinfo {volume} {115}},\ \bibinfo {pages} {132001} (\bibinfo {year} {2015})}\BibitemShut {NoStop}%
\bibitem [{\citenamefont {Savage}\ \emph {et~al.}(2017)\citenamefont {Savage}, \citenamefont {Shanahan}, \citenamefont {Tiburzi}, \citenamefont {Wagman}, \citenamefont {Winter}, \citenamefont {Beane}, \citenamefont {Chang}, \citenamefont {Davoudi}, \citenamefont {Detmold},\ and\ \citenamefont {Orginos}}]{savage2017proton}%
  \BibitemOpen
  \bibfield  {author} {\bibinfo {author} {\bibfnamefont {M.~J.}\ \bibnamefont {Savage}}, \bibinfo {author} {\bibfnamefont {P.~E.}\ \bibnamefont {Shanahan}}, \bibinfo {author} {\bibfnamefont {B.~C.}\ \bibnamefont {Tiburzi}}, \bibinfo {author} {\bibfnamefont {M.~L.}\ \bibnamefont {Wagman}}, \bibinfo {author} {\bibfnamefont {F.}~\bibnamefont {Winter}}, \bibinfo {author} {\bibfnamefont {S.~R.}\ \bibnamefont {Beane}}, \bibinfo {author} {\bibfnamefont {E.}~\bibnamefont {Chang}}, \bibinfo {author} {\bibfnamefont {Z.}~\bibnamefont {Davoudi}}, \bibinfo {author} {\bibfnamefont {W.}~\bibnamefont {Detmold}},\ and\ \bibinfo {author} {\bibfnamefont {K.}~\bibnamefont {Orginos}} (\bibinfo {collaboration} {NPLQCD Collaboration}),\ }\bibfield  {title} {\bibinfo {title} {Proton-proton fusion and tritium $\ensuremath{\beta}$ decay from lattice quantum chromodynamics},\ }\href {https://doi.org/10.1103/PhysRevLett.119.062002} {\bibfield  {journal} {\bibinfo  {journal} {Phys. Rev. Lett.}\ }\textbf {\bibinfo {volume} {119}},\
  \bibinfo {pages} {062002} (\bibinfo {year} {2017})}\BibitemShut {NoStop}%
\bibitem [{\citenamefont {Moore}(2020)}]{moore2020shear}%
  \BibitemOpen
  \bibfield  {author} {\bibinfo {author} {\bibfnamefont {G.~D.}\ \bibnamefont {Moore}},\ }\href@noop {} {\bibinfo {title} {Shear viscosity in qcd and why it's hard to calculate}} (\bibinfo {year} {2020}),\ \Eprint {https://arxiv.org/abs/2010.15704} {arXiv:2010.15704 [hep-ph]} \BibitemShut {NoStop}%
\bibitem [{\citenamefont {Huang}\ \emph {et~al.}(2020)\citenamefont {Huang}, \citenamefont {Wu}, \citenamefont {Fan},\ and\ \citenamefont {Zhu}}]{huang2020superconducting}%
  \BibitemOpen
  \bibfield  {author} {\bibinfo {author} {\bibfnamefont {H.-L.}\ \bibnamefont {Huang}}, \bibinfo {author} {\bibfnamefont {D.}~\bibnamefont {Wu}}, \bibinfo {author} {\bibfnamefont {D.}~\bibnamefont {Fan}},\ and\ \bibinfo {author} {\bibfnamefont {X.}~\bibnamefont {Zhu}},\ }\bibfield  {title} {\bibinfo {title} {Superconducting quantum computing: a review},\ }\bibfield  {journal} {\bibinfo  {journal} {Science China Information Sciences}\ }\textbf {\bibinfo {volume} {63}},\ \href {https://doi.org/10.1007/s11432-020-2881-9} {10.1007/s11432-020-2881-9} (\bibinfo {year} {2020})\BibitemShut {NoStop}%
\bibitem [{\citenamefont {Bianchetti}\ \emph {et~al.}(2010)\citenamefont {Bianchetti}, \citenamefont {Filipp}, \citenamefont {Baur}, \citenamefont {Fink}, \citenamefont {Lang}, \citenamefont {Steffen}, \citenamefont {Boissonneault}, \citenamefont {Blais},\ and\ \citenamefont {Wallraff}}]{bianchetti2010control}%
  \BibitemOpen
  \bibfield  {author} {\bibinfo {author} {\bibfnamefont {R.}~\bibnamefont {Bianchetti}}, \bibinfo {author} {\bibfnamefont {S.}~\bibnamefont {Filipp}}, \bibinfo {author} {\bibfnamefont {M.}~\bibnamefont {Baur}}, \bibinfo {author} {\bibfnamefont {J.~M.}\ \bibnamefont {Fink}}, \bibinfo {author} {\bibfnamefont {C.}~\bibnamefont {Lang}}, \bibinfo {author} {\bibfnamefont {L.}~\bibnamefont {Steffen}}, \bibinfo {author} {\bibfnamefont {M.}~\bibnamefont {Boissonneault}}, \bibinfo {author} {\bibfnamefont {A.}~\bibnamefont {Blais}},\ and\ \bibinfo {author} {\bibfnamefont {A.}~\bibnamefont {Wallraff}},\ }\bibfield  {title} {\bibinfo {title} {Control and tomography of a three level superconducting artificial atom},\ }\href {https://doi.org/10.1103/PhysRevLett.105.223601} {\bibfield  {journal} {\bibinfo  {journal} {Phys. Rev. Lett.}\ }\textbf {\bibinfo {volume} {105}},\ \bibinfo {pages} {223601} (\bibinfo {year} {2010})}\BibitemShut {NoStop}%
\bibitem [{\citenamefont {Wang}\ \emph {et~al.}(2022)\citenamefont {Wang}, \citenamefont {Gonin}, \citenamefont {Grassellino}, \citenamefont {Kazakov}, \citenamefont {Romanenko}, \citenamefont {Yakovlev},\ and\ \citenamefont {Zorzetti}}]{wang2022high}%
  \BibitemOpen
  \bibfield  {author} {\bibinfo {author} {\bibfnamefont {C.}~\bibnamefont {Wang}}, \bibinfo {author} {\bibfnamefont {I.}~\bibnamefont {Gonin}}, \bibinfo {author} {\bibfnamefont {A.}~\bibnamefont {Grassellino}}, \bibinfo {author} {\bibfnamefont {S.}~\bibnamefont {Kazakov}}, \bibinfo {author} {\bibfnamefont {A.}~\bibnamefont {Romanenko}}, \bibinfo {author} {\bibfnamefont {V.~P.}\ \bibnamefont {Yakovlev}},\ and\ \bibinfo {author} {\bibfnamefont {S.}~\bibnamefont {Zorzetti}},\ }\bibfield  {title} {\bibinfo {title} {High-efficiency microwave-optical quantum transduction based on a cavity electro-optic superconducting system with long coherence time},\ }\href@noop {} {\bibfield  {journal} {\bibinfo  {journal} {npj Quantum Information}\ }\textbf {\bibinfo {volume} {8}},\ \bibinfo {pages} {149} (\bibinfo {year} {2022})}\BibitemShut {NoStop}%
\bibitem [{\citenamefont {Wallraff}\ \emph {et~al.}(2004)\citenamefont {Wallraff}, \citenamefont {Schuster}, \citenamefont {Blais}, \citenamefont {Frunzio}, \citenamefont {Huang}, \citenamefont {Majer}, \citenamefont {Kumar}, \citenamefont {Girvin},\ and\ \citenamefont {Schoelkopf}}]{wallraff2004strong}%
  \BibitemOpen
  \bibfield  {author} {\bibinfo {author} {\bibfnamefont {A.}~\bibnamefont {Wallraff}}, \bibinfo {author} {\bibfnamefont {D.~I.}\ \bibnamefont {Schuster}}, \bibinfo {author} {\bibfnamefont {A.}~\bibnamefont {Blais}}, \bibinfo {author} {\bibfnamefont {L.}~\bibnamefont {Frunzio}}, \bibinfo {author} {\bibfnamefont {R.-S.}\ \bibnamefont {Huang}}, \bibinfo {author} {\bibfnamefont {J.}~\bibnamefont {Majer}}, \bibinfo {author} {\bibfnamefont {S.}~\bibnamefont {Kumar}}, \bibinfo {author} {\bibfnamefont {S.~M.}\ \bibnamefont {Girvin}},\ and\ \bibinfo {author} {\bibfnamefont {R.~J.}\ \bibnamefont {Schoelkopf}},\ }\bibfield  {title} {\bibinfo {title} {Strong coupling of a single photon to a superconducting qubit using circuit quantum electrodynamics},\ }\href@noop {} {\bibfield  {journal} {\bibinfo  {journal} {Nature}\ }\textbf {\bibinfo {volume} {431}},\ \bibinfo {pages} {162} (\bibinfo {year} {2004})}\BibitemShut {NoStop}%
\bibitem [{\citenamefont {Chiorescu}\ \emph {et~al.}(2004)\citenamefont {Chiorescu}, \citenamefont {Bertet}, \citenamefont {Semba}, \citenamefont {Nakamura}, \citenamefont {Harmans},\ and\ \citenamefont {Mooij}}]{chiorescu2004coherent}%
  \BibitemOpen
  \bibfield  {author} {\bibinfo {author} {\bibfnamefont {I.}~\bibnamefont {Chiorescu}}, \bibinfo {author} {\bibfnamefont {P.}~\bibnamefont {Bertet}}, \bibinfo {author} {\bibfnamefont {K.}~\bibnamefont {Semba}}, \bibinfo {author} {\bibfnamefont {Y.}~\bibnamefont {Nakamura}}, \bibinfo {author} {\bibfnamefont {C.~J. P.~M.}\ \bibnamefont {Harmans}},\ and\ \bibinfo {author} {\bibfnamefont {J.~E.}\ \bibnamefont {Mooij}},\ }\bibfield  {title} {\bibinfo {title} {Coherent dynamics of a flux qubit coupled to a harmonic oscillator},\ }\href {https://doi.org/10.1038/nature02831} {\bibfield  {journal} {\bibinfo  {journal} {Nature}\ }\textbf {\bibinfo {volume} {431}},\ \bibinfo {pages} {159} (\bibinfo {year} {2004})}\BibitemShut {NoStop}%
\bibitem [{\citenamefont {Bruzewicz}\ \emph {et~al.}(2019)\citenamefont {Bruzewicz}, \citenamefont {Chiaverini}, \citenamefont {McConnell},\ and\ \citenamefont {Sage}}]{bruzewicz2019trapped}%
  \BibitemOpen
  \bibfield  {author} {\bibinfo {author} {\bibfnamefont {C.~D.}\ \bibnamefont {Bruzewicz}}, \bibinfo {author} {\bibfnamefont {J.}~\bibnamefont {Chiaverini}}, \bibinfo {author} {\bibfnamefont {R.}~\bibnamefont {McConnell}},\ and\ \bibinfo {author} {\bibfnamefont {J.~M.}\ \bibnamefont {Sage}},\ }\bibfield  {title} {\bibinfo {title} {Trapped-ion quantum computing: Progress and challenges},\ }\href {https://doi.org/10.1063/1.5088164} {\bibfield  {journal} {\bibinfo  {journal} {Applied Physics Reviews}\ }\textbf {\bibinfo {volume} {6}},\ \bibinfo {pages} {021314} (\bibinfo {year} {2019})}\BibitemShut {NoStop}%
\bibitem [{\citenamefont {Henriet}\ \emph {et~al.}(2020)\citenamefont {Henriet}, \citenamefont {Beguin}, \citenamefont {Signoles}, \citenamefont {Lahaye}, \citenamefont {Browaeys}, \citenamefont {Reymond},\ and\ \citenamefont {Jurczak}}]{Henriet_2020}%
  \BibitemOpen
  \bibfield  {author} {\bibinfo {author} {\bibfnamefont {L.}~\bibnamefont {Henriet}}, \bibinfo {author} {\bibfnamefont {L.}~\bibnamefont {Beguin}}, \bibinfo {author} {\bibfnamefont {A.}~\bibnamefont {Signoles}}, \bibinfo {author} {\bibfnamefont {T.}~\bibnamefont {Lahaye}}, \bibinfo {author} {\bibfnamefont {A.}~\bibnamefont {Browaeys}}, \bibinfo {author} {\bibfnamefont {G.-O.}\ \bibnamefont {Reymond}},\ and\ \bibinfo {author} {\bibfnamefont {C.}~\bibnamefont {Jurczak}},\ }\bibfield  {title} {\bibinfo {title} {Quantum computing with neutral atoms},\ }\href {https://doi.org/10.22331/q-2020-09-21-327} {\bibfield  {journal} {\bibinfo  {journal} {Quantum}\ }\textbf {\bibinfo {volume} {4}},\ \bibinfo {pages} {327} (\bibinfo {year} {2020})}\BibitemShut {NoStop}%
\bibitem [{\citenamefont {Browaeys}\ and\ \citenamefont {Lahaye}(2020)}]{Browaeys_2020}%
  \BibitemOpen
  \bibfield  {author} {\bibinfo {author} {\bibfnamefont {A.}~\bibnamefont {Browaeys}}\ and\ \bibinfo {author} {\bibfnamefont {T.}~\bibnamefont {Lahaye}},\ }\bibfield  {title} {\bibinfo {title} {Many-body physics with individually controlled rydberg atoms},\ }\href {https://doi.org/10.1038/s41567-019-0733-z} {\bibfield  {journal} {\bibinfo  {journal} {Nature Physics}\ }\textbf {\bibinfo {volume} {16}},\ \bibinfo {pages} {132} (\bibinfo {year} {2020})}\BibitemShut {NoStop}%
\bibitem [{\citenamefont {Barredo}\ \emph {et~al.}(2020)\citenamefont {Barredo}, \citenamefont {Lienhard}, \citenamefont {Scholl}, \citenamefont {de~L{\'{e} }s{\'{e}}leuc}, \citenamefont {Boulier}, \citenamefont {Browaeys},\ and\ \citenamefont {Lahaye}}]{Barredo_2020}%
  \BibitemOpen
  \bibfield  {author} {\bibinfo {author} {\bibfnamefont {D.}~\bibnamefont {Barredo}}, \bibinfo {author} {\bibfnamefont {V.}~\bibnamefont {Lienhard}}, \bibinfo {author} {\bibfnamefont {P.}~\bibnamefont {Scholl}}, \bibinfo {author} {\bibfnamefont {S.}~\bibnamefont {de~L{\'{e} }s{\'{e}}leuc}}, \bibinfo {author} {\bibfnamefont {T.}~\bibnamefont {Boulier}}, \bibinfo {author} {\bibfnamefont {A.}~\bibnamefont {Browaeys}},\ and\ \bibinfo {author} {\bibfnamefont {T.}~\bibnamefont {Lahaye}},\ }\bibfield  {title} {\bibinfo {title} {Three-dimensional trapping of individual rydberg atoms in ponderomotive bottle beam traps},\ }\bibfield  {journal} {\bibinfo  {journal} {Physical Review Letters}\ }\textbf {\bibinfo {volume} {124}},\ \href {https://doi.org/10.1103/physrevlett.124.023201} {10.1103/physrevlett.124.023201} (\bibinfo {year} {2020})\BibitemShut {NoStop}%
\bibitem [{\citenamefont {Bluvstein}\ \emph {et~al.}(2022)\citenamefont {Bluvstein}, \citenamefont {Levine}, \citenamefont {Semeghini}, \citenamefont {Wang}, \citenamefont {Ebadi}, \citenamefont {Kalinowski}, \citenamefont {Keesling}, \citenamefont {Maskara}, \citenamefont {Pichler}, \citenamefont {Greiner}, \citenamefont {Vuleti{\'{c}}},\ and\ \citenamefont {Lukin}}]{bluvstein2022quantum}%
  \BibitemOpen
  \bibfield  {author} {\bibinfo {author} {\bibfnamefont {D.}~\bibnamefont {Bluvstein}}, \bibinfo {author} {\bibfnamefont {H.}~\bibnamefont {Levine}}, \bibinfo {author} {\bibfnamefont {G.}~\bibnamefont {Semeghini}}, \bibinfo {author} {\bibfnamefont {T.~T.}\ \bibnamefont {Wang}}, \bibinfo {author} {\bibfnamefont {S.}~\bibnamefont {Ebadi}}, \bibinfo {author} {\bibfnamefont {M.}~\bibnamefont {Kalinowski}}, \bibinfo {author} {\bibfnamefont {A.}~\bibnamefont {Keesling}}, \bibinfo {author} {\bibfnamefont {N.}~\bibnamefont {Maskara}}, \bibinfo {author} {\bibfnamefont {H.}~\bibnamefont {Pichler}}, \bibinfo {author} {\bibfnamefont {M.}~\bibnamefont {Greiner}}, \bibinfo {author} {\bibfnamefont {V.}~\bibnamefont {Vuleti{\'{c}}}},\ and\ \bibinfo {author} {\bibfnamefont {M.~D.}\ \bibnamefont {Lukin}},\ }\bibfield  {title} {\bibinfo {title} {A quantum processor based on coherent transport of entangled atom arrays},\ }\href {https://doi.org/10.1038/s41586-022-04592-6} {\bibfield  {journal} {\bibinfo  {journal} {Nature}\
  }\textbf {\bibinfo {volume} {604}},\ \bibinfo {pages} {451} (\bibinfo {year} {2022})}\BibitemShut {NoStop}%
\bibitem [{\citenamefont {Bluvstein}\ \emph {et~al.}(2023)\citenamefont {Bluvstein}, \citenamefont {Evered}, \citenamefont {Geim}, \citenamefont {Li}, \citenamefont {Zhou}, \citenamefont {Manovitz}, \citenamefont {Ebadi}, \citenamefont {Cain}, \citenamefont {Kalinowski}, \citenamefont {Hangleiter} \emph {et~al.}}]{bluvstein2023logical}%
  \BibitemOpen
  \bibfield  {author} {\bibinfo {author} {\bibfnamefont {D.}~\bibnamefont {Bluvstein}}, \bibinfo {author} {\bibfnamefont {S.~J.}\ \bibnamefont {Evered}}, \bibinfo {author} {\bibfnamefont {A.~A.}\ \bibnamefont {Geim}}, \bibinfo {author} {\bibfnamefont {S.~H.}\ \bibnamefont {Li}}, \bibinfo {author} {\bibfnamefont {H.}~\bibnamefont {Zhou}}, \bibinfo {author} {\bibfnamefont {T.}~\bibnamefont {Manovitz}}, \bibinfo {author} {\bibfnamefont {S.}~\bibnamefont {Ebadi}}, \bibinfo {author} {\bibfnamefont {M.}~\bibnamefont {Cain}}, \bibinfo {author} {\bibfnamefont {M.}~\bibnamefont {Kalinowski}}, \bibinfo {author} {\bibfnamefont {D.}~\bibnamefont {Hangleiter}}, \emph {et~al.},\ }\bibfield  {title} {\bibinfo {title} {Logical quantum processor based on reconfigurable atom arrays},\ }\href {https://doi.org/10.1038/s41586-023-06927-3} {\bibfield  {journal} {\bibinfo  {journal} {Nature}\ ,\ \bibinfo {pages} {1}} (\bibinfo {year} {2023})}\BibitemShut {NoStop}%
\bibitem [{\citenamefont {Feynman}(1981)}]{feynman1981simulating}%
  \BibitemOpen
  \bibfield  {author} {\bibinfo {author} {\bibfnamefont {R.~P.}\ \bibnamefont {Feynman}},\ }\bibfield  {title} {\bibinfo {title} {Simulating physics with computers, 1981},\ }\href@noop {} {\bibfield  {journal} {\bibinfo  {journal} {International Journal of Theoretical Physics}\ }\textbf {\bibinfo {volume} {21}} (\bibinfo {year} {1981})}\BibitemShut {NoStop}%
\bibitem [{\citenamefont {Bauer}\ \emph {et~al.}(2023)\citenamefont {Bauer}, \citenamefont {Davoudi}, \citenamefont {Balantekin}, \citenamefont {Bhattacharya}, \citenamefont {Carena}, \citenamefont {de~Jong}, \citenamefont {Draper}, \citenamefont {El-Khadra}, \citenamefont {Gemelke}, \citenamefont {Hanada}, \citenamefont {Kharzeev}, \citenamefont {Lamm}, \citenamefont {Li}, \citenamefont {Liu}, \citenamefont {Lukin}, \citenamefont {Meurice}, \citenamefont {Monroe}, \citenamefont {Nachman}, \citenamefont {Pagano}, \citenamefont {Preskill}, \citenamefont {Rinaldi}, \citenamefont {Roggero}, \citenamefont {Santiago}, \citenamefont {Savage}, \citenamefont {Siddiqi}, \citenamefont {Siopsis}, \citenamefont {Van~Zanten}, \citenamefont {Wiebe}, \citenamefont {Yamauchi}, \citenamefont {Yeter-Aydeniz},\ and\ \citenamefont {Zorzetti}}]{Bauer_2023}%
  \BibitemOpen
  \bibfield  {author} {\bibinfo {author} {\bibfnamefont {C.~W.}\ \bibnamefont {Bauer}}, \bibinfo {author} {\bibfnamefont {Z.}~\bibnamefont {Davoudi}}, \bibinfo {author} {\bibfnamefont {A.~B.}\ \bibnamefont {Balantekin}}, \bibinfo {author} {\bibfnamefont {T.}~\bibnamefont {Bhattacharya}}, \bibinfo {author} {\bibfnamefont {M.}~\bibnamefont {Carena}}, \bibinfo {author} {\bibfnamefont {W.~A.}\ \bibnamefont {de~Jong}}, \bibinfo {author} {\bibfnamefont {P.}~\bibnamefont {Draper}}, \bibinfo {author} {\bibfnamefont {A.}~\bibnamefont {El-Khadra}}, \bibinfo {author} {\bibfnamefont {N.}~\bibnamefont {Gemelke}}, \bibinfo {author} {\bibfnamefont {M.}~\bibnamefont {Hanada}}, \bibinfo {author} {\bibfnamefont {D.}~\bibnamefont {Kharzeev}}, \bibinfo {author} {\bibfnamefont {H.}~\bibnamefont {Lamm}}, \bibinfo {author} {\bibfnamefont {Y.-Y.}\ \bibnamefont {Li}}, \bibinfo {author} {\bibfnamefont {J.}~\bibnamefont {Liu}}, \bibinfo {author} {\bibfnamefont {M.}~\bibnamefont {Lukin}}, \bibinfo {author} {\bibfnamefont
  {Y.}~\bibnamefont {Meurice}}, \bibinfo {author} {\bibfnamefont {C.}~\bibnamefont {Monroe}}, \bibinfo {author} {\bibfnamefont {B.}~\bibnamefont {Nachman}}, \bibinfo {author} {\bibfnamefont {G.}~\bibnamefont {Pagano}}, \bibinfo {author} {\bibfnamefont {J.}~\bibnamefont {Preskill}}, \bibinfo {author} {\bibfnamefont {E.}~\bibnamefont {Rinaldi}}, \bibinfo {author} {\bibfnamefont {A.}~\bibnamefont {Roggero}}, \bibinfo {author} {\bibfnamefont {D.~I.}\ \bibnamefont {Santiago}}, \bibinfo {author} {\bibfnamefont {M.~J.}\ \bibnamefont {Savage}}, \bibinfo {author} {\bibfnamefont {I.}~\bibnamefont {Siddiqi}}, \bibinfo {author} {\bibfnamefont {G.}~\bibnamefont {Siopsis}}, \bibinfo {author} {\bibfnamefont {D.}~\bibnamefont {Van~Zanten}}, \bibinfo {author} {\bibfnamefont {N.}~\bibnamefont {Wiebe}}, \bibinfo {author} {\bibfnamefont {Y.}~\bibnamefont {Yamauchi}}, \bibinfo {author} {\bibfnamefont {K.}~\bibnamefont {Yeter-Aydeniz}},\ and\ \bibinfo {author} {\bibfnamefont {S.}~\bibnamefont {Zorzetti}},\ }\bibfield  {title}
  {\bibinfo {title} {Quantum simulation for high-energy physics},\ }\bibfield  {journal} {\bibinfo  {journal} {PRX Quantum}\ }\textbf {\bibinfo {volume} {4}},\ \href {https://doi.org/10.1103/prxquantum.4.027001} {10.1103/prxquantum.4.027001} (\bibinfo {year} {2023})\BibitemShut {NoStop}%
\bibitem [{\citenamefont {Humble}\ \emph {et~al.}(2022{\natexlab{a}})\citenamefont {Humble}, \citenamefont {Delgado}, \citenamefont {Pooser}, \citenamefont {Seck}, \citenamefont {Bennink}, \citenamefont {Leyton-Ortega}, \citenamefont {Wang}, \citenamefont {Dumitrescu}, \citenamefont {Morris}, \citenamefont {Hamilton}, \citenamefont {Lyakh}, \citenamefont {Date}, \citenamefont {Wang}, \citenamefont {Peters}, \citenamefont {Evans}, \citenamefont {Demarteau}, \citenamefont {McCaskey}, \citenamefont {Nguyen}, \citenamefont {Clark}, \citenamefont {Reville}, \citenamefont {Meglio}, \citenamefont {Grossi}, \citenamefont {Vallecorsa}, \citenamefont {Borras}, \citenamefont {Jansen},\ and\ \citenamefont {Krücker}}]{humble2022snowmass}%
  \BibitemOpen
  \bibfield  {author} {\bibinfo {author} {\bibfnamefont {T.~S.}\ \bibnamefont {Humble}}, \bibinfo {author} {\bibfnamefont {A.}~\bibnamefont {Delgado}}, \bibinfo {author} {\bibfnamefont {R.}~\bibnamefont {Pooser}}, \bibinfo {author} {\bibfnamefont {C.}~\bibnamefont {Seck}}, \bibinfo {author} {\bibfnamefont {R.}~\bibnamefont {Bennink}}, \bibinfo {author} {\bibfnamefont {V.}~\bibnamefont {Leyton-Ortega}}, \bibinfo {author} {\bibfnamefont {C.~C.~J.}\ \bibnamefont {Wang}}, \bibinfo {author} {\bibfnamefont {E.}~\bibnamefont {Dumitrescu}}, \bibinfo {author} {\bibfnamefont {T.}~\bibnamefont {Morris}}, \bibinfo {author} {\bibfnamefont {K.}~\bibnamefont {Hamilton}}, \bibinfo {author} {\bibfnamefont {D.}~\bibnamefont {Lyakh}}, \bibinfo {author} {\bibfnamefont {P.}~\bibnamefont {Date}}, \bibinfo {author} {\bibfnamefont {Y.}~\bibnamefont {Wang}}, \bibinfo {author} {\bibfnamefont {N.~A.}\ \bibnamefont {Peters}}, \bibinfo {author} {\bibfnamefont {K.~J.}\ \bibnamefont {Evans}}, \bibinfo {author} {\bibfnamefont
  {M.}~\bibnamefont {Demarteau}}, \bibinfo {author} {\bibfnamefont {A.}~\bibnamefont {McCaskey}}, \bibinfo {author} {\bibfnamefont {T.}~\bibnamefont {Nguyen}}, \bibinfo {author} {\bibfnamefont {S.}~\bibnamefont {Clark}}, \bibinfo {author} {\bibfnamefont {M.}~\bibnamefont {Reville}}, \bibinfo {author} {\bibfnamefont {A.~D.}\ \bibnamefont {Meglio}}, \bibinfo {author} {\bibfnamefont {M.}~\bibnamefont {Grossi}}, \bibinfo {author} {\bibfnamefont {S.}~\bibnamefont {Vallecorsa}}, \bibinfo {author} {\bibfnamefont {K.}~\bibnamefont {Borras}}, \bibinfo {author} {\bibfnamefont {K.}~\bibnamefont {Jansen}},\ and\ \bibinfo {author} {\bibfnamefont {D.}~\bibnamefont {Krücker}},\ }\href@noop {} {\bibinfo {title} {Snowmass white paper: Quantum computing systems and software for high-energy physics research}} (\bibinfo {year} {2022}{\natexlab{a}}),\ \Eprint {https://arxiv.org/abs/2203.07091} {arXiv:2203.07091 [quant-ph]} \BibitemShut {NoStop}%
\bibitem [{\citenamefont {Humble}\ \emph {et~al.}(2022{\natexlab{b}})\citenamefont {Humble}, \citenamefont {Perdue},\ and\ \citenamefont {Savage}}]{humble2022snowmass2}%
  \BibitemOpen
  \bibfield  {author} {\bibinfo {author} {\bibfnamefont {T.~S.}\ \bibnamefont {Humble}}, \bibinfo {author} {\bibfnamefont {G.~N.}\ \bibnamefont {Perdue}},\ and\ \bibinfo {author} {\bibfnamefont {M.~J.}\ \bibnamefont {Savage}},\ }\href@noop {} {\bibinfo {title} {Snowmass computational frontier: Topical group report on quantum computing}} (\bibinfo {year} {2022}{\natexlab{b}}),\ \Eprint {https://arxiv.org/abs/2209.06786} {arXiv:2209.06786 [quant-ph]} \BibitemShut {NoStop}%
\bibitem [{\citenamefont {Beck}\ \emph {et~al.}(2023)\citenamefont {Beck}, \citenamefont {Carlson}, \citenamefont {Davoudi}, \citenamefont {Formaggio}, \citenamefont {Quaglioni}, \citenamefont {Savage}, \citenamefont {Barata}, \citenamefont {Bhattacharya}, \citenamefont {Bishof}, \citenamefont {Cloet}, \citenamefont {Delgado}, \citenamefont {DeMarco}, \citenamefont {Fink}, \citenamefont {Florio}, \citenamefont {Francois}, \citenamefont {Grabowska}, \citenamefont {Hoogerheide}, \citenamefont {Huang}, \citenamefont {Ikeda}, \citenamefont {Illa}, \citenamefont {Joo}, \citenamefont {Kharzeev}, \citenamefont {Kowalski}, \citenamefont {Lai}, \citenamefont {Leach}, \citenamefont {Loer}, \citenamefont {Low}, \citenamefont {Martin}, \citenamefont {Moore}, \citenamefont {Mehen}, \citenamefont {Mueller}, \citenamefont {Mulligan}, \citenamefont {Mumm}, \citenamefont {Pederiva}, \citenamefont {Pisarski}, \citenamefont {Ploskon}, \citenamefont {Reddy}, \citenamefont {Rupak}, \citenamefont {Singh}, \citenamefont {Singh},
  \citenamefont {Stetcu}, \citenamefont {Stryker}, \citenamefont {Szypryt}, \citenamefont {Valgushev}, \citenamefont {VanDevender}, \citenamefont {Watkins}, \citenamefont {Wilson}, \citenamefont {Yao}, \citenamefont {Afanasev}, \citenamefont {Balantekin}, \citenamefont {Baroni}, \citenamefont {Bunker}, \citenamefont {Chakraborty}, \citenamefont {Chernyshev}, \citenamefont {Cirigliano}, \citenamefont {Clark}, \citenamefont {Dhiman}, \citenamefont {Du}, \citenamefont {Dutta}, \citenamefont {Edwards}, \citenamefont {Flores}, \citenamefont {Galindo-Uribarri}, \citenamefont {Ruiz}, \citenamefont {Gueorguiev}, \citenamefont {Guo}, \citenamefont {Hansen}, \citenamefont {Hernandez}, \citenamefont {Hattori}, \citenamefont {Hauke}, \citenamefont {Hjorth-Jensen}, \citenamefont {Jankowski}, \citenamefont {Johnson}, \citenamefont {Lacroix}, \citenamefont {Lee}, \citenamefont {Lin}, \citenamefont {Liu}, \citenamefont {Llanes-Estrada}, \citenamefont {Looney}, \citenamefont {Lukin}, \citenamefont {Mercenne}, \citenamefont
  {Miller}, \citenamefont {Mottola}, \citenamefont {Mueller}, \citenamefont {Nachman}, \citenamefont {Negele}, \citenamefont {Orrell}, \citenamefont {Patwardhan}, \citenamefont {Phillips}, \citenamefont {Poole}, \citenamefont {Qualters}, \citenamefont {Rumore}, \citenamefont {Schaefer}, \citenamefont {Scott}, \citenamefont {Singh}, \citenamefont {Vary}, \citenamefont {Galvez-Viruet}, \citenamefont {Wendt}, \citenamefont {Xing}, \citenamefont {Yang}, \citenamefont {Young},\ and\ \citenamefont {Zhao}}]{beck2023quantum}%
  \BibitemOpen
  \bibfield  {author} {\bibinfo {author} {\bibfnamefont {D.}~\bibnamefont {Beck}}, \bibinfo {author} {\bibfnamefont {J.}~\bibnamefont {Carlson}}, \bibinfo {author} {\bibfnamefont {Z.}~\bibnamefont {Davoudi}}, \bibinfo {author} {\bibfnamefont {J.}~\bibnamefont {Formaggio}}, \bibinfo {author} {\bibfnamefont {S.}~\bibnamefont {Quaglioni}}, \bibinfo {author} {\bibfnamefont {M.}~\bibnamefont {Savage}}, \bibinfo {author} {\bibfnamefont {J.}~\bibnamefont {Barata}}, \bibinfo {author} {\bibfnamefont {T.}~\bibnamefont {Bhattacharya}}, \bibinfo {author} {\bibfnamefont {M.}~\bibnamefont {Bishof}}, \bibinfo {author} {\bibfnamefont {I.}~\bibnamefont {Cloet}}, \bibinfo {author} {\bibfnamefont {A.}~\bibnamefont {Delgado}}, \bibinfo {author} {\bibfnamefont {M.}~\bibnamefont {DeMarco}}, \bibinfo {author} {\bibfnamefont {C.}~\bibnamefont {Fink}}, \bibinfo {author} {\bibfnamefont {A.}~\bibnamefont {Florio}}, \bibinfo {author} {\bibfnamefont {M.}~\bibnamefont {Francois}}, \bibinfo {author} {\bibfnamefont {D.}~\bibnamefont
  {Grabowska}}, \bibinfo {author} {\bibfnamefont {S.}~\bibnamefont {Hoogerheide}}, \bibinfo {author} {\bibfnamefont {M.}~\bibnamefont {Huang}}, \bibinfo {author} {\bibfnamefont {K.}~\bibnamefont {Ikeda}}, \bibinfo {author} {\bibfnamefont {M.}~\bibnamefont {Illa}}, \bibinfo {author} {\bibfnamefont {K.}~\bibnamefont {Joo}}, \bibinfo {author} {\bibfnamefont {D.}~\bibnamefont {Kharzeev}}, \bibinfo {author} {\bibfnamefont {K.}~\bibnamefont {Kowalski}}, \bibinfo {author} {\bibfnamefont {W.~K.}\ \bibnamefont {Lai}}, \bibinfo {author} {\bibfnamefont {K.}~\bibnamefont {Leach}}, \bibinfo {author} {\bibfnamefont {B.}~\bibnamefont {Loer}}, \bibinfo {author} {\bibfnamefont {I.}~\bibnamefont {Low}}, \bibinfo {author} {\bibfnamefont {J.}~\bibnamefont {Martin}}, \bibinfo {author} {\bibfnamefont {D.}~\bibnamefont {Moore}}, \bibinfo {author} {\bibfnamefont {T.}~\bibnamefont {Mehen}}, \bibinfo {author} {\bibfnamefont {N.}~\bibnamefont {Mueller}}, \bibinfo {author} {\bibfnamefont {J.}~\bibnamefont {Mulligan}}, \bibinfo {author}
  {\bibfnamefont {P.}~\bibnamefont {Mumm}}, \bibinfo {author} {\bibfnamefont {F.}~\bibnamefont {Pederiva}}, \bibinfo {author} {\bibfnamefont {R.}~\bibnamefont {Pisarski}}, \bibinfo {author} {\bibfnamefont {M.}~\bibnamefont {Ploskon}}, \bibinfo {author} {\bibfnamefont {S.}~\bibnamefont {Reddy}}, \bibinfo {author} {\bibfnamefont {G.}~\bibnamefont {Rupak}}, \bibinfo {author} {\bibfnamefont {H.}~\bibnamefont {Singh}}, \bibinfo {author} {\bibfnamefont {M.}~\bibnamefont {Singh}}, \bibinfo {author} {\bibfnamefont {I.}~\bibnamefont {Stetcu}}, \bibinfo {author} {\bibfnamefont {J.}~\bibnamefont {Stryker}}, \bibinfo {author} {\bibfnamefont {P.}~\bibnamefont {Szypryt}}, \bibinfo {author} {\bibfnamefont {S.}~\bibnamefont {Valgushev}}, \bibinfo {author} {\bibfnamefont {B.}~\bibnamefont {VanDevender}}, \bibinfo {author} {\bibfnamefont {S.}~\bibnamefont {Watkins}}, \bibinfo {author} {\bibfnamefont {C.}~\bibnamefont {Wilson}}, \bibinfo {author} {\bibfnamefont {X.}~\bibnamefont {Yao}}, \bibinfo {author} {\bibfnamefont
  {A.}~\bibnamefont {Afanasev}}, \bibinfo {author} {\bibfnamefont {A.~B.}\ \bibnamefont {Balantekin}}, \bibinfo {author} {\bibfnamefont {A.}~\bibnamefont {Baroni}}, \bibinfo {author} {\bibfnamefont {R.}~\bibnamefont {Bunker}}, \bibinfo {author} {\bibfnamefont {B.}~\bibnamefont {Chakraborty}}, \bibinfo {author} {\bibfnamefont {I.}~\bibnamefont {Chernyshev}}, \bibinfo {author} {\bibfnamefont {V.}~\bibnamefont {Cirigliano}}, \bibinfo {author} {\bibfnamefont {B.}~\bibnamefont {Clark}}, \bibinfo {author} {\bibfnamefont {S.~K.}\ \bibnamefont {Dhiman}}, \bibinfo {author} {\bibfnamefont {W.}~\bibnamefont {Du}}, \bibinfo {author} {\bibfnamefont {D.}~\bibnamefont {Dutta}}, \bibinfo {author} {\bibfnamefont {R.}~\bibnamefont {Edwards}}, \bibinfo {author} {\bibfnamefont {A.}~\bibnamefont {Flores}}, \bibinfo {author} {\bibfnamefont {A.}~\bibnamefont {Galindo-Uribarri}}, \bibinfo {author} {\bibfnamefont {R.~F.~G.}\ \bibnamefont {Ruiz}}, \bibinfo {author} {\bibfnamefont {V.}~\bibnamefont {Gueorguiev}}, \bibinfo {author}
  {\bibfnamefont {F.}~\bibnamefont {Guo}}, \bibinfo {author} {\bibfnamefont {E.}~\bibnamefont {Hansen}}, \bibinfo {author} {\bibfnamefont {H.}~\bibnamefont {Hernandez}}, \bibinfo {author} {\bibfnamefont {K.}~\bibnamefont {Hattori}}, \bibinfo {author} {\bibfnamefont {P.}~\bibnamefont {Hauke}}, \bibinfo {author} {\bibfnamefont {M.}~\bibnamefont {Hjorth-Jensen}}, \bibinfo {author} {\bibfnamefont {K.}~\bibnamefont {Jankowski}}, \bibinfo {author} {\bibfnamefont {C.}~\bibnamefont {Johnson}}, \bibinfo {author} {\bibfnamefont {D.}~\bibnamefont {Lacroix}}, \bibinfo {author} {\bibfnamefont {D.}~\bibnamefont {Lee}}, \bibinfo {author} {\bibfnamefont {H.-W.}\ \bibnamefont {Lin}}, \bibinfo {author} {\bibfnamefont {X.}~\bibnamefont {Liu}}, \bibinfo {author} {\bibfnamefont {F.~J.}\ \bibnamefont {Llanes-Estrada}}, \bibinfo {author} {\bibfnamefont {J.}~\bibnamefont {Looney}}, \bibinfo {author} {\bibfnamefont {M.}~\bibnamefont {Lukin}}, \bibinfo {author} {\bibfnamefont {A.}~\bibnamefont {Mercenne}}, \bibinfo {author}
  {\bibfnamefont {J.}~\bibnamefont {Miller}}, \bibinfo {author} {\bibfnamefont {E.}~\bibnamefont {Mottola}}, \bibinfo {author} {\bibfnamefont {B.}~\bibnamefont {Mueller}}, \bibinfo {author} {\bibfnamefont {B.}~\bibnamefont {Nachman}}, \bibinfo {author} {\bibfnamefont {J.}~\bibnamefont {Negele}}, \bibinfo {author} {\bibfnamefont {J.}~\bibnamefont {Orrell}}, \bibinfo {author} {\bibfnamefont {A.}~\bibnamefont {Patwardhan}}, \bibinfo {author} {\bibfnamefont {D.}~\bibnamefont {Phillips}}, \bibinfo {author} {\bibfnamefont {S.}~\bibnamefont {Poole}}, \bibinfo {author} {\bibfnamefont {I.}~\bibnamefont {Qualters}}, \bibinfo {author} {\bibfnamefont {M.}~\bibnamefont {Rumore}}, \bibinfo {author} {\bibfnamefont {T.}~\bibnamefont {Schaefer}}, \bibinfo {author} {\bibfnamefont {J.}~\bibnamefont {Scott}}, \bibinfo {author} {\bibfnamefont {R.}~\bibnamefont {Singh}}, \bibinfo {author} {\bibfnamefont {J.}~\bibnamefont {Vary}}, \bibinfo {author} {\bibfnamefont {J.-J.}\ \bibnamefont {Galvez-Viruet}}, \bibinfo {author}
  {\bibfnamefont {K.}~\bibnamefont {Wendt}}, \bibinfo {author} {\bibfnamefont {H.}~\bibnamefont {Xing}}, \bibinfo {author} {\bibfnamefont {L.}~\bibnamefont {Yang}}, \bibinfo {author} {\bibfnamefont {G.}~\bibnamefont {Young}},\ and\ \bibinfo {author} {\bibfnamefont {F.}~\bibnamefont {Zhao}},\ }\href@noop {} {\bibinfo {title} {Quantum information science and technology for nuclear physics. input into u.s. long-range planning, 2023}} (\bibinfo {year} {2023}),\ \Eprint {https://arxiv.org/abs/2303.00113} {arXiv:2303.00113 [nucl-ex]} \BibitemShut {NoStop}%
\bibitem [{\citenamefont {Di~Meglio}\ \emph {et~al.}(2024)\citenamefont {Di~Meglio}, \citenamefont {Jansen}, \citenamefont {Tavernelli}, \citenamefont {Alexandrou}, \citenamefont {Arunachalam}, \citenamefont {Bauer}, \citenamefont {Borras}, \citenamefont {Carrazza}, \citenamefont {Crippa}, \citenamefont {Croft}, \citenamefont {de~Putter}, \citenamefont {Delgado}, \citenamefont {Dunjko}, \citenamefont {Egger}, \citenamefont {Fern\'andez-Combarro}, \citenamefont {Fuchs}, \citenamefont {Funcke}, \citenamefont {Gonz\'alez-Cuadra}, \citenamefont {Grossi}, \citenamefont {Halimeh}, \citenamefont {Holmes}, \citenamefont {K\"uhn}, \citenamefont {Lacroix}, \citenamefont {Lewis}, \citenamefont {Lucchesi}, \citenamefont {Martinez}, \citenamefont {Meloni}, \citenamefont {Mezzacapo}, \citenamefont {Montangero}, \citenamefont {Nagano}, \citenamefont {Pascuzzi}, \citenamefont {Radescu}, \citenamefont {Ortega}, \citenamefont {Roggero}, \citenamefont {Schuhmacher}, \citenamefont {Seixas}, \citenamefont {Silvi}, \citenamefont
  {Spentzouris}, \citenamefont {Tacchino}, \citenamefont {Temme}, \citenamefont {Terashi}, \citenamefont {Tura}, \citenamefont {T\"uys\"uz}, \citenamefont {Vallecorsa}, \citenamefont {Wiese}, \citenamefont {Yoo},\ and\ \citenamefont {Zhang}}]{dimeglio2023quantum}%
  \BibitemOpen
  \bibfield  {author} {\bibinfo {author} {\bibfnamefont {A.}~\bibnamefont {Di~Meglio}}, \bibinfo {author} {\bibfnamefont {K.}~\bibnamefont {Jansen}}, \bibinfo {author} {\bibfnamefont {I.}~\bibnamefont {Tavernelli}}, \bibinfo {author} {\bibfnamefont {C.}~\bibnamefont {Alexandrou}}, \bibinfo {author} {\bibfnamefont {S.}~\bibnamefont {Arunachalam}}, \bibinfo {author} {\bibfnamefont {C.~W.}\ \bibnamefont {Bauer}}, \bibinfo {author} {\bibfnamefont {K.}~\bibnamefont {Borras}}, \bibinfo {author} {\bibfnamefont {S.}~\bibnamefont {Carrazza}}, \bibinfo {author} {\bibfnamefont {A.}~\bibnamefont {Crippa}}, \bibinfo {author} {\bibfnamefont {V.}~\bibnamefont {Croft}}, \bibinfo {author} {\bibfnamefont {R.}~\bibnamefont {de~Putter}}, \bibinfo {author} {\bibfnamefont {A.}~\bibnamefont {Delgado}}, \bibinfo {author} {\bibfnamefont {V.}~\bibnamefont {Dunjko}}, \bibinfo {author} {\bibfnamefont {D.~J.}\ \bibnamefont {Egger}}, \bibinfo {author} {\bibfnamefont {E.}~\bibnamefont {Fern\'andez-Combarro}}, \bibinfo {author}
  {\bibfnamefont {E.}~\bibnamefont {Fuchs}}, \bibinfo {author} {\bibfnamefont {L.}~\bibnamefont {Funcke}}, \bibinfo {author} {\bibfnamefont {D.}~\bibnamefont {Gonz\'alez-Cuadra}}, \bibinfo {author} {\bibfnamefont {M.}~\bibnamefont {Grossi}}, \bibinfo {author} {\bibfnamefont {J.~C.}\ \bibnamefont {Halimeh}}, \bibinfo {author} {\bibfnamefont {Z.}~\bibnamefont {Holmes}}, \bibinfo {author} {\bibfnamefont {S.}~\bibnamefont {K\"uhn}}, \bibinfo {author} {\bibfnamefont {D.}~\bibnamefont {Lacroix}}, \bibinfo {author} {\bibfnamefont {R.}~\bibnamefont {Lewis}}, \bibinfo {author} {\bibfnamefont {D.}~\bibnamefont {Lucchesi}}, \bibinfo {author} {\bibfnamefont {M.~L.}\ \bibnamefont {Martinez}}, \bibinfo {author} {\bibfnamefont {F.}~\bibnamefont {Meloni}}, \bibinfo {author} {\bibfnamefont {A.}~\bibnamefont {Mezzacapo}}, \bibinfo {author} {\bibfnamefont {S.}~\bibnamefont {Montangero}}, \bibinfo {author} {\bibfnamefont {L.}~\bibnamefont {Nagano}}, \bibinfo {author} {\bibfnamefont {V.~R.}\ \bibnamefont {Pascuzzi}}, \bibinfo
  {author} {\bibfnamefont {V.}~\bibnamefont {Radescu}}, \bibinfo {author} {\bibfnamefont {E.~R.}\ \bibnamefont {Ortega}}, \bibinfo {author} {\bibfnamefont {A.}~\bibnamefont {Roggero}}, \bibinfo {author} {\bibfnamefont {J.}~\bibnamefont {Schuhmacher}}, \bibinfo {author} {\bibfnamefont {J.}~\bibnamefont {Seixas}}, \bibinfo {author} {\bibfnamefont {P.}~\bibnamefont {Silvi}}, \bibinfo {author} {\bibfnamefont {P.}~\bibnamefont {Spentzouris}}, \bibinfo {author} {\bibfnamefont {F.}~\bibnamefont {Tacchino}}, \bibinfo {author} {\bibfnamefont {K.}~\bibnamefont {Temme}}, \bibinfo {author} {\bibfnamefont {K.}~\bibnamefont {Terashi}}, \bibinfo {author} {\bibfnamefont {J.}~\bibnamefont {Tura}}, \bibinfo {author} {\bibfnamefont {C.}~\bibnamefont {T\"uys\"uz}}, \bibinfo {author} {\bibfnamefont {S.}~\bibnamefont {Vallecorsa}}, \bibinfo {author} {\bibfnamefont {U.-J.}\ \bibnamefont {Wiese}}, \bibinfo {author} {\bibfnamefont {S.}~\bibnamefont {Yoo}},\ and\ \bibinfo {author} {\bibfnamefont {J.}~\bibnamefont {Zhang}},\ }\bibfield
   {title} {\bibinfo {title} {Quantum computing for high-energy physics: State of the art and challenges},\ }\href {https://doi.org/10.1103/PRXQuantum.5.037001} {\bibfield  {journal} {\bibinfo  {journal} {PRX Quantum}\ }\textbf {\bibinfo {volume} {5}},\ \bibinfo {pages} {037001} (\bibinfo {year} {2024})}\BibitemShut {NoStop}%
\bibitem [{\citenamefont {Nielsen}\ and\ \citenamefont {Chuang}(2011)}]{nielsen2001quantum}%
  \BibitemOpen
  \bibfield  {author} {\bibinfo {author} {\bibfnamefont {M.~A.}\ \bibnamefont {Nielsen}}\ and\ \bibinfo {author} {\bibfnamefont {I.~L.}\ \bibnamefont {Chuang}},\ }\href@noop {} {\emph {\bibinfo {title} {Quantum Computation and Quantum Information: 10th Anniversary Edition}}},\ \bibinfo {edition} {10th}\ ed.\ (\bibinfo  {publisher} {Cambridge University Press},\ \bibinfo {address} {New York, NY, USA},\ \bibinfo {year} {2011})\BibitemShut {NoStop}%
\bibitem [{\citenamefont {Alexandru}\ \emph {et~al.}(2019)\citenamefont {Alexandru}, \citenamefont {Bedaque}, \citenamefont {Harmalkar}, \citenamefont {Lamm}, \citenamefont {Lawrence},\ and\ \citenamefont {Warrington}}]{alexandru2019gluon}%
  \BibitemOpen
  \bibfield  {author} {\bibinfo {author} {\bibfnamefont {A.}~\bibnamefont {Alexandru}}, \bibinfo {author} {\bibfnamefont {P.~F.}\ \bibnamefont {Bedaque}}, \bibinfo {author} {\bibfnamefont {S.}~\bibnamefont {Harmalkar}}, \bibinfo {author} {\bibfnamefont {H.}~\bibnamefont {Lamm}}, \bibinfo {author} {\bibfnamefont {S.}~\bibnamefont {Lawrence}},\ and\ \bibinfo {author} {\bibfnamefont {N.~C.}\ \bibnamefont {Warrington}} (\bibinfo {collaboration} {NuQS Collaboration}),\ }\bibfield  {title} {\bibinfo {title} {Gluon field digitization for quantum computers},\ }\href {https://doi.org/10.1103/PhysRevD.100.114501} {\bibfield  {journal} {\bibinfo  {journal} {Phys. Rev. D}\ }\textbf {\bibinfo {volume} {100}},\ \bibinfo {pages} {114501} (\bibinfo {year} {2019})}\BibitemShut {NoStop}%
\bibitem [{\citenamefont {Lamm}\ \emph {et~al.}(2019)\citenamefont {Lamm}, \citenamefont {Lawrence},\ and\ \citenamefont {Yamauchi}}]{lamm2019general}%
  \BibitemOpen
  \bibfield  {author} {\bibinfo {author} {\bibfnamefont {H.}~\bibnamefont {Lamm}}, \bibinfo {author} {\bibfnamefont {S.}~\bibnamefont {Lawrence}},\ and\ \bibinfo {author} {\bibfnamefont {Y.}~\bibnamefont {Yamauchi}} (\bibinfo {collaboration} {NuQS Collaboration}),\ }\bibfield  {title} {\bibinfo {title} {General methods for digital quantum simulation of gauge theories},\ }\href {https://doi.org/10.1103/PhysRevD.100.034518} {\bibfield  {journal} {\bibinfo  {journal} {Phys. Rev. D}\ }\textbf {\bibinfo {volume} {100}},\ \bibinfo {pages} {034518} (\bibinfo {year} {2019})}\BibitemShut {NoStop}%
\bibitem [{\citenamefont {Ji}\ \emph {et~al.}(2020)\citenamefont {Ji}, \citenamefont {Lamm},\ and\ \citenamefont {Zhu}}]{ji2020gluon}%
  \BibitemOpen
  \bibfield  {author} {\bibinfo {author} {\bibfnamefont {Y.}~\bibnamefont {Ji}}, \bibinfo {author} {\bibfnamefont {H.}~\bibnamefont {Lamm}},\ and\ \bibinfo {author} {\bibfnamefont {S.}~\bibnamefont {Zhu}} (\bibinfo {collaboration} {NuQS Collaboration}),\ }\bibfield  {title} {\bibinfo {title} {Gluon field digitization via group space decimation for quantum computers},\ }\href {https://doi.org/10.1103/PhysRevD.102.114513} {\bibfield  {journal} {\bibinfo  {journal} {Phys. Rev. D}\ }\textbf {\bibinfo {volume} {102}},\ \bibinfo {pages} {114513} (\bibinfo {year} {2020})}\BibitemShut {NoStop}%
\bibitem [{\citenamefont {Alexandru}\ \emph {et~al.}(2022)\citenamefont {Alexandru}, \citenamefont {Bedaque}, \citenamefont {Brett},\ and\ \citenamefont {Lamm}}]{alexandru2022spectrum}%
  \BibitemOpen
  \bibfield  {author} {\bibinfo {author} {\bibfnamefont {A.}~\bibnamefont {Alexandru}}, \bibinfo {author} {\bibfnamefont {P.~F.}\ \bibnamefont {Bedaque}}, \bibinfo {author} {\bibfnamefont {R.}~\bibnamefont {Brett}},\ and\ \bibinfo {author} {\bibfnamefont {H.}~\bibnamefont {Lamm}},\ }\bibfield  {title} {\bibinfo {title} {Spectrum of digitized qcd: Glueballs in a $s(1080)$ gauge theory},\ }\href {https://doi.org/10.1103/PhysRevD.105.114508} {\bibfield  {journal} {\bibinfo  {journal} {Phys. Rev. D}\ }\textbf {\bibinfo {volume} {105}},\ \bibinfo {pages} {114508} (\bibinfo {year} {2022})}\BibitemShut {NoStop}%
\bibitem [{\citenamefont {Alam}\ \emph {et~al.}(2022)\citenamefont {Alam}, \citenamefont {Hadfield}, \citenamefont {Lamm},\ and\ \citenamefont {Li}}]{alam2022primitive}%
  \BibitemOpen
  \bibfield  {author} {\bibinfo {author} {\bibfnamefont {M.~S.}\ \bibnamefont {Alam}}, \bibinfo {author} {\bibfnamefont {S.}~\bibnamefont {Hadfield}}, \bibinfo {author} {\bibfnamefont {H.}~\bibnamefont {Lamm}},\ and\ \bibinfo {author} {\bibfnamefont {A.~C.~Y.}\ \bibnamefont {Li}} (\bibinfo {collaboration} {SQMS Collaboration}),\ }\bibfield  {title} {\bibinfo {title} {Primitive quantum gates for dihedral gauge theories},\ }\href {https://doi.org/10.1103/PhysRevD.105.114501} {\bibfield  {journal} {\bibinfo  {journal} {Phys. Rev. D}\ }\textbf {\bibinfo {volume} {105}},\ \bibinfo {pages} {114501} (\bibinfo {year} {2022})}\BibitemShut {NoStop}%
\bibitem [{\citenamefont {Gustafson}\ \emph {et~al.}(2022)\citenamefont {Gustafson}, \citenamefont {Lamm}, \citenamefont {Lovelace},\ and\ \citenamefont {Musk}}]{Gustafson_2022}%
  \BibitemOpen
  \bibfield  {author} {\bibinfo {author} {\bibfnamefont {E.~J.}\ \bibnamefont {Gustafson}}, \bibinfo {author} {\bibfnamefont {H.}~\bibnamefont {Lamm}}, \bibinfo {author} {\bibfnamefont {F.}~\bibnamefont {Lovelace}},\ and\ \bibinfo {author} {\bibfnamefont {D.}~\bibnamefont {Musk}},\ }\bibfield  {title} {\bibinfo {title} {Primitive quantum gates for an $su(2)$ discrete subgroup: Binary tetrahedral},\ }\href {https://doi.org/10.1103/PhysRevD.106.114501} {\bibfield  {journal} {\bibinfo  {journal} {Phys. Rev. D}\ }\textbf {\bibinfo {volume} {106}},\ \bibinfo {pages} {114501} (\bibinfo {year} {2022})}\BibitemShut {NoStop}%
\bibitem [{\citenamefont {Zache}\ \emph {et~al.}(2023{\natexlab{a}})\citenamefont {Zache}, \citenamefont {González-Cuadra},\ and\ \citenamefont {Zoller}}]{zache2023fermionqudit}%
  \BibitemOpen
  \bibfield  {author} {\bibinfo {author} {\bibfnamefont {T.~V.}\ \bibnamefont {Zache}}, \bibinfo {author} {\bibfnamefont {D.}~\bibnamefont {González-Cuadra}},\ and\ \bibinfo {author} {\bibfnamefont {P.}~\bibnamefont {Zoller}},\ }\bibfield  {title} {\bibinfo {title} {Fermion-qudit quantum processors for simulating lattice gauge theories with matter},\ }\href {https://doi.org/10.22331/q-2023-10-16-1140} {\bibfield  {journal} {\bibinfo  {journal} {Quantum}\ }\textbf {\bibinfo {volume} {7}},\ \bibinfo {pages} {1140} (\bibinfo {year} {2023}{\natexlab{a}})}\BibitemShut {NoStop}%
\bibitem [{\citenamefont {Gonz\'alez-Cuadra}\ \emph {et~al.}(2022)\citenamefont {Gonz\'alez-Cuadra}, \citenamefont {Zache}, \citenamefont {Carrasco}, \citenamefont {Kraus},\ and\ \citenamefont {Zoller}}]{gonzalez2022hardware}%
  \BibitemOpen
  \bibfield  {author} {\bibinfo {author} {\bibfnamefont {D.}~\bibnamefont {Gonz\'alez-Cuadra}}, \bibinfo {author} {\bibfnamefont {T.~V.}\ \bibnamefont {Zache}}, \bibinfo {author} {\bibfnamefont {J.}~\bibnamefont {Carrasco}}, \bibinfo {author} {\bibfnamefont {B.}~\bibnamefont {Kraus}},\ and\ \bibinfo {author} {\bibfnamefont {P.}~\bibnamefont {Zoller}},\ }\bibfield  {title} {\bibinfo {title} {Hardware efficient quantum simulation of non-abelian gauge theories with qudits on rydberg platforms},\ }\href {https://doi.org/10.1103/PhysRevLett.129.160501} {\bibfield  {journal} {\bibinfo  {journal} {Phys. Rev. Lett.}\ }\textbf {\bibinfo {volume} {129}},\ \bibinfo {pages} {160501} (\bibinfo {year} {2022})}\BibitemShut {NoStop}%
\bibitem [{\citenamefont {Byrnes}\ and\ \citenamefont {Yamamoto}(2006)}]{byrnes2006simulating}%
  \BibitemOpen
  \bibfield  {author} {\bibinfo {author} {\bibfnamefont {T.}~\bibnamefont {Byrnes}}\ and\ \bibinfo {author} {\bibfnamefont {Y.}~\bibnamefont {Yamamoto}},\ }\bibfield  {title} {\bibinfo {title} {Simulating lattice gauge theories on a quantum computer},\ }\href {https://doi.org/10.1103/PhysRevA.73.022328} {\bibfield  {journal} {\bibinfo  {journal} {Phys. Rev. A}\ }\textbf {\bibinfo {volume} {73}},\ \bibinfo {pages} {022328} (\bibinfo {year} {2006})}\BibitemShut {NoStop}%
\bibitem [{\citenamefont {Zohar}\ \emph {et~al.}(2012)\citenamefont {Zohar}, \citenamefont {Cirac},\ and\ \citenamefont {Reznik}}]{zohar2012simulating}%
  \BibitemOpen
  \bibfield  {author} {\bibinfo {author} {\bibfnamefont {E.}~\bibnamefont {Zohar}}, \bibinfo {author} {\bibfnamefont {J.~I.}\ \bibnamefont {Cirac}},\ and\ \bibinfo {author} {\bibfnamefont {B.}~\bibnamefont {Reznik}},\ }\bibfield  {title} {\bibinfo {title} {Simulating compact quantum electrodynamics with ultracold atoms: Probing confinement and nonperturbative effects},\ }\href {https://doi.org/10.1103/PhysRevLett.109.125302} {\bibfield  {journal} {\bibinfo  {journal} {Phys. Rev. Lett.}\ }\textbf {\bibinfo {volume} {109}},\ \bibinfo {pages} {125302} (\bibinfo {year} {2012})}\BibitemShut {NoStop}%
\bibitem [{\citenamefont {Zohar}\ \emph {et~al.}(2013{\natexlab{a}})\citenamefont {Zohar}, \citenamefont {Cirac},\ and\ \citenamefont {Reznik}}]{zohar2013quantum}%
  \BibitemOpen
  \bibfield  {author} {\bibinfo {author} {\bibfnamefont {E.}~\bibnamefont {Zohar}}, \bibinfo {author} {\bibfnamefont {J.~I.}\ \bibnamefont {Cirac}},\ and\ \bibinfo {author} {\bibfnamefont {B.}~\bibnamefont {Reznik}},\ }\bibfield  {title} {\bibinfo {title} {Quantum simulations of gauge theories with ultracold atoms: Local gauge invariance from angular-momentum conservation},\ }\href {https://doi.org/10.1103/PhysRevA.88.023617} {\bibfield  {journal} {\bibinfo  {journal} {Phys. Rev. A}\ }\textbf {\bibinfo {volume} {88}},\ \bibinfo {pages} {023617} (\bibinfo {year} {2013}{\natexlab{a}})}\BibitemShut {NoStop}%
\bibitem [{\citenamefont {Zohar}\ \emph {et~al.}(2013{\natexlab{b}})\citenamefont {Zohar}, \citenamefont {Cirac},\ and\ \citenamefont {Reznik}}]{zohar2013cold}%
  \BibitemOpen
  \bibfield  {author} {\bibinfo {author} {\bibfnamefont {E.}~\bibnamefont {Zohar}}, \bibinfo {author} {\bibfnamefont {J.~I.}\ \bibnamefont {Cirac}},\ and\ \bibinfo {author} {\bibfnamefont {B.}~\bibnamefont {Reznik}},\ }\bibfield  {title} {\bibinfo {title} {Cold-atom quantum simulator for su(2) yang-mills lattice gauge theory},\ }\href {https://doi.org/10.1103/PhysRevLett.110.125304} {\bibfield  {journal} {\bibinfo  {journal} {Phys. Rev. Lett.}\ }\textbf {\bibinfo {volume} {110}},\ \bibinfo {pages} {125304} (\bibinfo {year} {2013}{\natexlab{b}})}\BibitemShut {NoStop}%
\bibitem [{\citenamefont {Zohar}\ and\ \citenamefont {Burrello}(2015)}]{zohar2015formulation}%
  \BibitemOpen
  \bibfield  {author} {\bibinfo {author} {\bibfnamefont {E.}~\bibnamefont {Zohar}}\ and\ \bibinfo {author} {\bibfnamefont {M.}~\bibnamefont {Burrello}},\ }\bibfield  {title} {\bibinfo {title} {Formulation of lattice gauge theories for quantum simulations},\ }\href {https://doi.org/10.1103/PhysRevD.91.054506} {\bibfield  {journal} {\bibinfo  {journal} {Phys. Rev. D}\ }\textbf {\bibinfo {volume} {91}},\ \bibinfo {pages} {054506} (\bibinfo {year} {2015})}\BibitemShut {NoStop}%
\bibitem [{\citenamefont {Zohar}\ \emph {et~al.}(2015)\citenamefont {Zohar}, \citenamefont {Cirac},\ and\ \citenamefont {Reznik}}]{zohar2015quantum}%
  \BibitemOpen
  \bibfield  {author} {\bibinfo {author} {\bibfnamefont {E.}~\bibnamefont {Zohar}}, \bibinfo {author} {\bibfnamefont {J.~I.}\ \bibnamefont {Cirac}},\ and\ \bibinfo {author} {\bibfnamefont {B.}~\bibnamefont {Reznik}},\ }\bibfield  {title} {\bibinfo {title} {Quantum simulations of lattice gauge theories using ultracold atoms in optical lattices},\ }\href {https://doi.org/10.1088/0034-4885/79/1/014401} {\bibfield  {journal} {\bibinfo  {journal} {Reports on Progress in Physics}\ }\textbf {\bibinfo {volume} {79}},\ \bibinfo {pages} {014401} (\bibinfo {year} {2015})}\BibitemShut {NoStop}%
\bibitem [{\citenamefont {Zohar}(2021)}]{zohar2022quantum}%
  \BibitemOpen
  \bibfield  {author} {\bibinfo {author} {\bibfnamefont {E.}~\bibnamefont {Zohar}},\ }\bibfield  {title} {\bibinfo {title} {Quantum simulation of lattice gauge theories in more than one space dimension{\textemdash}requirements, challenges and methods},\ }\bibfield  {journal} {\bibinfo  {journal} {Philosophical Transactions of the Royal Society A: Mathematical, Physical and Engineering Sciences}\ }\textbf {\bibinfo {volume} {380}},\ \href {https://doi.org/10.1098/rsta.2021.0069} {10.1098/rsta.2021.0069} (\bibinfo {year} {2021})\BibitemShut {NoStop}%
\bibitem [{\citenamefont {Raychowdhury}\ and\ \citenamefont {Stryker}(2020{\natexlab{a}})}]{raychowdhury2020solving}%
  \BibitemOpen
  \bibfield  {author} {\bibinfo {author} {\bibfnamefont {I.}~\bibnamefont {Raychowdhury}}\ and\ \bibinfo {author} {\bibfnamefont {J.~R.}\ \bibnamefont {Stryker}},\ }\bibfield  {title} {\bibinfo {title} {Solving gauss's law on digital quantum computers with loop-string-hadron digitization},\ }\href {https://doi.org/10.1103/PhysRevResearch.2.033039} {\bibfield  {journal} {\bibinfo  {journal} {Phys. Rev. Res.}\ }\textbf {\bibinfo {volume} {2}},\ \bibinfo {pages} {033039} (\bibinfo {year} {2020}{\natexlab{a}})}\BibitemShut {NoStop}%
\bibitem [{\citenamefont {Raychowdhury}\ and\ \citenamefont {Stryker}(2020{\natexlab{b}})}]{raychowdhury2020loop}%
  \BibitemOpen
  \bibfield  {author} {\bibinfo {author} {\bibfnamefont {I.}~\bibnamefont {Raychowdhury}}\ and\ \bibinfo {author} {\bibfnamefont {J.~R.}\ \bibnamefont {Stryker}},\ }\bibfield  {title} {\bibinfo {title} {Loop, string, and hadron dynamics in su(2) hamiltonian lattice gauge theories},\ }\href {https://doi.org/10.1103/PhysRevD.101.114502} {\bibfield  {journal} {\bibinfo  {journal} {Phys. Rev. D}\ }\textbf {\bibinfo {volume} {101}},\ \bibinfo {pages} {114502} (\bibinfo {year} {2020}{\natexlab{b}})}\BibitemShut {NoStop}%
\bibitem [{\citenamefont {Kadam}\ \emph {et~al.}(2023)\citenamefont {Kadam}, \citenamefont {Raychowdhury},\ and\ \citenamefont {Stryker}}]{kadam2023loop}%
  \BibitemOpen
  \bibfield  {author} {\bibinfo {author} {\bibfnamefont {S.~V.}\ \bibnamefont {Kadam}}, \bibinfo {author} {\bibfnamefont {I.}~\bibnamefont {Raychowdhury}},\ and\ \bibinfo {author} {\bibfnamefont {J.~R.}\ \bibnamefont {Stryker}},\ }\bibfield  {title} {\bibinfo {title} {Loop-string-hadron formulation of an su(3) gauge theory with dynamical quarks},\ }\href {https://doi.org/10.1103/PhysRevD.107.094513} {\bibfield  {journal} {\bibinfo  {journal} {Phys. Rev. D}\ }\textbf {\bibinfo {volume} {107}},\ \bibinfo {pages} {094513} (\bibinfo {year} {2023})}\BibitemShut {NoStop}%
\bibitem [{\citenamefont {Shaw}\ \emph {et~al.}(2020)\citenamefont {Shaw}, \citenamefont {Lougovski}, \citenamefont {Stryker},\ and\ \citenamefont {Wiebe}}]{shaw2020quantum}%
  \BibitemOpen
  \bibfield  {author} {\bibinfo {author} {\bibfnamefont {A.~F.}\ \bibnamefont {Shaw}}, \bibinfo {author} {\bibfnamefont {P.}~\bibnamefont {Lougovski}}, \bibinfo {author} {\bibfnamefont {J.~R.}\ \bibnamefont {Stryker}},\ and\ \bibinfo {author} {\bibfnamefont {N.}~\bibnamefont {Wiebe}},\ }\bibfield  {title} {\bibinfo {title} {Quantum algorithms for simulating the lattice schwinger model},\ }\href {https://doi.org/10.22331/q-2020-08-10-306} {\bibfield  {journal} {\bibinfo  {journal} {Quantum}\ }\textbf {\bibinfo {volume} {4}},\ \bibinfo {pages} {306} (\bibinfo {year} {2020})}\BibitemShut {NoStop}%
\bibitem [{\citenamefont {Ciavarella}\ \emph {et~al.}(2022)\citenamefont {Ciavarella}, \citenamefont {Klco},\ and\ \citenamefont {Savage}}]{ciavarella2022conceptual}%
  \BibitemOpen
  \bibfield  {author} {\bibinfo {author} {\bibfnamefont {A.}~\bibnamefont {Ciavarella}}, \bibinfo {author} {\bibfnamefont {N.}~\bibnamefont {Klco}},\ and\ \bibinfo {author} {\bibfnamefont {M.~J.}\ \bibnamefont {Savage}},\ }\href@noop {} {\bibinfo {title} {Some conceptual aspects of operator design for quantum simulations of non-abelian lattice gauge theories}} (\bibinfo {year} {2022}),\ \Eprint {https://arxiv.org/abs/2203.11988} {arXiv:2203.11988 [quant-ph]} \BibitemShut {NoStop}%
\bibitem [{\citenamefont {Ciavarella}\ and\ \citenamefont {Chernyshev}(2022)}]{ciavarella2022preparation}%
  \BibitemOpen
  \bibfield  {author} {\bibinfo {author} {\bibfnamefont {A.~N.}\ \bibnamefont {Ciavarella}}\ and\ \bibinfo {author} {\bibfnamefont {I.~A.}\ \bibnamefont {Chernyshev}},\ }\bibfield  {title} {\bibinfo {title} {Preparation of the su(3) lattice yang-mills vacuum with variational quantum methods},\ }\href {https://doi.org/10.1103/PhysRevD.105.074504} {\bibfield  {journal} {\bibinfo  {journal} {Phys. Rev. D}\ }\textbf {\bibinfo {volume} {105}},\ \bibinfo {pages} {074504} (\bibinfo {year} {2022})}\BibitemShut {NoStop}%
\bibitem [{\citenamefont {Klco}\ \emph {et~al.}(2020)\citenamefont {Klco}, \citenamefont {Savage},\ and\ \citenamefont {Stryker}}]{klco20202}%
  \BibitemOpen
  \bibfield  {author} {\bibinfo {author} {\bibfnamefont {N.}~\bibnamefont {Klco}}, \bibinfo {author} {\bibfnamefont {M.~J.}\ \bibnamefont {Savage}},\ and\ \bibinfo {author} {\bibfnamefont {J.~R.}\ \bibnamefont {Stryker}},\ }\bibfield  {title} {\bibinfo {title} {Su(2) non-abelian gauge field theory in one dimension on digital quantum computers},\ }\href {https://doi.org/10.1103/PhysRevD.101.074512} {\bibfield  {journal} {\bibinfo  {journal} {Phys. Rev. D}\ }\textbf {\bibinfo {volume} {101}},\ \bibinfo {pages} {074512} (\bibinfo {year} {2020})}\BibitemShut {NoStop}%
\bibitem [{\citenamefont {Ciavarella}\ \emph {et~al.}(2021)\citenamefont {Ciavarella}, \citenamefont {Klco},\ and\ \citenamefont {Savage}}]{ciavarella2021trailhead}%
  \BibitemOpen
  \bibfield  {author} {\bibinfo {author} {\bibfnamefont {A.}~\bibnamefont {Ciavarella}}, \bibinfo {author} {\bibfnamefont {N.}~\bibnamefont {Klco}},\ and\ \bibinfo {author} {\bibfnamefont {M.~J.}\ \bibnamefont {Savage}},\ }\bibfield  {title} {\bibinfo {title} {Trailhead for quantum simulation of su(3) yang-mills lattice gauge theory in the local multiplet basis},\ }\href {https://doi.org/10.1103/PhysRevD.103.094501} {\bibfield  {journal} {\bibinfo  {journal} {Phys. Rev. D}\ }\textbf {\bibinfo {volume} {103}},\ \bibinfo {pages} {094501} (\bibinfo {year} {2021})}\BibitemShut {NoStop}%
\bibitem [{\citenamefont {Kavaki}\ and\ \citenamefont {Lewis}(2024)}]{kavaki2024square}%
  \BibitemOpen
  \bibfield  {author} {\bibinfo {author} {\bibfnamefont {A.~H.~Z.}\ \bibnamefont {Kavaki}}\ and\ \bibinfo {author} {\bibfnamefont {R.}~\bibnamefont {Lewis}},\ }\bibfield  {title} {\bibinfo {title} {{From square plaquettes to triamond lattices for SU(2) gauge theory}},\ }\href {https://doi.org/10.1038/s42005-024-01697-4} {\bibfield  {journal} {\bibinfo  {journal} {Commun. Phys.}\ }\textbf {\bibinfo {volume} {7}},\ \bibinfo {pages} {208} (\bibinfo {year} {2024})},\ \Eprint {https://arxiv.org/abs/2401.14570} {arXiv:2401.14570 [hep-lat]} \BibitemShut {NoStop}%
\bibitem [{\citenamefont {Mendicelli}\ \emph {et~al.}(2023)\citenamefont {Mendicelli}, \citenamefont {Lewis}, \citenamefont {Rahman},\ and\ \citenamefont {Powell}}]{rahman2022real}%
  \BibitemOpen
  \bibfield  {author} {\bibinfo {author} {\bibfnamefont {E.}~\bibnamefont {Mendicelli}}, \bibinfo {author} {\bibfnamefont {R.}~\bibnamefont {Lewis}}, \bibinfo {author} {\bibfnamefont {S.~A.}\ \bibnamefont {Rahman}},\ and\ \bibinfo {author} {\bibfnamefont {S.}~\bibnamefont {Powell}},\ }\bibfield  {title} {\bibinfo {title} {{Real time evolution and a traveling excitation in SU(2) pure gauge theory on a quantum computer.}},\ }\href {https://doi.org/10.22323/1.430.0025} {\bibfield  {journal} {\bibinfo  {journal} {PoS}\ }\textbf {\bibinfo {volume} {LATTICE2022}},\ \bibinfo {pages} {025} (\bibinfo {year} {2023})},\ \Eprint {https://arxiv.org/abs/2210.11606} {arXiv:2210.11606 [hep-lat]} \BibitemShut {NoStop}%
\bibitem [{\citenamefont {Atas}\ \emph {et~al.}(2021)\citenamefont {Atas}, \citenamefont {Zhang}, \citenamefont {Lewis}, \citenamefont {Jahanpour}, \citenamefont {Haase},\ and\ \citenamefont {Muschik}}]{Atas_2021}%
  \BibitemOpen
  \bibfield  {author} {\bibinfo {author} {\bibfnamefont {Y.~Y.}\ \bibnamefont {Atas}}, \bibinfo {author} {\bibfnamefont {J.}~\bibnamefont {Zhang}}, \bibinfo {author} {\bibfnamefont {R.}~\bibnamefont {Lewis}}, \bibinfo {author} {\bibfnamefont {A.}~\bibnamefont {Jahanpour}}, \bibinfo {author} {\bibfnamefont {J.~F.}\ \bibnamefont {Haase}},\ and\ \bibinfo {author} {\bibfnamefont {C.~A.}\ \bibnamefont {Muschik}},\ }\bibfield  {title} {\bibinfo {title} {{SU}(2) hadrons on a quantum computer via a variational approach},\ }\bibfield  {journal} {\bibinfo  {journal} {Nature Communications}\ }\textbf {\bibinfo {volume} {12}},\ \href {https://doi.org/10.1038/s41467-021-26825-4} {10.1038/s41467-021-26825-4} (\bibinfo {year} {2021})\BibitemShut {NoStop}%
\bibitem [{\citenamefont {A~Rahman}\ \emph {et~al.}(2022)\citenamefont {A~Rahman}, \citenamefont {Lewis}, \citenamefont {Mendicelli},\ and\ \citenamefont {Powell}}]{Rahman:2022rlg}%
  \BibitemOpen
  \bibfield  {author} {\bibinfo {author} {\bibfnamefont {S.}~\bibnamefont {A~Rahman}}, \bibinfo {author} {\bibfnamefont {R.}~\bibnamefont {Lewis}}, \bibinfo {author} {\bibfnamefont {E.}~\bibnamefont {Mendicelli}},\ and\ \bibinfo {author} {\bibfnamefont {S.}~\bibnamefont {Powell}},\ }\bibfield  {title} {\bibinfo {title} {Self-mitigating trotter circuits for su(2) lattice gauge theory on a quantum computer},\ }\href {https://doi.org/10.1103/PhysRevD.106.074502} {\bibfield  {journal} {\bibinfo  {journal} {Phys. Rev. D}\ }\textbf {\bibinfo {volume} {106}},\ \bibinfo {pages} {074502} (\bibinfo {year} {2022})}\BibitemShut {NoStop}%
\bibitem [{\citenamefont {Paulson}\ \emph {et~al.}(2021)\citenamefont {Paulson}, \citenamefont {Dellantonio}, \citenamefont {Haase}, \citenamefont {Celi}, \citenamefont {Kan}, \citenamefont {Jena}, \citenamefont {Kokail}, \citenamefont {van Bijnen}, \citenamefont {Jansen}, \citenamefont {Zoller},\ and\ \citenamefont {Muschik}}]{paulson2021simulating}%
  \BibitemOpen
  \bibfield  {author} {\bibinfo {author} {\bibfnamefont {D.}~\bibnamefont {Paulson}}, \bibinfo {author} {\bibfnamefont {L.}~\bibnamefont {Dellantonio}}, \bibinfo {author} {\bibfnamefont {J.~F.}\ \bibnamefont {Haase}}, \bibinfo {author} {\bibfnamefont {A.}~\bibnamefont {Celi}}, \bibinfo {author} {\bibfnamefont {A.}~\bibnamefont {Kan}}, \bibinfo {author} {\bibfnamefont {A.}~\bibnamefont {Jena}}, \bibinfo {author} {\bibfnamefont {C.}~\bibnamefont {Kokail}}, \bibinfo {author} {\bibfnamefont {R.}~\bibnamefont {van Bijnen}}, \bibinfo {author} {\bibfnamefont {K.}~\bibnamefont {Jansen}}, \bibinfo {author} {\bibfnamefont {P.}~\bibnamefont {Zoller}},\ and\ \bibinfo {author} {\bibfnamefont {C.~A.}\ \bibnamefont {Muschik}},\ }\bibfield  {title} {\bibinfo {title} {Simulating 2d effects in lattice gauge theories on a quantum computer},\ }\href {https://doi.org/10.1103/PRXQuantum.2.030334} {\bibfield  {journal} {\bibinfo  {journal} {PRX Quantum}\ }\textbf {\bibinfo {volume} {2}},\ \bibinfo {pages} {030334} (\bibinfo {year}
  {2021})}\BibitemShut {NoStop}%
\bibitem [{\citenamefont {Halimeh}\ \emph {et~al.}(2023)\citenamefont {Halimeh}, \citenamefont {Homeier}, \citenamefont {Bohrdt},\ and\ \citenamefont {Grusdt}}]{halimeh2023spin}%
  \BibitemOpen
  \bibfield  {author} {\bibinfo {author} {\bibfnamefont {J.~C.}\ \bibnamefont {Halimeh}}, \bibinfo {author} {\bibfnamefont {L.}~\bibnamefont {Homeier}}, \bibinfo {author} {\bibfnamefont {A.}~\bibnamefont {Bohrdt}},\ and\ \bibinfo {author} {\bibfnamefont {F.}~\bibnamefont {Grusdt}},\ }\href@noop {} {\bibinfo {title} {Spin exchange-enabled quantum simulator for large-scale non-abelian gauge theories}} (\bibinfo {year} {2023}),\ \Eprint {https://arxiv.org/abs/2305.06373} {arXiv:2305.06373 [cond-mat.quant-gas]} \BibitemShut {NoStop}%
\bibitem [{\citenamefont {Meurice}(2021)}]{meurice2021theoretical}%
  \BibitemOpen
  \bibfield  {author} {\bibinfo {author} {\bibfnamefont {Y.}~\bibnamefont {Meurice}},\ }\bibfield  {title} {\bibinfo {title} {Theoretical methods to design and test quantum simulators for the compact abelian higgs model},\ }\href {https://doi.org/10.1103/PhysRevD.104.094513} {\bibfield  {journal} {\bibinfo  {journal} {Phys. Rev. D}\ }\textbf {\bibinfo {volume} {104}},\ \bibinfo {pages} {094513} (\bibinfo {year} {2021})}\BibitemShut {NoStop}%
\bibitem [{\citenamefont {Davoudi}\ \emph {et~al.}(2020)\citenamefont {Davoudi}, \citenamefont {Hafezi}, \citenamefont {Monroe}, \citenamefont {Pagano}, \citenamefont {Seif},\ and\ \citenamefont {Shaw}}]{davoudi2020towards}%
  \BibitemOpen
  \bibfield  {author} {\bibinfo {author} {\bibfnamefont {Z.}~\bibnamefont {Davoudi}}, \bibinfo {author} {\bibfnamefont {M.}~\bibnamefont {Hafezi}}, \bibinfo {author} {\bibfnamefont {C.}~\bibnamefont {Monroe}}, \bibinfo {author} {\bibfnamefont {G.}~\bibnamefont {Pagano}}, \bibinfo {author} {\bibfnamefont {A.}~\bibnamefont {Seif}},\ and\ \bibinfo {author} {\bibfnamefont {A.}~\bibnamefont {Shaw}},\ }\bibfield  {title} {\bibinfo {title} {Towards analog quantum simulations of lattice gauge theories with trapped ions},\ }\href {https://doi.org/10.1103/PhysRevResearch.2.023015} {\bibfield  {journal} {\bibinfo  {journal} {Phys. Rev. Res.}\ }\textbf {\bibinfo {volume} {2}},\ \bibinfo {pages} {023015} (\bibinfo {year} {2020})}\BibitemShut {NoStop}%
\bibitem [{\citenamefont {Davoudi}\ \emph {et~al.}(2021)\citenamefont {Davoudi}, \citenamefont {Raychowdhury},\ and\ \citenamefont {Shaw}}]{davoudi2021search}%
  \BibitemOpen
  \bibfield  {author} {\bibinfo {author} {\bibfnamefont {Z.}~\bibnamefont {Davoudi}}, \bibinfo {author} {\bibfnamefont {I.}~\bibnamefont {Raychowdhury}},\ and\ \bibinfo {author} {\bibfnamefont {A.}~\bibnamefont {Shaw}},\ }\bibfield  {title} {\bibinfo {title} {Search for efficient formulations for hamiltonian simulation of non-abelian lattice gauge theories},\ }\href {https://doi.org/10.1103/PhysRevD.104.074505} {\bibfield  {journal} {\bibinfo  {journal} {Phys. Rev. D}\ }\textbf {\bibinfo {volume} {104}},\ \bibinfo {pages} {074505} (\bibinfo {year} {2021})}\BibitemShut {NoStop}%
\bibitem [{\citenamefont {Belyansky}\ \emph {et~al.}(2024)\citenamefont {Belyansky}, \citenamefont {Whitsitt}, \citenamefont {Mueller}, \citenamefont {Fahimniya}, \citenamefont {Bennewitz}, \citenamefont {Davoudi},\ and\ \citenamefont {Gorshkov}}]{belyansky2023highenergy}%
  \BibitemOpen
  \bibfield  {author} {\bibinfo {author} {\bibfnamefont {R.}~\bibnamefont {Belyansky}}, \bibinfo {author} {\bibfnamefont {S.}~\bibnamefont {Whitsitt}}, \bibinfo {author} {\bibfnamefont {N.}~\bibnamefont {Mueller}}, \bibinfo {author} {\bibfnamefont {A.}~\bibnamefont {Fahimniya}}, \bibinfo {author} {\bibfnamefont {E.~R.}\ \bibnamefont {Bennewitz}}, \bibinfo {author} {\bibfnamefont {Z.}~\bibnamefont {Davoudi}},\ and\ \bibinfo {author} {\bibfnamefont {A.~V.}\ \bibnamefont {Gorshkov}},\ }\bibfield  {title} {\bibinfo {title} {High-energy collision of quarks and mesons in the schwinger model: From tensor networks to circuit qed},\ }\href {https://doi.org/10.1103/PhysRevLett.132.091903} {\bibfield  {journal} {\bibinfo  {journal} {Phys. Rev. Lett.}\ }\textbf {\bibinfo {volume} {132}},\ \bibinfo {pages} {091903} (\bibinfo {year} {2024})}\BibitemShut {NoStop}%
\bibitem [{\citenamefont {Berenstein}\ and\ \citenamefont {Kawai}(2023)}]{berenstein2023integrable}%
  \BibitemOpen
  \bibfield  {author} {\bibinfo {author} {\bibfnamefont {D.}~\bibnamefont {Berenstein}}\ and\ \bibinfo {author} {\bibfnamefont {H.}~\bibnamefont {Kawai}},\ }\href@noop {} {\bibinfo {title} {Integrable spin chains from large-$n$ qcd at strong coupling}} (\bibinfo {year} {2023}),\ \Eprint {https://arxiv.org/abs/2308.11716} {arXiv:2308.11716 [hep-th]} \BibitemShut {NoStop}%
\bibitem [{\citenamefont {Rigobello}\ \emph {et~al.}(2023)\citenamefont {Rigobello}, \citenamefont {Magnifico}, \citenamefont {Silvi},\ and\ \citenamefont {Montangero}}]{rigobello2023hadrons}%
  \BibitemOpen
  \bibfield  {author} {\bibinfo {author} {\bibfnamefont {M.}~\bibnamefont {Rigobello}}, \bibinfo {author} {\bibfnamefont {G.}~\bibnamefont {Magnifico}}, \bibinfo {author} {\bibfnamefont {P.}~\bibnamefont {Silvi}},\ and\ \bibinfo {author} {\bibfnamefont {S.}~\bibnamefont {Montangero}},\ }\href@noop {} {\bibinfo {title} {Hadrons in (1+1)d hamiltonian hardcore lattice qcd}} (\bibinfo {year} {2023}),\ \Eprint {https://arxiv.org/abs/2308.04488} {arXiv:2308.04488 [hep-lat]} \BibitemShut {NoStop}%
\bibitem [{\citenamefont {Kane}\ \emph {et~al.}(2024)\citenamefont {Kane}, \citenamefont {Gomes},\ and\ \citenamefont {Kreshchuk}}]{kane2024nearlyoptimal}%
  \BibitemOpen
  \bibfield  {author} {\bibinfo {author} {\bibfnamefont {C.~F.}\ \bibnamefont {Kane}}, \bibinfo {author} {\bibfnamefont {N.}~\bibnamefont {Gomes}},\ and\ \bibinfo {author} {\bibfnamefont {M.}~\bibnamefont {Kreshchuk}},\ }\bibfield  {title} {\bibinfo {title} {Nearly optimal state preparation for quantum simulations of lattice gauge theories},\ }\href {https://doi.org/10.1103/PhysRevA.110.012455} {\bibfield  {journal} {\bibinfo  {journal} {Phys. Rev. A}\ }\textbf {\bibinfo {volume} {110}},\ \bibinfo {pages} {012455} (\bibinfo {year} {2024})}\BibitemShut {NoStop}%
\bibitem [{\citenamefont {Hariprakash}\ \emph {et~al.}(2023)\citenamefont {Hariprakash}, \citenamefont {Modi}, \citenamefont {Kreshchuk}, \citenamefont {Kane},\ and\ \citenamefont {Bauer}}]{hariprakash2023strategies}%
  \BibitemOpen
  \bibfield  {author} {\bibinfo {author} {\bibfnamefont {S.}~\bibnamefont {Hariprakash}}, \bibinfo {author} {\bibfnamefont {N.~S.}\ \bibnamefont {Modi}}, \bibinfo {author} {\bibfnamefont {M.}~\bibnamefont {Kreshchuk}}, \bibinfo {author} {\bibfnamefont {C.~F.}\ \bibnamefont {Kane}},\ and\ \bibinfo {author} {\bibfnamefont {C.~W.}\ \bibnamefont {Bauer}},\ }\href@noop {} {\bibinfo {title} {Strategies for simulating time evolution of hamiltonian lattice field theories}} (\bibinfo {year} {2023}),\ \Eprint {https://arxiv.org/abs/2312.11637} {arXiv:2312.11637 [quant-ph]} \BibitemShut {NoStop}%
\bibitem [{\citenamefont {Su}\ \emph {et~al.}(2024)\citenamefont {Su}, \citenamefont {Osborne},\ and\ \citenamefont {Halimeh}}]{su2024coldatom}%
  \BibitemOpen
  \bibfield  {author} {\bibinfo {author} {\bibfnamefont {G.-X.}\ \bibnamefont {Su}}, \bibinfo {author} {\bibfnamefont {J.}~\bibnamefont {Osborne}},\ and\ \bibinfo {author} {\bibfnamefont {J.~C.}\ \bibnamefont {Halimeh}},\ }\href@noop {} {\bibinfo {title} {A cold-atom particle collider}} (\bibinfo {year} {2024}),\ \Eprint {https://arxiv.org/abs/2401.05489} {arXiv:2401.05489 [cond-mat.quant-gas]} \BibitemShut {NoStop}%
\bibitem [{\citenamefont {Grabowska}\ \emph {et~al.}(2023)\citenamefont {Grabowska}, \citenamefont {Kane}, \citenamefont {Nachman},\ and\ \citenamefont {Bauer}}]{grabowska2023overcoming}%
  \BibitemOpen
  \bibfield  {author} {\bibinfo {author} {\bibfnamefont {D.~M.}\ \bibnamefont {Grabowska}}, \bibinfo {author} {\bibfnamefont {C.}~\bibnamefont {Kane}}, \bibinfo {author} {\bibfnamefont {B.}~\bibnamefont {Nachman}},\ and\ \bibinfo {author} {\bibfnamefont {C.~W.}\ \bibnamefont {Bauer}},\ }\href@noop {} {\bibinfo {title} {Overcoming exponential scaling with system size in trotter-suzuki implementations of constrained hamiltonians: 2+1 u(1) lattice gauge theories}} (\bibinfo {year} {2023}),\ \Eprint {https://arxiv.org/abs/2208.03333} {arXiv:2208.03333 [quant-ph]} \BibitemShut {NoStop}%
\bibitem [{\citenamefont {Bauer}\ and\ \citenamefont {Grabowska}(2023)}]{bauer2023efficient}%
  \BibitemOpen
  \bibfield  {author} {\bibinfo {author} {\bibfnamefont {C.~W.}\ \bibnamefont {Bauer}}\ and\ \bibinfo {author} {\bibfnamefont {D.~M.}\ \bibnamefont {Grabowska}},\ }\bibfield  {title} {\bibinfo {title} {Efficient representation for simulating u(1) gauge theories on digital quantum computers at all values of the coupling},\ }\href {https://doi.org/10.1103/PhysRevD.107.L031503} {\bibfield  {journal} {\bibinfo  {journal} {Phys. Rev. D}\ }\textbf {\bibinfo {volume} {107}},\ \bibinfo {pages} {L031503} (\bibinfo {year} {2023})}\BibitemShut {NoStop}%
\bibitem [{\citenamefont {Kane}\ \emph {et~al.}(2022)\citenamefont {Kane}, \citenamefont {Grabowska}, \citenamefont {Nachman},\ and\ \citenamefont {Bauer}}]{kane2022efficient}%
  \BibitemOpen
  \bibfield  {author} {\bibinfo {author} {\bibfnamefont {C.}~\bibnamefont {Kane}}, \bibinfo {author} {\bibfnamefont {D.~M.}\ \bibnamefont {Grabowska}}, \bibinfo {author} {\bibfnamefont {B.}~\bibnamefont {Nachman}},\ and\ \bibinfo {author} {\bibfnamefont {C.~W.}\ \bibnamefont {Bauer}},\ }\href@noop {} {\bibinfo {title} {Efficient quantum implementation of 2+1 u(1) lattice gauge theories with gauss law constraints}} (\bibinfo {year} {2022}),\ \Eprint {https://arxiv.org/abs/2211.10497} {arXiv:2211.10497 [quant-ph]} \BibitemShut {NoStop}%
\bibitem [{\citenamefont {D'Andrea}\ \emph {et~al.}(2024)\citenamefont {D'Andrea}, \citenamefont {Bauer}, \citenamefont {Grabowska},\ and\ \citenamefont {Freytsis}}]{bauer2023new}%
  \BibitemOpen
  \bibfield  {author} {\bibinfo {author} {\bibfnamefont {I.}~\bibnamefont {D'Andrea}}, \bibinfo {author} {\bibfnamefont {C.~W.}\ \bibnamefont {Bauer}}, \bibinfo {author} {\bibfnamefont {D.~M.}\ \bibnamefont {Grabowska}},\ and\ \bibinfo {author} {\bibfnamefont {M.}~\bibnamefont {Freytsis}},\ }\bibfield  {title} {\bibinfo {title} {New basis for hamiltonian su(2) simulations},\ }\href {https://doi.org/10.1103/PhysRevD.109.074501} {\bibfield  {journal} {\bibinfo  {journal} {Phys. Rev. D}\ }\textbf {\bibinfo {volume} {109}},\ \bibinfo {pages} {074501} (\bibinfo {year} {2024})}\BibitemShut {NoStop}%
\bibitem [{\citenamefont {Martinez}\ \emph {et~al.}(2016)\citenamefont {Martinez}, \citenamefont {Muschik}, \citenamefont {Schindler}, \citenamefont {Nigg}, \citenamefont {Erhard}, \citenamefont {Heyl}, \citenamefont {Hauke}, \citenamefont {Dalmonte}, \citenamefont {Monz}, \citenamefont {Zoller},\ and\ \citenamefont {Blatt}}]{martinez2016real}%
  \BibitemOpen
  \bibfield  {author} {\bibinfo {author} {\bibfnamefont {E.~A.}\ \bibnamefont {Martinez}}, \bibinfo {author} {\bibfnamefont {C.~A.}\ \bibnamefont {Muschik}}, \bibinfo {author} {\bibfnamefont {P.}~\bibnamefont {Schindler}}, \bibinfo {author} {\bibfnamefont {D.}~\bibnamefont {Nigg}}, \bibinfo {author} {\bibfnamefont {A.}~\bibnamefont {Erhard}}, \bibinfo {author} {\bibfnamefont {M.}~\bibnamefont {Heyl}}, \bibinfo {author} {\bibfnamefont {P.}~\bibnamefont {Hauke}}, \bibinfo {author} {\bibfnamefont {M.}~\bibnamefont {Dalmonte}}, \bibinfo {author} {\bibfnamefont {T.}~\bibnamefont {Monz}}, \bibinfo {author} {\bibfnamefont {P.}~\bibnamefont {Zoller}},\ and\ \bibinfo {author} {\bibfnamefont {R.}~\bibnamefont {Blatt}},\ }\bibfield  {title} {\bibinfo {title} {Real-time dynamics of lattice gauge theories with a few-qubit quantum computer},\ }\href {https://doi.org/10.1038/nature18318} {\bibfield  {journal} {\bibinfo  {journal} {Nature}\ }\textbf {\bibinfo {volume} {534}},\ \bibinfo {pages} {516} (\bibinfo {year}
  {2016})}\BibitemShut {NoStop}%
\bibitem [{\citenamefont {Illa}\ and\ \citenamefont {Savage}(2022)}]{Illa:2022jqb}%
  \BibitemOpen
  \bibfield  {author} {\bibinfo {author} {\bibfnamefont {M.}~\bibnamefont {Illa}}\ and\ \bibinfo {author} {\bibfnamefont {M.~J.}\ \bibnamefont {Savage}},\ }\bibfield  {title} {\bibinfo {title} {{Basic Elements for Simulations of Standard Model Physics with Quantum Annealers: Multigrid and Clock States}},\ }\href {https://doi.org/10.1103/PhysRevA.106.052605} {\bibfield  {journal} {\bibinfo  {journal} {Phys. Rev. A}\ }\textbf {\bibinfo {volume} {106}},\ \bibinfo {pages} {052605} (\bibinfo {year} {2022})},\ \Eprint {https://arxiv.org/abs/2202.12340} {arXiv:2202.12340 [quant-ph]} \BibitemShut {NoStop}%
\bibitem [{\citenamefont {Farrell}\ \emph {et~al.}(2023{\natexlab{a}})\citenamefont {Farrell}, \citenamefont {Chernyshev}, \citenamefont {Powell}, \citenamefont {Zemlevskiy}, \citenamefont {Illa},\ and\ \citenamefont {Savage}}]{farrell2023preparations}%
  \BibitemOpen
  \bibfield  {author} {\bibinfo {author} {\bibfnamefont {R.~C.}\ \bibnamefont {Farrell}}, \bibinfo {author} {\bibfnamefont {I.~A.}\ \bibnamefont {Chernyshev}}, \bibinfo {author} {\bibfnamefont {S.~J.~M.}\ \bibnamefont {Powell}}, \bibinfo {author} {\bibfnamefont {N.~A.}\ \bibnamefont {Zemlevskiy}}, \bibinfo {author} {\bibfnamefont {M.}~\bibnamefont {Illa}},\ and\ \bibinfo {author} {\bibfnamefont {M.~J.}\ \bibnamefont {Savage}},\ }\bibfield  {title} {\bibinfo {title} {Preparations for quantum simulations of quantum chromodynamics in $1+1$ dimensions. i. axial gauge},\ }\href {https://doi.org/10.1103/PhysRevD.107.054512} {\bibfield  {journal} {\bibinfo  {journal} {Phys. Rev. D}\ }\textbf {\bibinfo {volume} {107}},\ \bibinfo {pages} {054512} (\bibinfo {year} {2023}{\natexlab{a}})}\BibitemShut {NoStop}%
\bibitem [{\citenamefont {Farrell}\ \emph {et~al.}(2023{\natexlab{b}})\citenamefont {Farrell}, \citenamefont {Chernyshev}, \citenamefont {Powell}, \citenamefont {Zemlevskiy}, \citenamefont {Illa},\ and\ \citenamefont {Savage}}]{farrell2023preparations2}%
  \BibitemOpen
  \bibfield  {author} {\bibinfo {author} {\bibfnamefont {R.~C.}\ \bibnamefont {Farrell}}, \bibinfo {author} {\bibfnamefont {I.~A.}\ \bibnamefont {Chernyshev}}, \bibinfo {author} {\bibfnamefont {S.~J.~M.}\ \bibnamefont {Powell}}, \bibinfo {author} {\bibfnamefont {N.~A.}\ \bibnamefont {Zemlevskiy}}, \bibinfo {author} {\bibfnamefont {M.}~\bibnamefont {Illa}},\ and\ \bibinfo {author} {\bibfnamefont {M.~J.}\ \bibnamefont {Savage}},\ }\bibfield  {title} {\bibinfo {title} {Preparations for quantum simulations of quantum chromodynamics in $1+1$ dimensions. ii. single-baryon $\ensuremath{\beta}$-decay in real time},\ }\href {https://doi.org/10.1103/PhysRevD.107.054513} {\bibfield  {journal} {\bibinfo  {journal} {Phys. Rev. D}\ }\textbf {\bibinfo {volume} {107}},\ \bibinfo {pages} {054513} (\bibinfo {year} {2023}{\natexlab{b}})}\BibitemShut {NoStop}%
\bibitem [{\citenamefont {Atas}\ \emph {et~al.}(2022)\citenamefont {Atas}, \citenamefont {Haase}, \citenamefont {Zhang}, \citenamefont {Wei}, \citenamefont {Pfaendler}, \citenamefont {Lewis},\ and\ \citenamefont {Muschik}}]{Atas:2022dqm}%
  \BibitemOpen
  \bibfield  {author} {\bibinfo {author} {\bibfnamefont {Y.~Y.}\ \bibnamefont {Atas}}, \bibinfo {author} {\bibfnamefont {J.~F.}\ \bibnamefont {Haase}}, \bibinfo {author} {\bibfnamefont {J.}~\bibnamefont {Zhang}}, \bibinfo {author} {\bibfnamefont {V.}~\bibnamefont {Wei}}, \bibinfo {author} {\bibfnamefont {S.~M.~L.}\ \bibnamefont {Pfaendler}}, \bibinfo {author} {\bibfnamefont {R.}~\bibnamefont {Lewis}},\ and\ \bibinfo {author} {\bibfnamefont {C.~A.}\ \bibnamefont {Muschik}},\ }\href@noop {} {\bibinfo {title} {{Real-time evolution of SU(3) hadrons on a quantum computer}}} (\bibinfo {year} {2022}),\ \Eprint {https://arxiv.org/abs/2207.03473} {arXiv:2207.03473 [quant-ph]} \BibitemShut {NoStop}%
\bibitem [{\citenamefont {Yang}\ \emph {et~al.}(2020)\citenamefont {Yang}, \citenamefont {Sun}, \citenamefont {Ott}, \citenamefont {Wang}, \citenamefont {Zache}, \citenamefont {Halimeh}, \citenamefont {Yuan}, \citenamefont {Hauke},\ and\ \citenamefont {Pan}}]{yang2020observation}%
  \BibitemOpen
  \bibfield  {author} {\bibinfo {author} {\bibfnamefont {B.}~\bibnamefont {Yang}}, \bibinfo {author} {\bibfnamefont {H.}~\bibnamefont {Sun}}, \bibinfo {author} {\bibfnamefont {R.}~\bibnamefont {Ott}}, \bibinfo {author} {\bibfnamefont {H.-Y.}\ \bibnamefont {Wang}}, \bibinfo {author} {\bibfnamefont {T.~V.}\ \bibnamefont {Zache}}, \bibinfo {author} {\bibfnamefont {J.~C.}\ \bibnamefont {Halimeh}}, \bibinfo {author} {\bibfnamefont {Z.-S.}\ \bibnamefont {Yuan}}, \bibinfo {author} {\bibfnamefont {P.}~\bibnamefont {Hauke}},\ and\ \bibinfo {author} {\bibfnamefont {J.-W.}\ \bibnamefont {Pan}},\ }\bibfield  {title} {\bibinfo {title} {Observation of gauge invariance in a 71-site bose--hubbard quantum simulator},\ }\href {https://doi.org/10.1038/s41586-020-2910-8} {\bibfield  {journal} {\bibinfo  {journal} {Nature}\ }\textbf {\bibinfo {volume} {587}},\ \bibinfo {pages} {392} (\bibinfo {year} {2020})}\BibitemShut {NoStop}%
\bibitem [{\citenamefont {Zhou}\ \emph {et~al.}(2022)\citenamefont {Zhou}, \citenamefont {Su}, \citenamefont {Halimeh}, \citenamefont {Ott}, \citenamefont {Sun}, \citenamefont {Hauke}, \citenamefont {Yang}, \citenamefont {Yuan}, \citenamefont {Berges},\ and\ \citenamefont {Pan}}]{Zhou_2022}%
  \BibitemOpen
  \bibfield  {author} {\bibinfo {author} {\bibfnamefont {Z.-Y.}\ \bibnamefont {Zhou}}, \bibinfo {author} {\bibfnamefont {G.-X.}\ \bibnamefont {Su}}, \bibinfo {author} {\bibfnamefont {J.~C.}\ \bibnamefont {Halimeh}}, \bibinfo {author} {\bibfnamefont {R.}~\bibnamefont {Ott}}, \bibinfo {author} {\bibfnamefont {H.}~\bibnamefont {Sun}}, \bibinfo {author} {\bibfnamefont {P.}~\bibnamefont {Hauke}}, \bibinfo {author} {\bibfnamefont {B.}~\bibnamefont {Yang}}, \bibinfo {author} {\bibfnamefont {Z.-S.}\ \bibnamefont {Yuan}}, \bibinfo {author} {\bibfnamefont {J.}~\bibnamefont {Berges}},\ and\ \bibinfo {author} {\bibfnamefont {J.-W.}\ \bibnamefont {Pan}},\ }\bibfield  {title} {\bibinfo {title} {Thermalization dynamics of a gauge theory on a quantum simulator},\ }\href {https://doi.org/10.1126/science.abl6277} {\bibfield  {journal} {\bibinfo  {journal} {Science}\ }\textbf {\bibinfo {volume} {377}},\ \bibinfo {pages} {311} (\bibinfo {year} {2022})}\BibitemShut {NoStop}%
\bibitem [{\citenamefont {Su}\ \emph {et~al.}(2023)\citenamefont {Su}, \citenamefont {Sun}, \citenamefont {Hudomal}, \citenamefont {Desaules}, \citenamefont {Zhou}, \citenamefont {Yang}, \citenamefont {Halimeh}, \citenamefont {Yuan}, \citenamefont {Papi{\'{c} }},\ and\ \citenamefont {Pan}}]{Su_2023}%
  \BibitemOpen
  \bibfield  {author} {\bibinfo {author} {\bibfnamefont {G.-X.}\ \bibnamefont {Su}}, \bibinfo {author} {\bibfnamefont {H.}~\bibnamefont {Sun}}, \bibinfo {author} {\bibfnamefont {A.}~\bibnamefont {Hudomal}}, \bibinfo {author} {\bibfnamefont {J.-Y.}\ \bibnamefont {Desaules}}, \bibinfo {author} {\bibfnamefont {Z.-Y.}\ \bibnamefont {Zhou}}, \bibinfo {author} {\bibfnamefont {B.}~\bibnamefont {Yang}}, \bibinfo {author} {\bibfnamefont {J.~C.}\ \bibnamefont {Halimeh}}, \bibinfo {author} {\bibfnamefont {Z.-S.}\ \bibnamefont {Yuan}}, \bibinfo {author} {\bibfnamefont {Z.}~\bibnamefont {Papi{\'{c} }}},\ and\ \bibinfo {author} {\bibfnamefont {J.-W.}\ \bibnamefont {Pan}},\ }\bibfield  {title} {\bibinfo {title} {Observation of many-body scarring in a bose-hubbard quantum simulator},\ }\bibfield  {journal} {\bibinfo  {journal} {Physical Review Research}\ }\textbf {\bibinfo {volume} {5}},\ \href {https://doi.org/10.1103/physrevresearch.5.023010} {10.1103/physrevresearch.5.023010} (\bibinfo {year} {2023})\BibitemShut {NoStop}%
\bibitem [{\citenamefont {Zhang}\ \emph {et~al.}(2023)\citenamefont {Zhang}, \citenamefont {Liu}, \citenamefont {Cheng}, \citenamefont {He}, \citenamefont {Wang}, \citenamefont {Wang}, \citenamefont {Zhu}, \citenamefont {Su}, \citenamefont {Zhou}, \citenamefont {Zheng}, \citenamefont {Sun}, \citenamefont {Yang}, \citenamefont {Hauke}, \citenamefont {Zheng}, \citenamefont {Halimeh}, \citenamefont {Yuan},\ and\ \citenamefont {Pan}}]{zhang2023observation}%
  \BibitemOpen
  \bibfield  {author} {\bibinfo {author} {\bibfnamefont {W.-Y.}\ \bibnamefont {Zhang}}, \bibinfo {author} {\bibfnamefont {Y.}~\bibnamefont {Liu}}, \bibinfo {author} {\bibfnamefont {Y.}~\bibnamefont {Cheng}}, \bibinfo {author} {\bibfnamefont {M.-G.}\ \bibnamefont {He}}, \bibinfo {author} {\bibfnamefont {H.-Y.}\ \bibnamefont {Wang}}, \bibinfo {author} {\bibfnamefont {T.-Y.}\ \bibnamefont {Wang}}, \bibinfo {author} {\bibfnamefont {Z.-H.}\ \bibnamefont {Zhu}}, \bibinfo {author} {\bibfnamefont {G.-X.}\ \bibnamefont {Su}}, \bibinfo {author} {\bibfnamefont {Z.-Y.}\ \bibnamefont {Zhou}}, \bibinfo {author} {\bibfnamefont {Y.-G.}\ \bibnamefont {Zheng}}, \bibinfo {author} {\bibfnamefont {H.}~\bibnamefont {Sun}}, \bibinfo {author} {\bibfnamefont {B.}~\bibnamefont {Yang}}, \bibinfo {author} {\bibfnamefont {P.}~\bibnamefont {Hauke}}, \bibinfo {author} {\bibfnamefont {W.}~\bibnamefont {Zheng}}, \bibinfo {author} {\bibfnamefont {J.~C.}\ \bibnamefont {Halimeh}}, \bibinfo {author} {\bibfnamefont {Z.-S.}\ \bibnamefont {Yuan}},\
  and\ \bibinfo {author} {\bibfnamefont {J.-W.}\ \bibnamefont {Pan}},\ }\href@noop {} {\bibinfo {title} {Observation of microscopic confinement dynamics by a tunable topological $\theta$-angle}} (\bibinfo {year} {2023}),\ \Eprint {https://arxiv.org/abs/2306.11794} {arXiv:2306.11794 [cond-mat.quant-gas]} \BibitemShut {NoStop}%
\bibitem [{\citenamefont {Mildenberger}\ \emph {et~al.}(2022)\citenamefont {Mildenberger}, \citenamefont {Mruczkiewicz}, \citenamefont {Halimeh}, \citenamefont {Jiang},\ and\ \citenamefont {Hauke}}]{mildenberger2022probing}%
  \BibitemOpen
  \bibfield  {author} {\bibinfo {author} {\bibfnamefont {J.}~\bibnamefont {Mildenberger}}, \bibinfo {author} {\bibfnamefont {W.}~\bibnamefont {Mruczkiewicz}}, \bibinfo {author} {\bibfnamefont {J.~C.}\ \bibnamefont {Halimeh}}, \bibinfo {author} {\bibfnamefont {Z.}~\bibnamefont {Jiang}},\ and\ \bibinfo {author} {\bibfnamefont {P.}~\bibnamefont {Hauke}},\ }\href@noop {} {\bibinfo {title} {Probing confinement in a $\mathbb{Z}_2$ lattice gauge theory on a quantum computer}} (\bibinfo {year} {2022}),\ \Eprint {https://arxiv.org/abs/2203.08905} {arXiv:2203.08905 [quant-ph]} \BibitemShut {NoStop}%
\bibitem [{\citenamefont {Ciavarella}(2023)}]{ciavarella2023quantum}%
  \BibitemOpen
  \bibfield  {author} {\bibinfo {author} {\bibfnamefont {A.~N.}\ \bibnamefont {Ciavarella}},\ }\bibfield  {title} {\bibinfo {title} {Quantum simulation of lattice qcd with improved hamiltonians},\ }\href {https://doi.org/10.1103/PhysRevD.108.094513} {\bibfield  {journal} {\bibinfo  {journal} {Phys. Rev. D}\ }\textbf {\bibinfo {volume} {108}},\ \bibinfo {pages} {094513} (\bibinfo {year} {2023})}\BibitemShut {NoStop}%
\bibitem [{\citenamefont {Farrell}\ \emph {et~al.}(2024{\natexlab{a}})\citenamefont {Farrell}, \citenamefont {Illa}, \citenamefont {Ciavarella},\ and\ \citenamefont {Savage}}]{farrell2023scalable}%
  \BibitemOpen
  \bibfield  {author} {\bibinfo {author} {\bibfnamefont {R.~C.}\ \bibnamefont {Farrell}}, \bibinfo {author} {\bibfnamefont {M.}~\bibnamefont {Illa}}, \bibinfo {author} {\bibfnamefont {A.~N.}\ \bibnamefont {Ciavarella}},\ and\ \bibinfo {author} {\bibfnamefont {M.~J.}\ \bibnamefont {Savage}},\ }\bibfield  {title} {\bibinfo {title} {Scalable circuits for preparing ground states on digital quantum computers: The schwinger model vacuum on 100 qubits},\ }\href {https://doi.org/10.1103/PRXQuantum.5.020315} {\bibfield  {journal} {\bibinfo  {journal} {PRX Quantum}\ }\textbf {\bibinfo {volume} {5}},\ \bibinfo {pages} {020315} (\bibinfo {year} {2024}{\natexlab{a}})}\BibitemShut {NoStop}%
\bibitem [{\citenamefont {Farrell}\ \emph {et~al.}(2024{\natexlab{b}})\citenamefont {Farrell}, \citenamefont {Illa}, \citenamefont {Ciavarella},\ and\ \citenamefont {Savage}}]{farrell2024quantum}%
  \BibitemOpen
  \bibfield  {author} {\bibinfo {author} {\bibfnamefont {R.~C.}\ \bibnamefont {Farrell}}, \bibinfo {author} {\bibfnamefont {M.}~\bibnamefont {Illa}}, \bibinfo {author} {\bibfnamefont {A.~N.}\ \bibnamefont {Ciavarella}},\ and\ \bibinfo {author} {\bibfnamefont {M.~J.}\ \bibnamefont {Savage}},\ }\bibfield  {title} {\bibinfo {title} {Quantum simulations of hadron dynamics in the schwinger model using 112 qubits},\ }\href {https://doi.org/10.1103/PhysRevD.109.114510} {\bibfield  {journal} {\bibinfo  {journal} {Phys. Rev. D}\ }\textbf {\bibinfo {volume} {109}},\ \bibinfo {pages} {114510} (\bibinfo {year} {2024}{\natexlab{b}})}\BibitemShut {NoStop}%
\bibitem [{\citenamefont {Charles}\ \emph {et~al.}(2024)\citenamefont {Charles}, \citenamefont {Gustafson}, \citenamefont {Hardt}, \citenamefont {Herren}, \citenamefont {Hogan}, \citenamefont {Lamm}, \citenamefont {Starecheski}, \citenamefont {Van~de Water},\ and\ \citenamefont {Wagman}}]{charles2023simulating}%
  \BibitemOpen
  \bibfield  {author} {\bibinfo {author} {\bibfnamefont {C.}~\bibnamefont {Charles}}, \bibinfo {author} {\bibfnamefont {E.~J.}\ \bibnamefont {Gustafson}}, \bibinfo {author} {\bibfnamefont {E.}~\bibnamefont {Hardt}}, \bibinfo {author} {\bibfnamefont {F.}~\bibnamefont {Herren}}, \bibinfo {author} {\bibfnamefont {N.}~\bibnamefont {Hogan}}, \bibinfo {author} {\bibfnamefont {H.}~\bibnamefont {Lamm}}, \bibinfo {author} {\bibfnamefont {S.}~\bibnamefont {Starecheski}}, \bibinfo {author} {\bibfnamefont {R.~S.}\ \bibnamefont {Van~de Water}},\ and\ \bibinfo {author} {\bibfnamefont {M.~L.}\ \bibnamefont {Wagman}},\ }\bibfield  {title} {\bibinfo {title} {Simulating ${\mathbb{z}}_{2}$ lattice gauge theory on a quantum computer},\ }\href {https://doi.org/10.1103/PhysRevE.109.015307} {\bibfield  {journal} {\bibinfo  {journal} {Phys. Rev. E}\ }\textbf {\bibinfo {volume} {109}},\ \bibinfo {pages} {015307} (\bibinfo {year} {2024})}\BibitemShut {NoStop}%
\bibitem [{\citenamefont {Mueller}\ \emph {et~al.}(2023)\citenamefont {Mueller}, \citenamefont {Carolan}, \citenamefont {Connelly}, \citenamefont {Davoudi}, \citenamefont {Dumitrescu},\ and\ \citenamefont {Yeter-Aydeniz}}]{mueller2022quantum}%
  \BibitemOpen
  \bibfield  {author} {\bibinfo {author} {\bibfnamefont {N.}~\bibnamefont {Mueller}}, \bibinfo {author} {\bibfnamefont {J.~A.}\ \bibnamefont {Carolan}}, \bibinfo {author} {\bibfnamefont {A.}~\bibnamefont {Connelly}}, \bibinfo {author} {\bibfnamefont {Z.}~\bibnamefont {Davoudi}}, \bibinfo {author} {\bibfnamefont {E.~F.}\ \bibnamefont {Dumitrescu}},\ and\ \bibinfo {author} {\bibfnamefont {K.}~\bibnamefont {Yeter-Aydeniz}},\ }\bibfield  {title} {\bibinfo {title} {Quantum computation of dynamical quantum phase transitions and entanglement tomography in a lattice gauge theory},\ }\href {https://doi.org/10.1103/PRXQuantum.4.030323} {\bibfield  {journal} {\bibinfo  {journal} {PRX Quantum}\ }\textbf {\bibinfo {volume} {4}},\ \bibinfo {pages} {030323} (\bibinfo {year} {2023})}\BibitemShut {NoStop}%
\bibitem [{\citenamefont {Lucini}\ and\ \citenamefont {Panero}(2013)}]{LUCINI201393}%
  \BibitemOpen
  \bibfield  {author} {\bibinfo {author} {\bibfnamefont {B.}~\bibnamefont {Lucini}}\ and\ \bibinfo {author} {\bibfnamefont {M.}~\bibnamefont {Panero}},\ }\bibfield  {title} {\bibinfo {title} {Su(n) gauge theories at large n},\ }\href {https://doi.org/https://doi.org/10.1016/j.physrep.2013.01.001} {\bibfield  {journal} {\bibinfo  {journal} {Physics Reports}\ }\textbf {\bibinfo {volume} {526}},\ \bibinfo {pages} {93} (\bibinfo {year} {2013})},\ \bibinfo {note} {sU(N) gauge theories at large N}\BibitemShut {NoStop}%
\bibitem [{\citenamefont {Manohar}(1998)}]{Manohar:1998xv}%
  \BibitemOpen
  \bibfield  {author} {\bibinfo {author} {\bibfnamefont {A.~V.}\ \bibnamefont {Manohar}},\ }\bibfield  {title} {\bibinfo {title} {{Large N QCD}},\ }in\ \href@noop {} {\emph {\bibinfo {booktitle} {{Les Houches Summer School in Theoretical Physics, Session 68: Probing the Standard Model of Particle Interactions}}}}\ (\bibinfo {year} {1998})\ pp.\ \bibinfo {pages} {1091--1169},\ \Eprint {https://arxiv.org/abs/hep-ph/9802419} {arXiv:hep-ph/9802419} \BibitemShut {NoStop}%
\bibitem [{\citenamefont {Sjostrand}\ \emph {et~al.}(2006)\citenamefont {Sjostrand}, \citenamefont {Mrenna},\ and\ \citenamefont {Skands}}]{Sjostrand:2006za}%
  \BibitemOpen
  \bibfield  {author} {\bibinfo {author} {\bibfnamefont {T.}~\bibnamefont {Sjostrand}}, \bibinfo {author} {\bibfnamefont {S.}~\bibnamefont {Mrenna}},\ and\ \bibinfo {author} {\bibfnamefont {P.~Z.}\ \bibnamefont {Skands}},\ }\bibfield  {title} {\bibinfo {title} {{PYTHIA 6.4 Physics and Manual}},\ }\href {https://doi.org/10.1088/1126-6708/2006/05/026} {\bibfield  {journal} {\bibinfo  {journal} {JHEP}\ }\textbf {\bibinfo {volume} {05}},\ \bibinfo {pages} {026}},\ \Eprint {https://arxiv.org/abs/hep-ph/0603175} {arXiv:hep-ph/0603175} \BibitemShut {NoStop}%
\bibitem [{\citenamefont {Bahr}\ \emph {et~al.}(2008)\citenamefont {Bahr} \emph {et~al.}}]{Bahr:2008pv}%
  \BibitemOpen
  \bibfield  {author} {\bibinfo {author} {\bibfnamefont {M.}~\bibnamefont {Bahr}} \emph {et~al.},\ }\bibfield  {title} {\bibinfo {title} {{Herwig++ Physics and Manual}},\ }\href {https://doi.org/10.1140/epjc/s10052-008-0798-9} {\bibfield  {journal} {\bibinfo  {journal} {Eur. Phys. J. C}\ }\textbf {\bibinfo {volume} {58}},\ \bibinfo {pages} {639} (\bibinfo {year} {2008})},\ \Eprint {https://arxiv.org/abs/0803.0883} {arXiv:0803.0883 [hep-ph]} \BibitemShut {NoStop}%
\bibitem [{\citenamefont {'t~Hooft}(1974)}]{tHooft:1973alw}%
  \BibitemOpen
  \bibfield  {author} {\bibinfo {author} {\bibfnamefont {G.}~\bibnamefont {'t~Hooft}},\ }\bibfield  {title} {\bibinfo {title} {{A Planar Diagram Theory for Strong Interactions}},\ }\href {https://doi.org/10.1016/0550-3213(74)90154-0} {\bibfield  {journal} {\bibinfo  {journal} {Nucl. Phys. B}\ }\textbf {\bibinfo {volume} {72}},\ \bibinfo {pages} {461} (\bibinfo {year} {1974})}\BibitemShut {NoStop}%
\bibitem [{\citenamefont {PICH}(2002)}]{PICH_2002}%
  \BibitemOpen
  \bibfield  {author} {\bibinfo {author} {\bibfnamefont {A.}~\bibnamefont {PICH}},\ }\bibfield  {title} {\bibinfo {title} {Colourless mesons in a polychromatic world},\ }in\ \href {https://doi.org/10.1142/9789812776914_0023} {\emph {\bibinfo {booktitle} {Phenomenology of Large Nc QCD}}}\ (\bibinfo  {publisher} {WORLD SCIENTIFIC},\ \bibinfo {year} {2002})\BibitemShut {NoStop}%
\bibitem [{\citenamefont {Kaplan}\ and\ \citenamefont {Savage}(1996)}]{KAPLAN1996244}%
  \BibitemOpen
  \bibfield  {author} {\bibinfo {author} {\bibfnamefont {D.~B.}\ \bibnamefont {Kaplan}}\ and\ \bibinfo {author} {\bibfnamefont {M.~J.}\ \bibnamefont {Savage}},\ }\bibfield  {title} {\bibinfo {title} {The spin-flavor dependence of nuclear forces from large-n qcd},\ }\href {https://doi.org/https://doi.org/10.1016/0370-2693(95)01277-X} {\bibfield  {journal} {\bibinfo  {journal} {Physics Letters B}\ }\textbf {\bibinfo {volume} {365}},\ \bibinfo {pages} {244} (\bibinfo {year} {1996})}\BibitemShut {NoStop}%
\bibitem [{\citenamefont {Maldacena}(1999)}]{maldacena1999large}%
  \BibitemOpen
  \bibfield  {author} {\bibinfo {author} {\bibfnamefont {J.}~\bibnamefont {Maldacena}},\ }\bibfield  {title} {\bibinfo {title} {The large-n limit of superconformal field theories and supergravity},\ }\href {https://doi.org/10.1023/A:1026654312961} {\bibfield  {journal} {\bibinfo  {journal} {International journal of theoretical physics}\ }\textbf {\bibinfo {volume} {38}},\ \bibinfo {pages} {1113} (\bibinfo {year} {1999})}\BibitemShut {NoStop}%
\bibitem [{\citenamefont {Witten}(1980)}]{Witten1980}%
  \BibitemOpen
  \bibfield  {author} {\bibinfo {author} {\bibfnamefont {E.}~\bibnamefont {Witten}},\ }\bibinfo {title} {The 1/n expansion in atomic and particle physics},\ in\ \href {https://doi.org/10.1007/978-1-4684-7571-5_21} {\emph {\bibinfo {booktitle} {Recent Developments in Gauge Theories}}},\ \bibinfo {editor} {edited by\ \bibinfo {editor} {\bibfnamefont {G.}~\bibnamefont {Hooft}}, \bibinfo {editor} {\bibfnamefont {C.}~\bibnamefont {Itzykson}}, \bibinfo {editor} {\bibfnamefont {A.}~\bibnamefont {Jaffe}}, \bibinfo {editor} {\bibfnamefont {H.}~\bibnamefont {Lehmann}}, \bibinfo {editor} {\bibfnamefont {P.~K.}\ \bibnamefont {Mitter}}, \bibinfo {editor} {\bibfnamefont {I.~M.}\ \bibnamefont {Singer}},\ and\ \bibinfo {editor} {\bibfnamefont {R.}~\bibnamefont {Stora}}}\ (\bibinfo  {publisher} {Springer US},\ \bibinfo {address} {Boston, MA},\ \bibinfo {year} {1980})\ pp.\ \bibinfo {pages} {403--419}\BibitemShut {NoStop}%
\bibitem [{\citenamefont {Yaffe}(1982)}]{yaffe1982large}%
  \BibitemOpen
  \bibfield  {author} {\bibinfo {author} {\bibfnamefont {L.~G.}\ \bibnamefont {Yaffe}},\ }\bibfield  {title} {\bibinfo {title} {Large $n$ limits as classical mechanics},\ }\href {https://doi.org/10.1103/RevModPhys.54.407} {\bibfield  {journal} {\bibinfo  {journal} {Rev. Mod. Phys.}\ }\textbf {\bibinfo {volume} {54}},\ \bibinfo {pages} {407} (\bibinfo {year} {1982})}\BibitemShut {NoStop}%
\bibitem [{\citenamefont {Jevicki}\ \emph {et~al.}(1983)\citenamefont {Jevicki}, \citenamefont {Karim}, \citenamefont {Rodrigues},\ and\ \citenamefont {Levine}}]{JEVICKI1983169}%
  \BibitemOpen
  \bibfield  {author} {\bibinfo {author} {\bibfnamefont {A.}~\bibnamefont {Jevicki}}, \bibinfo {author} {\bibfnamefont {O.}~\bibnamefont {Karim}}, \bibinfo {author} {\bibfnamefont {J.}~\bibnamefont {Rodrigues}},\ and\ \bibinfo {author} {\bibfnamefont {H.}~\bibnamefont {Levine}},\ }\bibfield  {title} {\bibinfo {title} {Loop space hamiltonians and numerical methods for large-n gauge theories},\ }\href {https://doi.org/https://doi.org/10.1016/0550-3213(83)90180-3} {\bibfield  {journal} {\bibinfo  {journal} {Nuclear Physics B}\ }\textbf {\bibinfo {volume} {213}},\ \bibinfo {pages} {169} (\bibinfo {year} {1983})}\BibitemShut {NoStop}%
\bibitem [{\citenamefont {Jevicki}\ \emph {et~al.}(1984)\citenamefont {Jevicki}, \citenamefont {Karim}, \citenamefont {Rodrigues},\ and\ \citenamefont {Levine}}]{JEVICKI1984299}%
  \BibitemOpen
  \bibfield  {author} {\bibinfo {author} {\bibfnamefont {A.}~\bibnamefont {Jevicki}}, \bibinfo {author} {\bibfnamefont {O.}~\bibnamefont {Karim}}, \bibinfo {author} {\bibfnamefont {J.}~\bibnamefont {Rodrigues}},\ and\ \bibinfo {author} {\bibfnamefont {H.}~\bibnamefont {Levine}},\ }\bibfield  {title} {\bibinfo {title} {Loop-space hamiltonians and numerical methods for large-n gauge theories (ii)},\ }\href {https://doi.org/https://doi.org/10.1016/0550-3213(84)90215-3} {\bibfield  {journal} {\bibinfo  {journal} {Nuclear Physics B}\ }\textbf {\bibinfo {volume} {230}},\ \bibinfo {pages} {299} (\bibinfo {year} {1984})}\BibitemShut {NoStop}%
\bibitem [{\citenamefont {Kogut}\ and\ \citenamefont {Susskind}(1975)}]{kogut1975hamiltonian}%
  \BibitemOpen
  \bibfield  {author} {\bibinfo {author} {\bibfnamefont {J.}~\bibnamefont {Kogut}}\ and\ \bibinfo {author} {\bibfnamefont {L.}~\bibnamefont {Susskind}},\ }\bibfield  {title} {\bibinfo {title} {Hamiltonian formulation of wilson's lattice gauge theories},\ }\href {https://doi.org/10.1103/PhysRevD.11.395} {\bibfield  {journal} {\bibinfo  {journal} {Phys. Rev. D}\ }\textbf {\bibinfo {volume} {11}},\ \bibinfo {pages} {395} (\bibinfo {year} {1975})}\BibitemShut {NoStop}%
\bibitem [{\citenamefont {Kogut}(1979)}]{kogut1979introduction}%
  \BibitemOpen
  \bibfield  {author} {\bibinfo {author} {\bibfnamefont {J.~B.}\ \bibnamefont {Kogut}},\ }\bibfield  {title} {\bibinfo {title} {An introduction to lattice gauge theory and spin systems},\ }\href {https://doi.org/10.1103/RevModPhys.51.659} {\bibfield  {journal} {\bibinfo  {journal} {Rev. Mod. Phys.}\ }\textbf {\bibinfo {volume} {51}},\ \bibinfo {pages} {659} (\bibinfo {year} {1979})}\BibitemShut {NoStop}%
\bibitem [{\citenamefont {Banks}\ \emph {et~al.}(1977)\citenamefont {Banks}, \citenamefont {Raby}, \citenamefont {Susskind}, \citenamefont {Kogut}, \citenamefont {Jones}, \citenamefont {Scharbach},\ and\ \citenamefont {Sinclair}}]{banks1977strong}%
  \BibitemOpen
  \bibfield  {author} {\bibinfo {author} {\bibfnamefont {T.}~\bibnamefont {Banks}}, \bibinfo {author} {\bibfnamefont {S.}~\bibnamefont {Raby}}, \bibinfo {author} {\bibfnamefont {L.}~\bibnamefont {Susskind}}, \bibinfo {author} {\bibfnamefont {J.}~\bibnamefont {Kogut}}, \bibinfo {author} {\bibfnamefont {D.~R.~T.}\ \bibnamefont {Jones}}, \bibinfo {author} {\bibfnamefont {P.~N.}\ \bibnamefont {Scharbach}},\ and\ \bibinfo {author} {\bibfnamefont {D.~K.}\ \bibnamefont {Sinclair}},\ }\bibfield  {title} {\bibinfo {title} {Strong-coupling calculations of the hadron spectrum of quantum chromodynamics},\ }\href {https://doi.org/10.1103/PhysRevD.15.1111} {\bibfield  {journal} {\bibinfo  {journal} {Phys. Rev. D}\ }\textbf {\bibinfo {volume} {15}},\ \bibinfo {pages} {1111} (\bibinfo {year} {1977})}\BibitemShut {NoStop}%
\bibitem [{\citenamefont {Jones}\ \emph {et~al.}(1979)\citenamefont {Jones}, \citenamefont {Kenway}, \citenamefont {Kogut},\ and\ \citenamefont {Sinclair}}]{jones1979lattice}%
  \BibitemOpen
  \bibfield  {author} {\bibinfo {author} {\bibfnamefont {D.}~\bibnamefont {Jones}}, \bibinfo {author} {\bibfnamefont {R.}~\bibnamefont {Kenway}}, \bibinfo {author} {\bibfnamefont {J.}~\bibnamefont {Kogut}},\ and\ \bibinfo {author} {\bibfnamefont {D.}~\bibnamefont {Sinclair}},\ }\bibfield  {title} {\bibinfo {title} {Lattice gauge theory calculations using an improved strong-coupling expansion and matrix padé approximants},\ }\href {https://doi.org/https://doi.org/10.1016/0550-3213(79)90190-1} {\bibfield  {journal} {\bibinfo  {journal} {Nuclear Physics B}\ }\textbf {\bibinfo {volume} {158}},\ \bibinfo {pages} {102} (\bibinfo {year} {1979})}\BibitemShut {NoStop}%
\bibitem [{\citenamefont {Zache}\ \emph {et~al.}(2023{\natexlab{b}})\citenamefont {Zache}, \citenamefont {Gonz\'alez-Cuadra},\ and\ \citenamefont {Zoller}}]{zache2023quantum}%
  \BibitemOpen
  \bibfield  {author} {\bibinfo {author} {\bibfnamefont {T.~V.}\ \bibnamefont {Zache}}, \bibinfo {author} {\bibfnamefont {D.}~\bibnamefont {Gonz\'alez-Cuadra}},\ and\ \bibinfo {author} {\bibfnamefont {P.}~\bibnamefont {Zoller}},\ }\bibfield  {title} {\bibinfo {title} {Quantum and classical spin-network algorithms for $q$-deformed kogut-susskind gauge theories},\ }\href {https://doi.org/10.1103/PhysRevLett.131.171902} {\bibfield  {journal} {\bibinfo  {journal} {Phys. Rev. Lett.}\ }\textbf {\bibinfo {volume} {131}},\ \bibinfo {pages} {171902} (\bibinfo {year} {2023}{\natexlab{b}})}\BibitemShut {NoStop}%
\bibitem [{\citenamefont {Hayata}\ and\ \citenamefont {Hidaka}(2023)}]{hayata2023q}%
  \BibitemOpen
  \bibfield  {author} {\bibinfo {author} {\bibfnamefont {T.}~\bibnamefont {Hayata}}\ and\ \bibinfo {author} {\bibfnamefont {Y.}~\bibnamefont {Hidaka}},\ }\href@noop {} {\bibinfo {title} {$q$ deformed formulation of hamiltonian su(3) yang-mills theory}} (\bibinfo {year} {2023}),\ \Eprint {https://arxiv.org/abs/2306.12324} {arXiv:2306.12324 [hep-lat]} \BibitemShut {NoStop}%
\bibitem [{\citenamefont {Davoudi}\ \emph {et~al.}(2023)\citenamefont {Davoudi}, \citenamefont {Shaw},\ and\ \citenamefont {Stryker}}]{davoudi2023general}%
  \BibitemOpen
  \bibfield  {author} {\bibinfo {author} {\bibfnamefont {Z.}~\bibnamefont {Davoudi}}, \bibinfo {author} {\bibfnamefont {A.~F.}\ \bibnamefont {Shaw}},\ and\ \bibinfo {author} {\bibfnamefont {J.~R.}\ \bibnamefont {Stryker}},\ }\bibfield  {title} {\bibinfo {title} {General quantum algorithms for hamiltonian simulation with applications to a non-abelian lattice gauge theory},\ }\href {https://doi.org/10.22331/q-2023-12-20-1213} {\bibfield  {journal} {\bibinfo  {journal} {Quantum}\ }\textbf {\bibinfo {volume} {7}},\ \bibinfo {pages} {1213} (\bibinfo {year} {2023})}\BibitemShut {NoStop}%
\bibitem [{\citenamefont {Levin}\ and\ \citenamefont {Wen}(2005)}]{levin2005string}%
  \BibitemOpen
  \bibfield  {author} {\bibinfo {author} {\bibfnamefont {M.~A.}\ \bibnamefont {Levin}}\ and\ \bibinfo {author} {\bibfnamefont {X.-G.}\ \bibnamefont {Wen}},\ }\bibfield  {title} {\bibinfo {title} {String-net condensation: A physical mechanism for topological phases},\ }\href {https://doi.org/10.1103/PhysRevB.71.045110} {\bibfield  {journal} {\bibinfo  {journal} {Phys. Rev. B}\ }\textbf {\bibinfo {volume} {71}},\ \bibinfo {pages} {045110} (\bibinfo {year} {2005})}\BibitemShut {NoStop}%
\bibitem [{\citenamefont {Ba\~nuls}\ \emph {et~al.}(2017)\citenamefont {Ba\~nuls}, \citenamefont {Cichy}, \citenamefont {Cirac}, \citenamefont {Jansen},\ and\ \citenamefont {K\"uhn}}]{banuls2017efficient}%
  \BibitemOpen
  \bibfield  {author} {\bibinfo {author} {\bibfnamefont {M.~C.}\ \bibnamefont {Ba\~nuls}}, \bibinfo {author} {\bibfnamefont {K.}~\bibnamefont {Cichy}}, \bibinfo {author} {\bibfnamefont {J.~I.}\ \bibnamefont {Cirac}}, \bibinfo {author} {\bibfnamefont {K.}~\bibnamefont {Jansen}},\ and\ \bibinfo {author} {\bibfnamefont {S.}~\bibnamefont {K\"uhn}},\ }\bibfield  {title} {\bibinfo {title} {Efficient basis formulation for ($1+1$)-dimensional su(2) lattice gauge theory: Spectral calculations with matrix product states},\ }\href {https://doi.org/10.1103/PhysRevX.7.041046} {\bibfield  {journal} {\bibinfo  {journal} {Phys. Rev. X}\ }\textbf {\bibinfo {volume} {7}},\ \bibinfo {pages} {041046} (\bibinfo {year} {2017})}\BibitemShut {NoStop}%
\bibitem [{\citenamefont {Haase}\ \emph {et~al.}(2021)\citenamefont {Haase}, \citenamefont {Dellantonio}, \citenamefont {Celi}, \citenamefont {Paulson}, \citenamefont {Kan}, \citenamefont {Jansen},\ and\ \citenamefont {Muschik}}]{haase2021resource}%
  \BibitemOpen
  \bibfield  {author} {\bibinfo {author} {\bibfnamefont {J.~F.}\ \bibnamefont {Haase}}, \bibinfo {author} {\bibfnamefont {L.}~\bibnamefont {Dellantonio}}, \bibinfo {author} {\bibfnamefont {A.}~\bibnamefont {Celi}}, \bibinfo {author} {\bibfnamefont {D.}~\bibnamefont {Paulson}}, \bibinfo {author} {\bibfnamefont {A.}~\bibnamefont {Kan}}, \bibinfo {author} {\bibfnamefont {K.}~\bibnamefont {Jansen}},\ and\ \bibinfo {author} {\bibfnamefont {C.~A.}\ \bibnamefont {Muschik}},\ }\bibfield  {title} {\bibinfo {title} {A resource efficient approach for quantum and classical simulations of gauge theories in particle physics},\ }\href {https://doi.org/10.22331/q-2021-02-04-393} {\bibfield  {journal} {\bibinfo  {journal} {Quantum}\ }\textbf {\bibinfo {volume} {5}},\ \bibinfo {pages} {393} (\bibinfo {year} {2021})}\BibitemShut {NoStop}%
\bibitem [{\citenamefont {Bruckmann}\ \emph {et~al.}(2019)\citenamefont {Bruckmann}, \citenamefont {Jansen},\ and\ \citenamefont {K\"uhn}}]{bruckmann20193}%
  \BibitemOpen
  \bibfield  {author} {\bibinfo {author} {\bibfnamefont {F.}~\bibnamefont {Bruckmann}}, \bibinfo {author} {\bibfnamefont {K.}~\bibnamefont {Jansen}},\ and\ \bibinfo {author} {\bibfnamefont {S.}~\bibnamefont {K\"uhn}},\ }\bibfield  {title} {\bibinfo {title} {O(3) nonlinear sigma model in $1+1$ dimensions with matrix product states},\ }\href {https://doi.org/10.1103/PhysRevD.99.074501} {\bibfield  {journal} {\bibinfo  {journal} {Phys. Rev. D}\ }\textbf {\bibinfo {volume} {99}},\ \bibinfo {pages} {074501} (\bibinfo {year} {2019})}\BibitemShut {NoStop}%
\bibitem [{\citenamefont {Zache}\ \emph {et~al.}(2022)\citenamefont {Zache}, \citenamefont {Van~Damme}, \citenamefont {Halimeh}, \citenamefont {Hauke},\ and\ \citenamefont {Banerjee}}]{zache2022toward}%
  \BibitemOpen
  \bibfield  {author} {\bibinfo {author} {\bibfnamefont {T.~V.}\ \bibnamefont {Zache}}, \bibinfo {author} {\bibfnamefont {M.}~\bibnamefont {Van~Damme}}, \bibinfo {author} {\bibfnamefont {J.~C.}\ \bibnamefont {Halimeh}}, \bibinfo {author} {\bibfnamefont {P.}~\bibnamefont {Hauke}},\ and\ \bibinfo {author} {\bibfnamefont {D.}~\bibnamefont {Banerjee}},\ }\bibfield  {title} {\bibinfo {title} {Toward the continuum limit of a $(1+1)\mathrm{D}$ quantum link schwinger model},\ }\href {https://doi.org/10.1103/PhysRevD.106.L091502} {\bibfield  {journal} {\bibinfo  {journal} {Phys. Rev. D}\ }\textbf {\bibinfo {volume} {106}},\ \bibinfo {pages} {L091502} (\bibinfo {year} {2022})}\BibitemShut {NoStop}%
\bibitem [{\citenamefont {Alexandru}\ \emph {et~al.}(2023)\citenamefont {Alexandru}, \citenamefont {Bedaque}, \citenamefont {Carosso}, \citenamefont {Cervia},\ and\ \citenamefont {Sheng}}]{qubitboson}%
  \BibitemOpen
  \bibfield  {author} {\bibinfo {author} {\bibfnamefont {A.}~\bibnamefont {Alexandru}}, \bibinfo {author} {\bibfnamefont {P.~F.}\ \bibnamefont {Bedaque}}, \bibinfo {author} {\bibfnamefont {A.}~\bibnamefont {Carosso}}, \bibinfo {author} {\bibfnamefont {M.~J.}\ \bibnamefont {Cervia}},\ and\ \bibinfo {author} {\bibfnamefont {A.}~\bibnamefont {Sheng}},\ }\bibfield  {title} {\bibinfo {title} {Qubitization strategies for bosonic field theories},\ }\href {https://doi.org/10.1103/PhysRevD.107.034503} {\bibfield  {journal} {\bibinfo  {journal} {Phys. Rev. D}\ }\textbf {\bibinfo {volume} {107}},\ \bibinfo {pages} {034503} (\bibinfo {year} {2023})}\BibitemShut {NoStop}%
\bibitem [{\citenamefont {Araz}\ \emph {et~al.}(2023)\citenamefont {Araz}, \citenamefont {Schenk},\ and\ \citenamefont {Spannowsky}}]{araz2023toward}%
  \BibitemOpen
  \bibfield  {author} {\bibinfo {author} {\bibfnamefont {J.~Y.}\ \bibnamefont {Araz}}, \bibinfo {author} {\bibfnamefont {S.}~\bibnamefont {Schenk}},\ and\ \bibinfo {author} {\bibfnamefont {M.}~\bibnamefont {Spannowsky}},\ }\bibfield  {title} {\bibinfo {title} {Toward a quantum simulation of nonlinear sigma models with a topological term},\ }\href {https://doi.org/10.1103/PhysRevA.107.032619} {\bibfield  {journal} {\bibinfo  {journal} {Phys. Rev. A}\ }\textbf {\bibinfo {volume} {107}},\ \bibinfo {pages} {032619} (\bibinfo {year} {2023})}\BibitemShut {NoStop}%
\bibitem [{\citenamefont {Liu}\ \emph {et~al.}(2023)\citenamefont {Liu}, \citenamefont {Bhattacharya}, \citenamefont {Chandrasekharan},\ and\ \citenamefont {Gupta}}]{liu2023phases}%
  \BibitemOpen
  \bibfield  {author} {\bibinfo {author} {\bibfnamefont {H.}~\bibnamefont {Liu}}, \bibinfo {author} {\bibfnamefont {T.}~\bibnamefont {Bhattacharya}}, \bibinfo {author} {\bibfnamefont {S.}~\bibnamefont {Chandrasekharan}},\ and\ \bibinfo {author} {\bibfnamefont {R.}~\bibnamefont {Gupta}},\ }\href@noop {} {\bibinfo {title} {Phases of 2d massless qcd with qubit regularization}} (\bibinfo {year} {2023}),\ \Eprint {https://arxiv.org/abs/2312.17734} {arXiv:2312.17734 [hep-lat]} \BibitemShut {NoStop}%
\bibitem [{\citenamefont {M\"uller}\ and\ \citenamefont {Yao}(2023)}]{muller2023simple}%
  \BibitemOpen
  \bibfield  {author} {\bibinfo {author} {\bibfnamefont {B.}~\bibnamefont {M\"uller}}\ and\ \bibinfo {author} {\bibfnamefont {X.}~\bibnamefont {Yao}},\ }\bibfield  {title} {\bibinfo {title} {Simple hamiltonian for quantum simulation of strongly coupled $(2+1)\mathrm{D}$ su(2) lattice gauge theory on a honeycomb lattice},\ }\href {https://doi.org/10.1103/PhysRevD.108.094505} {\bibfield  {journal} {\bibinfo  {journal} {Phys. Rev. D}\ }\textbf {\bibinfo {volume} {108}},\ \bibinfo {pages} {094505} (\bibinfo {year} {2023})}\BibitemShut {NoStop}%
\bibitem [{\citenamefont {Ebner}\ \emph {et~al.}(2024{\natexlab{a}})\citenamefont {Ebner}, \citenamefont {Sch\"afer}, \citenamefont {Seidl}, \citenamefont {M\"uller},\ and\ \citenamefont {Yao}}]{ebner2024eigenstate}%
  \BibitemOpen
  \bibfield  {author} {\bibinfo {author} {\bibfnamefont {L.}~\bibnamefont {Ebner}}, \bibinfo {author} {\bibfnamefont {A.}~\bibnamefont {Sch\"afer}}, \bibinfo {author} {\bibfnamefont {C.}~\bibnamefont {Seidl}}, \bibinfo {author} {\bibfnamefont {B.}~\bibnamefont {M\"uller}},\ and\ \bibinfo {author} {\bibfnamefont {X.}~\bibnamefont {Yao}},\ }\bibfield  {title} {\bibinfo {title} {Eigenstate thermalization in ($2+1$)-dimensional su(2) lattice gauge theory},\ }\href {https://doi.org/10.1103/PhysRevD.109.014504} {\bibfield  {journal} {\bibinfo  {journal} {Phys. Rev. D}\ }\textbf {\bibinfo {volume} {109}},\ \bibinfo {pages} {014504} (\bibinfo {year} {2024}{\natexlab{a}})}\BibitemShut {NoStop}%
\bibitem [{\citenamefont {Yao}(2023)}]{yao20232}%
  \BibitemOpen
  \bibfield  {author} {\bibinfo {author} {\bibfnamefont {X.}~\bibnamefont {Yao}},\ }\bibfield  {title} {\bibinfo {title} {Su(2) gauge theory in $2+1$ dimensions on a plaquette chain obeys the eigenstate thermalization hypothesis},\ }\href {https://doi.org/10.1103/PhysRevD.108.L031504} {\bibfield  {journal} {\bibinfo  {journal} {Phys. Rev. D}\ }\textbf {\bibinfo {volume} {108}},\ \bibinfo {pages} {L031504} (\bibinfo {year} {2023})}\BibitemShut {NoStop}%
\bibitem [{\citenamefont {Yao}\ \emph {et~al.}(2023)\citenamefont {Yao}, \citenamefont {Ebner}, \citenamefont {Müller}, \citenamefont {Schäfer},\ and\ \citenamefont {Seidl}}]{yao2023testing}%
  \BibitemOpen
  \bibfield  {author} {\bibinfo {author} {\bibfnamefont {X.}~\bibnamefont {Yao}}, \bibinfo {author} {\bibfnamefont {L.}~\bibnamefont {Ebner}}, \bibinfo {author} {\bibfnamefont {B.}~\bibnamefont {Müller}}, \bibinfo {author} {\bibfnamefont {A.}~\bibnamefont {Schäfer}},\ and\ \bibinfo {author} {\bibfnamefont {C.}~\bibnamefont {Seidl}},\ }\href@noop {} {\bibinfo {title} {Testing eigenstate thermalization hypothesis for non-abelian gauge theories}} (\bibinfo {year} {2023}),\ \Eprint {https://arxiv.org/abs/2312.13408} {arXiv:2312.13408 [hep-lat]} \BibitemShut {NoStop}%
\bibitem [{\citenamefont {Ebner}\ \emph {et~al.}(2024{\natexlab{b}})\citenamefont {Ebner}, \citenamefont {Sch\"afer}, \citenamefont {Seidl}, \citenamefont {M\"uller},\ and\ \citenamefont {Yao}}]{ebner2024entanglement}%
  \BibitemOpen
  \bibfield  {author} {\bibinfo {author} {\bibfnamefont {L.}~\bibnamefont {Ebner}}, \bibinfo {author} {\bibfnamefont {A.}~\bibnamefont {Sch\"afer}}, \bibinfo {author} {\bibfnamefont {C.}~\bibnamefont {Seidl}}, \bibinfo {author} {\bibfnamefont {B.}~\bibnamefont {M\"uller}},\ and\ \bibinfo {author} {\bibfnamefont {X.}~\bibnamefont {Yao}},\ }\bibfield  {title} {\bibinfo {title} {Entanglement entropy of ($2+1$)-dimensional su(2) lattice gauge theory on plaquette chains},\ }\href {https://doi.org/10.1103/PhysRevD.110.014505} {\bibfield  {journal} {\bibinfo  {journal} {Phys. Rev. D}\ }\textbf {\bibinfo {volume} {110}},\ \bibinfo {pages} {014505} (\bibinfo {year} {2024}{\natexlab{b}})}\BibitemShut {NoStop}%
\bibitem [{\citenamefont {Turro}\ \emph {et~al.}(2024)\citenamefont {Turro}, \citenamefont {Ciavarella},\ and\ \citenamefont {Yao}}]{turro2024classical}%
  \BibitemOpen
  \bibfield  {author} {\bibinfo {author} {\bibfnamefont {F.}~\bibnamefont {Turro}}, \bibinfo {author} {\bibfnamefont {A.}~\bibnamefont {Ciavarella}},\ and\ \bibinfo {author} {\bibfnamefont {X.}~\bibnamefont {Yao}},\ }\bibfield  {title} {\bibinfo {title} {Classical and quantum computing of shear viscosity for $(2+1)d$ su(2) gauge theory},\ }\href {https://doi.org/10.1103/PhysRevD.109.114511} {\bibfield  {journal} {\bibinfo  {journal} {Phys. Rev. D}\ }\textbf {\bibinfo {volume} {109}},\ \bibinfo {pages} {114511} (\bibinfo {year} {2024})}\BibitemShut {NoStop}%
\bibitem [{\citenamefont {Ebadi}\ \emph {et~al.}(2021)\citenamefont {Ebadi}, \citenamefont {Wang}, \citenamefont {Levine}, \citenamefont {Keesling}, \citenamefont {Semeghini}, \citenamefont {Omran}, \citenamefont {Bluvstein}, \citenamefont {Samajdar}, \citenamefont {Pichler}, \citenamefont {Ho}, \citenamefont {Choi}, \citenamefont {Sachdev}, \citenamefont {Greiner}, \citenamefont {Vuleti{\'{c}}},\ and\ \citenamefont {Lukin}}]{ebadi2021quantum}%
  \BibitemOpen
  \bibfield  {author} {\bibinfo {author} {\bibfnamefont {S.}~\bibnamefont {Ebadi}}, \bibinfo {author} {\bibfnamefont {T.~T.}\ \bibnamefont {Wang}}, \bibinfo {author} {\bibfnamefont {H.}~\bibnamefont {Levine}}, \bibinfo {author} {\bibfnamefont {A.}~\bibnamefont {Keesling}}, \bibinfo {author} {\bibfnamefont {G.}~\bibnamefont {Semeghini}}, \bibinfo {author} {\bibfnamefont {A.}~\bibnamefont {Omran}}, \bibinfo {author} {\bibfnamefont {D.}~\bibnamefont {Bluvstein}}, \bibinfo {author} {\bibfnamefont {R.}~\bibnamefont {Samajdar}}, \bibinfo {author} {\bibfnamefont {H.}~\bibnamefont {Pichler}}, \bibinfo {author} {\bibfnamefont {W.~W.}\ \bibnamefont {Ho}}, \bibinfo {author} {\bibfnamefont {S.}~\bibnamefont {Choi}}, \bibinfo {author} {\bibfnamefont {S.}~\bibnamefont {Sachdev}}, \bibinfo {author} {\bibfnamefont {M.}~\bibnamefont {Greiner}}, \bibinfo {author} {\bibfnamefont {V.}~\bibnamefont {Vuleti{\'{c}}}},\ and\ \bibinfo {author} {\bibfnamefont {M.~D.}\ \bibnamefont {Lukin}},\ }\bibfield  {title} {\bibinfo {title}
  {Quantum phases of matter on a 256-atom programmable quantum simulator},\ }\href {https://doi.org/10.1038/s41586-021-03582-4} {\bibfield  {journal} {\bibinfo  {journal} {Nature}\ }\textbf {\bibinfo {volume} {595}},\ \bibinfo {pages} {227} (\bibinfo {year} {2021})}\BibitemShut {NoStop}%
\bibitem [{\citenamefont {Semeghini}\ \emph {et~al.}(2021)\citenamefont {Semeghini}, \citenamefont {Levine}, \citenamefont {Keesling}, \citenamefont {Ebadi}, \citenamefont {Wang}, \citenamefont {Bluvstein}, \citenamefont {Verresen}, \citenamefont {Pichler}, \citenamefont {Kalinowski}, \citenamefont {Samajdar}, \citenamefont {Omran}, \citenamefont {Sachdev}, \citenamefont {Vishwanath}, \citenamefont {Greiner}, \citenamefont {Vuletić},\ and\ \citenamefont {Lukin}}]{semeghini2021probing}%
  \BibitemOpen
  \bibfield  {author} {\bibinfo {author} {\bibfnamefont {G.}~\bibnamefont {Semeghini}}, \bibinfo {author} {\bibfnamefont {H.}~\bibnamefont {Levine}}, \bibinfo {author} {\bibfnamefont {A.}~\bibnamefont {Keesling}}, \bibinfo {author} {\bibfnamefont {S.}~\bibnamefont {Ebadi}}, \bibinfo {author} {\bibfnamefont {T.~T.}\ \bibnamefont {Wang}}, \bibinfo {author} {\bibfnamefont {D.}~\bibnamefont {Bluvstein}}, \bibinfo {author} {\bibfnamefont {R.}~\bibnamefont {Verresen}}, \bibinfo {author} {\bibfnamefont {H.}~\bibnamefont {Pichler}}, \bibinfo {author} {\bibfnamefont {M.}~\bibnamefont {Kalinowski}}, \bibinfo {author} {\bibfnamefont {R.}~\bibnamefont {Samajdar}}, \bibinfo {author} {\bibfnamefont {A.}~\bibnamefont {Omran}}, \bibinfo {author} {\bibfnamefont {S.}~\bibnamefont {Sachdev}}, \bibinfo {author} {\bibfnamefont {A.}~\bibnamefont {Vishwanath}}, \bibinfo {author} {\bibfnamefont {M.}~\bibnamefont {Greiner}}, \bibinfo {author} {\bibfnamefont {V.}~\bibnamefont {Vuletić}},\ and\ \bibinfo {author} {\bibfnamefont
  {M.~D.}\ \bibnamefont {Lukin}},\ }\bibfield  {title} {\bibinfo {title} {Probing topological spin liquids on a programmable quantum simulator},\ }\href {https://doi.org/10.1126/science.abi8794} {\bibfield  {journal} {\bibinfo  {journal} {Science}\ }\textbf {\bibinfo {volume} {374}},\ \bibinfo {pages} {1242} (\bibinfo {year} {2021})},\ \Eprint {https://arxiv.org/abs/https://www.science.org/doi/pdf/10.1126/science.abi8794} {https://www.science.org/doi/pdf/10.1126/science.abi8794} \BibitemShut {NoStop}%
\bibitem [{\citenamefont {Omran}\ \emph {et~al.}(2019)\citenamefont {Omran}, \citenamefont {Levine}, \citenamefont {Keesling}, \citenamefont {Semeghini}, \citenamefont {Wang}, \citenamefont {Ebadi}, \citenamefont {Bernien}, \citenamefont {Zibrov}, \citenamefont {Pichler}, \citenamefont {Choi}, \citenamefont {Cui}, \citenamefont {Rossignolo}, \citenamefont {Rembold}, \citenamefont {Montangero}, \citenamefont {Calarco}, \citenamefont {Endres}, \citenamefont {Greiner}, \citenamefont {Vuleti{\'{c} }},\ and\ \citenamefont {Lukin}}]{Omran_2019}%
  \BibitemOpen
  \bibfield  {author} {\bibinfo {author} {\bibfnamefont {A.}~\bibnamefont {Omran}}, \bibinfo {author} {\bibfnamefont {H.}~\bibnamefont {Levine}}, \bibinfo {author} {\bibfnamefont {A.}~\bibnamefont {Keesling}}, \bibinfo {author} {\bibfnamefont {G.}~\bibnamefont {Semeghini}}, \bibinfo {author} {\bibfnamefont {T.~T.}\ \bibnamefont {Wang}}, \bibinfo {author} {\bibfnamefont {S.}~\bibnamefont {Ebadi}}, \bibinfo {author} {\bibfnamefont {H.}~\bibnamefont {Bernien}}, \bibinfo {author} {\bibfnamefont {A.~S.}\ \bibnamefont {Zibrov}}, \bibinfo {author} {\bibfnamefont {H.}~\bibnamefont {Pichler}}, \bibinfo {author} {\bibfnamefont {S.}~\bibnamefont {Choi}}, \bibinfo {author} {\bibfnamefont {J.}~\bibnamefont {Cui}}, \bibinfo {author} {\bibfnamefont {M.}~\bibnamefont {Rossignolo}}, \bibinfo {author} {\bibfnamefont {P.}~\bibnamefont {Rembold}}, \bibinfo {author} {\bibfnamefont {S.}~\bibnamefont {Montangero}}, \bibinfo {author} {\bibfnamefont {T.}~\bibnamefont {Calarco}}, \bibinfo {author} {\bibfnamefont {M.}~\bibnamefont
  {Endres}}, \bibinfo {author} {\bibfnamefont {M.}~\bibnamefont {Greiner}}, \bibinfo {author} {\bibfnamefont {V.}~\bibnamefont {Vuleti{\'{c} }}},\ and\ \bibinfo {author} {\bibfnamefont {M.~D.}\ \bibnamefont {Lukin}},\ }\bibfield  {title} {\bibinfo {title} {Generation and manipulation of schrödinger cat states in rydberg atom arrays},\ }\href {https://doi.org/10.1126/science.aax9743} {\bibfield  {journal} {\bibinfo  {journal} {Science}\ }\textbf {\bibinfo {volume} {365}},\ \bibinfo {pages} {570} (\bibinfo {year} {2019})}\BibitemShut {NoStop}%
\bibitem [{\citenamefont {Nandkishore}\ and\ \citenamefont {Huse}(2015)}]{nandkishore2015many}%
  \BibitemOpen
  \bibfield  {author} {\bibinfo {author} {\bibfnamefont {R.}~\bibnamefont {Nandkishore}}\ and\ \bibinfo {author} {\bibfnamefont {D.~A.}\ \bibnamefont {Huse}},\ }\bibfield  {title} {\bibinfo {title} {Many-body localization and thermalization in quantum statistical mechanics},\ }\href {https://doi.org/10.1146/annurev-conmatphys-031214-014726} {\bibfield  {journal} {\bibinfo  {journal} {Annual Review of Condensed Matter Physics}\ }\textbf {\bibinfo {volume} {6}},\ \bibinfo {pages} {15} (\bibinfo {year} {2015})},\ \Eprint {https://arxiv.org/abs/https://doi.org/10.1146/annurev-conmatphys-031214-014726} {https://doi.org/10.1146/annurev-conmatphys-031214-014726} \BibitemShut {NoStop}%
\bibitem [{\citenamefont {Choi}\ \emph {et~al.}(2019)\citenamefont {Choi}, \citenamefont {Turner}, \citenamefont {Pichler}, \citenamefont {Ho}, \citenamefont {Michailidis}, \citenamefont {Papi\ifmmode~\acute{c}\else \'{c}\fi{}}, \citenamefont {Serbyn}, \citenamefont {Lukin},\ and\ \citenamefont {Abanin}}]{choi2019emergent}%
  \BibitemOpen
  \bibfield  {author} {\bibinfo {author} {\bibfnamefont {S.}~\bibnamefont {Choi}}, \bibinfo {author} {\bibfnamefont {C.~J.}\ \bibnamefont {Turner}}, \bibinfo {author} {\bibfnamefont {H.}~\bibnamefont {Pichler}}, \bibinfo {author} {\bibfnamefont {W.~W.}\ \bibnamefont {Ho}}, \bibinfo {author} {\bibfnamefont {A.~A.}\ \bibnamefont {Michailidis}}, \bibinfo {author} {\bibfnamefont {Z.}~\bibnamefont {Papi\ifmmode~\acute{c}\else \'{c}\fi{}}}, \bibinfo {author} {\bibfnamefont {M.}~\bibnamefont {Serbyn}}, \bibinfo {author} {\bibfnamefont {M.~D.}\ \bibnamefont {Lukin}},\ and\ \bibinfo {author} {\bibfnamefont {D.~A.}\ \bibnamefont {Abanin}},\ }\bibfield  {title} {\bibinfo {title} {Emergent su(2) dynamics and perfect quantum many-body scars},\ }\href {https://doi.org/10.1103/PhysRevLett.122.220603} {\bibfield  {journal} {\bibinfo  {journal} {Phys. Rev. Lett.}\ }\textbf {\bibinfo {volume} {122}},\ \bibinfo {pages} {220603} (\bibinfo {year} {2019})}\BibitemShut {NoStop}%
\bibitem [{\citenamefont {Moudgalya}\ \emph {et~al.}(2022)\citenamefont {Moudgalya}, \citenamefont {Bernevig},\ and\ \citenamefont {Regnault}}]{Moudgalya_2022}%
  \BibitemOpen
  \bibfield  {author} {\bibinfo {author} {\bibfnamefont {S.}~\bibnamefont {Moudgalya}}, \bibinfo {author} {\bibfnamefont {B.~A.}\ \bibnamefont {Bernevig}},\ and\ \bibinfo {author} {\bibfnamefont {N.}~\bibnamefont {Regnault}},\ }\bibfield  {title} {\bibinfo {title} {Quantum many-body scars and hilbert space fragmentation: a review of exact results},\ }\href {https://doi.org/10.1088/1361-6633/ac73a0} {\bibfield  {journal} {\bibinfo  {journal} {Reports on Progress in Physics}\ }\textbf {\bibinfo {volume} {85}},\ \bibinfo {pages} {086501} (\bibinfo {year} {2022})}\BibitemShut {NoStop}%
\bibitem [{\citenamefont {Chandran}\ \emph {et~al.}(2023)\citenamefont {Chandran}, \citenamefont {Iadecola}, \citenamefont {Khemani},\ and\ \citenamefont {Moessner}}]{chandran2023quantum}%
  \BibitemOpen
  \bibfield  {author} {\bibinfo {author} {\bibfnamefont {A.}~\bibnamefont {Chandran}}, \bibinfo {author} {\bibfnamefont {T.}~\bibnamefont {Iadecola}}, \bibinfo {author} {\bibfnamefont {V.}~\bibnamefont {Khemani}},\ and\ \bibinfo {author} {\bibfnamefont {R.}~\bibnamefont {Moessner}},\ }\bibfield  {title} {\bibinfo {title} {Quantum many-body scars: A quasiparticle perspective},\ }\href {https://doi.org/10.1146/annurev-conmatphys-031620-101617} {\bibfield  {journal} {\bibinfo  {journal} {Annual Review of Condensed Matter Physics}\ }\textbf {\bibinfo {volume} {14}},\ \bibinfo {pages} {443} (\bibinfo {year} {2023})},\ \Eprint {https://arxiv.org/abs/https://doi.org/10.1146/annurev-conmatphys-031620-101617} {https://doi.org/10.1146/annurev-conmatphys-031620-101617} \BibitemShut {NoStop}%
\bibitem [{\citenamefont {Surace}\ \emph {et~al.}(2020)\citenamefont {Surace}, \citenamefont {Mazza}, \citenamefont {Giudici}, \citenamefont {Lerose}, \citenamefont {Gambassi},\ and\ \citenamefont {Dalmonte}}]{surace2020lattice}%
  \BibitemOpen
  \bibfield  {author} {\bibinfo {author} {\bibfnamefont {F.~M.}\ \bibnamefont {Surace}}, \bibinfo {author} {\bibfnamefont {P.~P.}\ \bibnamefont {Mazza}}, \bibinfo {author} {\bibfnamefont {G.}~\bibnamefont {Giudici}}, \bibinfo {author} {\bibfnamefont {A.}~\bibnamefont {Lerose}}, \bibinfo {author} {\bibfnamefont {A.}~\bibnamefont {Gambassi}},\ and\ \bibinfo {author} {\bibfnamefont {M.}~\bibnamefont {Dalmonte}},\ }\bibfield  {title} {\bibinfo {title} {Lattice gauge theories and string dynamics in rydberg atom quantum simulators},\ }\href {https://doi.org/10.1103/PhysRevX.10.021041} {\bibfield  {journal} {\bibinfo  {journal} {Phys. Rev. X}\ }\textbf {\bibinfo {volume} {10}},\ \bibinfo {pages} {021041} (\bibinfo {year} {2020})}\BibitemShut {NoStop}%
\bibitem [{\citenamefont {Kormos}\ \emph {et~al.}(2016)\citenamefont {Kormos}, \citenamefont {Collura}, \citenamefont {Tak{\'{a}}cs},\ and\ \citenamefont {Calabrese}}]{kormos2017real}%
  \BibitemOpen
  \bibfield  {author} {\bibinfo {author} {\bibfnamefont {M.}~\bibnamefont {Kormos}}, \bibinfo {author} {\bibfnamefont {M.}~\bibnamefont {Collura}}, \bibinfo {author} {\bibfnamefont {G.}~\bibnamefont {Tak{\'{a}}cs}},\ and\ \bibinfo {author} {\bibfnamefont {P.}~\bibnamefont {Calabrese}},\ }\bibfield  {title} {\bibinfo {title} {Real-time confinement following a quantum quench to a non-integrable model},\ }\href {https://doi.org/10.1038/nphys3934} {\bibfield  {journal} {\bibinfo  {journal} {Nature Physics}\ }\textbf {\bibinfo {volume} {13}},\ \bibinfo {pages} {246} (\bibinfo {year} {2016})}\BibitemShut {NoStop}%
\bibitem [{\citenamefont {James}\ \emph {et~al.}(2019)\citenamefont {James}, \citenamefont {Konik},\ and\ \citenamefont {Robinson}}]{james2019nonthermal}%
  \BibitemOpen
  \bibfield  {author} {\bibinfo {author} {\bibfnamefont {A.~J.~A.}\ \bibnamefont {James}}, \bibinfo {author} {\bibfnamefont {R.~M.}\ \bibnamefont {Konik}},\ and\ \bibinfo {author} {\bibfnamefont {N.~J.}\ \bibnamefont {Robinson}},\ }\bibfield  {title} {\bibinfo {title} {Nonthermal states arising from confinement in one and two dimensions},\ }\href {https://doi.org/10.1103/PhysRevLett.122.130603} {\bibfield  {journal} {\bibinfo  {journal} {Phys. Rev. Lett.}\ }\textbf {\bibinfo {volume} {122}},\ \bibinfo {pages} {130603} (\bibinfo {year} {2019})}\BibitemShut {NoStop}%
\bibitem [{\citenamefont {Robinson}\ \emph {et~al.}(2019)\citenamefont {Robinson}, \citenamefont {James},\ and\ \citenamefont {Konik}}]{robinson2019signatures}%
  \BibitemOpen
  \bibfield  {author} {\bibinfo {author} {\bibfnamefont {N.~J.}\ \bibnamefont {Robinson}}, \bibinfo {author} {\bibfnamefont {A.~J.~A.}\ \bibnamefont {James}},\ and\ \bibinfo {author} {\bibfnamefont {R.~M.}\ \bibnamefont {Konik}},\ }\bibfield  {title} {\bibinfo {title} {Signatures of rare states and thermalization in a theory with confinement},\ }\href {https://doi.org/10.1103/PhysRevB.99.195108} {\bibfield  {journal} {\bibinfo  {journal} {Phys. Rev. B}\ }\textbf {\bibinfo {volume} {99}},\ \bibinfo {pages} {195108} (\bibinfo {year} {2019})}\BibitemShut {NoStop}%
\bibitem [{\citenamefont {Bravyi}\ \emph {et~al.}(2007)\citenamefont {Bravyi}, \citenamefont {DiVincenzo}, \citenamefont {Oliveira},\ and\ \citenamefont {Terhal}}]{bravyi2007complexity}%
  \BibitemOpen
  \bibfield  {author} {\bibinfo {author} {\bibfnamefont {S.}~\bibnamefont {Bravyi}}, \bibinfo {author} {\bibfnamefont {D.~P.}\ \bibnamefont {DiVincenzo}}, \bibinfo {author} {\bibfnamefont {R.~I.}\ \bibnamefont {Oliveira}},\ and\ \bibinfo {author} {\bibfnamefont {B.~M.}\ \bibnamefont {Terhal}},\ }\href {https://doi.org/10.48550/arXiv.quant-ph/0606140} {\bibinfo {title} {The complexity of stoquastic local hamiltonian problems}} (\bibinfo {year} {2007}),\ \Eprint {https://arxiv.org/abs/quant-ph/0606140} {arXiv:quant-ph/0606140 [quant-ph]} \BibitemShut {NoStop}%
\bibitem [{\citenamefont {Suzuki}\ \emph {et~al.}(2013)\citenamefont {Suzuki}, \citenamefont {Inoue}, \citenamefont {Chakrabarti}, \citenamefont {Suzuki}, \citenamefont {Inoue},\ and\ \citenamefont {Chakrabarti}}]{suzuki2013transverse}%
  \BibitemOpen
  \bibfield  {author} {\bibinfo {author} {\bibfnamefont {S.}~\bibnamefont {Suzuki}}, \bibinfo {author} {\bibfnamefont {J.-i.}\ \bibnamefont {Inoue}}, \bibinfo {author} {\bibfnamefont {B.~K.}\ \bibnamefont {Chakrabarti}}, \bibinfo {author} {\bibfnamefont {S.}~\bibnamefont {Suzuki}}, \bibinfo {author} {\bibfnamefont {J.-i.}\ \bibnamefont {Inoue}},\ and\ \bibinfo {author} {\bibfnamefont {B.~K.}\ \bibnamefont {Chakrabarti}},\ }\bibfield  {title} {\bibinfo {title} {Transverse ising system in higher dimensions (pure systems)},\ }\href@noop {} {\bibfield  {journal} {\bibinfo  {journal} {Quantum Ising Phases and Transitions in Transverse Ising Models}\ ,\ \bibinfo {pages} {47}} (\bibinfo {year} {2013})}\BibitemShut {NoStop}%
\bibitem [{\citenamefont {Lamm}\ and\ \citenamefont {Lawrence}(2018)}]{lamm2018simulation}%
  \BibitemOpen
  \bibfield  {author} {\bibinfo {author} {\bibfnamefont {H.}~\bibnamefont {Lamm}}\ and\ \bibinfo {author} {\bibfnamefont {S.}~\bibnamefont {Lawrence}},\ }\bibfield  {title} {\bibinfo {title} {Simulation of nonequilibrium dynamics on a quantum computer},\ }\href {https://doi.org/10.1103/PhysRevLett.121.170501} {\bibfield  {journal} {\bibinfo  {journal} {Phys. Rev. Lett.}\ }\textbf {\bibinfo {volume} {121}},\ \bibinfo {pages} {170501} (\bibinfo {year} {2018})}\BibitemShut {NoStop}%
\bibitem [{\citenamefont {Harmalkar}\ \emph {et~al.}(2020)\citenamefont {Harmalkar}, \citenamefont {Lamm},\ and\ \citenamefont {Lawrence}}]{harmalkar2020quantum}%
  \BibitemOpen
  \bibfield  {author} {\bibinfo {author} {\bibfnamefont {S.}~\bibnamefont {Harmalkar}}, \bibinfo {author} {\bibfnamefont {H.}~\bibnamefont {Lamm}},\ and\ \bibinfo {author} {\bibfnamefont {S.}~\bibnamefont {Lawrence}},\ }\href@noop {} {\bibinfo {title} {Quantum simulation of field theories without state preparation}} (\bibinfo {year} {2020}),\ \Eprint {https://arxiv.org/abs/2001.11490} {arXiv:2001.11490 [hep-lat]} \BibitemShut {NoStop}%
\bibitem [{\citenamefont {Gustafson}\ and\ \citenamefont {Lamm}(2021)}]{gustafson2021toward}%
  \BibitemOpen
  \bibfield  {author} {\bibinfo {author} {\bibfnamefont {E.~J.}\ \bibnamefont {Gustafson}}\ and\ \bibinfo {author} {\bibfnamefont {H.}~\bibnamefont {Lamm}},\ }\bibfield  {title} {\bibinfo {title} {Toward quantum simulations of ${\mathbb{z}}_{2}$ gauge theory without state preparation},\ }\href {https://doi.org/10.1103/PhysRevD.103.054507} {\bibfield  {journal} {\bibinfo  {journal} {Phys. Rev. D}\ }\textbf {\bibinfo {volume} {103}},\ \bibinfo {pages} {054507} (\bibinfo {year} {2021})}\BibitemShut {NoStop}%
\bibitem [{\citenamefont {Blunt}\ \emph {et~al.}(2014)\citenamefont {Blunt}, \citenamefont {Rogers}, \citenamefont {Spencer},\ and\ \citenamefont {Foulkes}}]{blunt2014density}%
  \BibitemOpen
  \bibfield  {author} {\bibinfo {author} {\bibfnamefont {N.~S.}\ \bibnamefont {Blunt}}, \bibinfo {author} {\bibfnamefont {T.~W.}\ \bibnamefont {Rogers}}, \bibinfo {author} {\bibfnamefont {J.~S.}\ \bibnamefont {Spencer}},\ and\ \bibinfo {author} {\bibfnamefont {W.~M.~C.}\ \bibnamefont {Foulkes}},\ }\bibfield  {title} {\bibinfo {title} {Density-matrix quantum monte carlo method},\ }\href {https://doi.org/10.1103/PhysRevB.89.245124} {\bibfield  {journal} {\bibinfo  {journal} {Phys. Rev. B}\ }\textbf {\bibinfo {volume} {89}},\ \bibinfo {pages} {245124} (\bibinfo {year} {2014})}\BibitemShut {NoStop}%
\bibitem [{\citenamefont {Saroni}\ \emph {et~al.}(2023)\citenamefont {Saroni}, \citenamefont {Lamm}, \citenamefont {Orth},\ and\ \citenamefont {Iadecola}}]{saroni2023reconstructing}%
  \BibitemOpen
  \bibfield  {author} {\bibinfo {author} {\bibfnamefont {J.}~\bibnamefont {Saroni}}, \bibinfo {author} {\bibfnamefont {H.}~\bibnamefont {Lamm}}, \bibinfo {author} {\bibfnamefont {P.~P.}\ \bibnamefont {Orth}},\ and\ \bibinfo {author} {\bibfnamefont {T.}~\bibnamefont {Iadecola}},\ }\bibfield  {title} {\bibinfo {title} {Reconstructing thermal quantum quench dynamics from pure states},\ }\href {https://doi.org/10.1103/PhysRevB.108.134301} {\bibfield  {journal} {\bibinfo  {journal} {Phys. Rev. B}\ }\textbf {\bibinfo {volume} {108}},\ \bibinfo {pages} {134301} (\bibinfo {year} {2023})}\BibitemShut {NoStop}%
\bibitem [{\citenamefont {Sommer}(2014)}]{sommer2014scale}%
  \BibitemOpen
  \bibfield  {author} {\bibinfo {author} {\bibfnamefont {R.}~\bibnamefont {Sommer}},\ }\bibfield  {title} {\bibinfo {title} {{Scale setting in lattice QCD}},\ }\href {https://doi.org/10.22323/1.187.0015} {\bibfield  {journal} {\bibinfo  {journal} {PoS}\ }\textbf {\bibinfo {volume} {LATTICE2013}},\ \bibinfo {pages} {015} (\bibinfo {year} {2014})},\ \Eprint {https://arxiv.org/abs/1401.3270} {arXiv:1401.3270 [hep-lat]} \BibitemShut {NoStop}%
\bibitem [{\citenamefont {Clemente}\ \emph {et~al.}(2022)\citenamefont {Clemente}, \citenamefont {Crippa},\ and\ \citenamefont {Jansen}}]{clemente2022strategies}%
  \BibitemOpen
  \bibfield  {author} {\bibinfo {author} {\bibfnamefont {G.}~\bibnamefont {Clemente}}, \bibinfo {author} {\bibfnamefont {A.}~\bibnamefont {Crippa}},\ and\ \bibinfo {author} {\bibfnamefont {K.}~\bibnamefont {Jansen}},\ }\bibfield  {title} {\bibinfo {title} {Strategies for the determination of the running coupling of ($2+1$)-dimensional qed with quantum computing},\ }\href {https://doi.org/10.1103/PhysRevD.106.114511} {\bibfield  {journal} {\bibinfo  {journal} {Phys. Rev. D}\ }\textbf {\bibinfo {volume} {106}},\ \bibinfo {pages} {114511} (\bibinfo {year} {2022})}\BibitemShut {NoStop}%
\bibitem [{\citenamefont {Ciavarella}\ \emph {et~al.}(2023)\citenamefont {Ciavarella}, \citenamefont {Caspar}, \citenamefont {Singh},\ and\ \citenamefont {Savage}}]{ciavarella2022preparationO3}%
  \BibitemOpen
  \bibfield  {author} {\bibinfo {author} {\bibfnamefont {A.~N.}\ \bibnamefont {Ciavarella}}, \bibinfo {author} {\bibfnamefont {S.}~\bibnamefont {Caspar}}, \bibinfo {author} {\bibfnamefont {H.}~\bibnamefont {Singh}},\ and\ \bibinfo {author} {\bibfnamefont {M.~J.}\ \bibnamefont {Savage}},\ }\bibfield  {title} {\bibinfo {title} {Preparation for quantum simulation of the $(1+1)$-dimensional o(3) nonlinear $\ensuremath{\sigma}$ model using cold atoms},\ }\href {https://doi.org/10.1103/PhysRevA.107.042404} {\bibfield  {journal} {\bibinfo  {journal} {Phys. Rev. A}\ }\textbf {\bibinfo {volume} {107}},\ \bibinfo {pages} {042404} (\bibinfo {year} {2023})}\BibitemShut {NoStop}%
\bibitem [{\citenamefont {Aleksandrowicz}\ \emph {et~al.}(2019)\citenamefont {Aleksandrowicz}, \citenamefont {Alexander}, \citenamefont {Barkoutsos}, \citenamefont {Bello}, \citenamefont {Ben-Haim}, \citenamefont {Bucher}, \citenamefont {Cabrera-Hern{\'a}ndez}, \citenamefont {Carballo-Franquis}, \citenamefont {Chen}, \citenamefont {Chen} \emph {et~al.}}]{aleksandrowicz2019qiskit}%
  \BibitemOpen
  \bibfield  {author} {\bibinfo {author} {\bibfnamefont {G.}~\bibnamefont {Aleksandrowicz}}, \bibinfo {author} {\bibfnamefont {T.}~\bibnamefont {Alexander}}, \bibinfo {author} {\bibfnamefont {P.}~\bibnamefont {Barkoutsos}}, \bibinfo {author} {\bibfnamefont {L.}~\bibnamefont {Bello}}, \bibinfo {author} {\bibfnamefont {Y.}~\bibnamefont {Ben-Haim}}, \bibinfo {author} {\bibfnamefont {D.}~\bibnamefont {Bucher}}, \bibinfo {author} {\bibfnamefont {F.~J.}\ \bibnamefont {Cabrera-Hern{\'a}ndez}}, \bibinfo {author} {\bibfnamefont {J.}~\bibnamefont {Carballo-Franquis}}, \bibinfo {author} {\bibfnamefont {A.}~\bibnamefont {Chen}}, \bibinfo {author} {\bibfnamefont {C.-F.}\ \bibnamefont {Chen}}, \emph {et~al.},\ }\bibfield  {title} {\bibinfo {title} {Qiskit: An open-source framework for quantum computing},\ }\href@noop {} {\bibfield  {journal} {\bibinfo  {journal} {Accessed on: Mar}\ }\textbf {\bibinfo {volume} {16}} (\bibinfo {year} {2019})}\BibitemShut {NoStop}%
\bibitem [{\citenamefont {{IBM Quantum Experience}}(2023)}]{ibmTorino}%
  \BibitemOpen
  \bibfield  {author} {\bibinfo {author} {\bibnamefont {{IBM Quantum Experience}}},\ }\href@noop {} {\bibinfo {title} {ibm\_torino v1.0.2}},\ \bibinfo {howpublished} {\url{https://quantum-computing.ibm.com}} (\bibinfo {year} {2023})\BibitemShut {NoStop}%
\bibitem [{\citenamefont {Viola}\ \emph {et~al.}(1999)\citenamefont {Viola}, \citenamefont {Knill},\ and\ \citenamefont {Lloyd}}]{viola1999dynamical}%
  \BibitemOpen
  \bibfield  {author} {\bibinfo {author} {\bibfnamefont {L.}~\bibnamefont {Viola}}, \bibinfo {author} {\bibfnamefont {E.}~\bibnamefont {Knill}},\ and\ \bibinfo {author} {\bibfnamefont {S.}~\bibnamefont {Lloyd}},\ }\bibfield  {title} {\bibinfo {title} {Dynamical decoupling of open quantum systems},\ }\href {https://doi.org/10.1103/PhysRevLett.82.2417} {\bibfield  {journal} {\bibinfo  {journal} {Phys. Rev. Lett.}\ }\textbf {\bibinfo {volume} {82}},\ \bibinfo {pages} {2417} (\bibinfo {year} {1999})}\BibitemShut {NoStop}%
\bibitem [{\citenamefont {Urbanek}\ \emph {et~al.}(2021)\citenamefont {Urbanek}, \citenamefont {Nachman}, \citenamefont {Pascuzzi}, \citenamefont {He}, \citenamefont {Bauer},\ and\ \citenamefont {de~Jong}}]{urbanek2021mitigating}%
  \BibitemOpen
  \bibfield  {author} {\bibinfo {author} {\bibfnamefont {M.}~\bibnamefont {Urbanek}}, \bibinfo {author} {\bibfnamefont {B.}~\bibnamefont {Nachman}}, \bibinfo {author} {\bibfnamefont {V.~R.}\ \bibnamefont {Pascuzzi}}, \bibinfo {author} {\bibfnamefont {A.}~\bibnamefont {He}}, \bibinfo {author} {\bibfnamefont {C.~W.}\ \bibnamefont {Bauer}},\ and\ \bibinfo {author} {\bibfnamefont {W.~A.}\ \bibnamefont {de~Jong}},\ }\bibfield  {title} {\bibinfo {title} {Mitigating depolarizing noise on quantum computers with noise-estimation circuits},\ }\href {https://doi.org/10.1103/PhysRevLett.127.270502} {\bibfield  {journal} {\bibinfo  {journal} {Phys. Rev. Lett.}\ }\textbf {\bibinfo {volume} {127}},\ \bibinfo {pages} {270502} (\bibinfo {year} {2021})}\BibitemShut {NoStop}%
\bibitem [{\citenamefont {Asaduzzaman}\ \emph {et~al.}(2024)\citenamefont {Asaduzzaman}, \citenamefont {Jha},\ and\ \citenamefont {Sambasivam}}]{asaduzzaman2024model}%
  \BibitemOpen
  \bibfield  {author} {\bibinfo {author} {\bibfnamefont {M.}~\bibnamefont {Asaduzzaman}}, \bibinfo {author} {\bibfnamefont {R.~G.}\ \bibnamefont {Jha}},\ and\ \bibinfo {author} {\bibfnamefont {B.}~\bibnamefont {Sambasivam}},\ }\bibfield  {title} {\bibinfo {title} {Sachdev-ye-kitaev model on a noisy quantum computer},\ }\href {https://doi.org/10.1103/PhysRevD.109.105002} {\bibfield  {journal} {\bibinfo  {journal} {Phys. Rev. D}\ }\textbf {\bibinfo {volume} {109}},\ \bibinfo {pages} {105002} (\bibinfo {year} {2024})}\BibitemShut {NoStop}%
\bibitem [{\citenamefont {Hidalgo}\ and\ \citenamefont {Draper}(2024)}]{hidalgo2023quantum}%
  \BibitemOpen
  \bibfield  {author} {\bibinfo {author} {\bibfnamefont {L.}~\bibnamefont {Hidalgo}}\ and\ \bibinfo {author} {\bibfnamefont {P.}~\bibnamefont {Draper}},\ }\bibfield  {title} {\bibinfo {title} {Quantum simulations for strong-field qed},\ }\href {https://doi.org/10.1103/PhysRevD.109.076004} {\bibfield  {journal} {\bibinfo  {journal} {Phys. Rev. D}\ }\textbf {\bibinfo {volume} {109}},\ \bibinfo {pages} {076004} (\bibinfo {year} {2024})}\BibitemShut {NoStop}%
\bibitem [{\citenamefont {Kiss}\ \emph {et~al.}(2024)\citenamefont {Kiss}, \citenamefont {Grossi},\ and\ \citenamefont {Roggero}}]{kiss2024quantum}%
  \BibitemOpen
  \bibfield  {author} {\bibinfo {author} {\bibfnamefont {O.}~\bibnamefont {Kiss}}, \bibinfo {author} {\bibfnamefont {M.}~\bibnamefont {Grossi}},\ and\ \bibinfo {author} {\bibfnamefont {A.}~\bibnamefont {Roggero}},\ }\href@noop {} {\bibinfo {title} {Quantum error mitigation for fourier moment computation}} (\bibinfo {year} {2024}),\ \Eprint {https://arxiv.org/abs/2401.13048} {arXiv:2401.13048 [quant-ph]} \BibitemShut {NoStop}%
\bibitem [{\citenamefont {He}\ \emph {et~al.}(2020)\citenamefont {He}, \citenamefont {Nachman}, \citenamefont {de~Jong},\ and\ \citenamefont {Bauer}}]{he2020zero}%
  \BibitemOpen
  \bibfield  {author} {\bibinfo {author} {\bibfnamefont {A.}~\bibnamefont {He}}, \bibinfo {author} {\bibfnamefont {B.}~\bibnamefont {Nachman}}, \bibinfo {author} {\bibfnamefont {W.~A.}\ \bibnamefont {de~Jong}},\ and\ \bibinfo {author} {\bibfnamefont {C.~W.}\ \bibnamefont {Bauer}},\ }\bibfield  {title} {\bibinfo {title} {Zero-noise extrapolation for quantum-gate error mitigation with identity insertions},\ }\href {https://doi.org/10.1103/PhysRevA.102.012426} {\bibfield  {journal} {\bibinfo  {journal} {Phys. Rev. A}\ }\textbf {\bibinfo {volume} {102}},\ \bibinfo {pages} {012426} (\bibinfo {year} {2020})}\BibitemShut {NoStop}%
\bibitem [{\citenamefont {Pascuzzi}\ \emph {et~al.}(2022)\citenamefont {Pascuzzi}, \citenamefont {He}, \citenamefont {Bauer}, \citenamefont {de~Jong},\ and\ \citenamefont {Nachman}}]{pascuzzi2022computationally}%
  \BibitemOpen
  \bibfield  {author} {\bibinfo {author} {\bibfnamefont {V.~R.}\ \bibnamefont {Pascuzzi}}, \bibinfo {author} {\bibfnamefont {A.}~\bibnamefont {He}}, \bibinfo {author} {\bibfnamefont {C.~W.}\ \bibnamefont {Bauer}}, \bibinfo {author} {\bibfnamefont {W.~A.}\ \bibnamefont {de~Jong}},\ and\ \bibinfo {author} {\bibfnamefont {B.}~\bibnamefont {Nachman}},\ }\bibfield  {title} {\bibinfo {title} {Computationally efficient zero-noise extrapolation for quantum-gate-error mitigation},\ }\href {https://doi.org/10.1103/PhysRevA.105.042406} {\bibfield  {journal} {\bibinfo  {journal} {Phys. Rev. A}\ }\textbf {\bibinfo {volume} {105}},\ \bibinfo {pages} {042406} (\bibinfo {year} {2022})}\BibitemShut {NoStop}%
\bibitem [{\citenamefont {van~den Berg}\ \emph {et~al.}(2022)\citenamefont {van~den Berg}, \citenamefont {Minev},\ and\ \citenamefont {Temme}}]{trexmit}%
  \BibitemOpen
  \bibfield  {author} {\bibinfo {author} {\bibfnamefont {E.}~\bibnamefont {van~den Berg}}, \bibinfo {author} {\bibfnamefont {Z.~K.}\ \bibnamefont {Minev}},\ and\ \bibinfo {author} {\bibfnamefont {K.}~\bibnamefont {Temme}},\ }\bibfield  {title} {\bibinfo {title} {Model-free readout-error mitigation for quantum expectation values},\ }\href {https://doi.org/10.1103/PhysRevA.105.032620} {\bibfield  {journal} {\bibinfo  {journal} {Phys. Rev. A}\ }\textbf {\bibinfo {volume} {105}},\ \bibinfo {pages} {032620} (\bibinfo {year} {2022})}\BibitemShut {NoStop}%
\bibitem [{Note1()}]{Note1}%
  \BibitemOpen
  \bibinfo {note} {The icons in the corners of plots indicate if classical or quantum compute resources were used to perform the calculation~\cite {klco2020minimally} and are available at {\protect \tt https://iqus.uw.edu/resources/icons/}}\BibitemShut {NoStop}%
\bibitem [{\citenamefont {Hayata}\ and\ \citenamefont {Hidaka}(2021)}]{hayata2021thermalization}%
  \BibitemOpen
  \bibfield  {author} {\bibinfo {author} {\bibfnamefont {T.}~\bibnamefont {Hayata}}\ and\ \bibinfo {author} {\bibfnamefont {Y.}~\bibnamefont {Hidaka}},\ }\bibfield  {title} {\bibinfo {title} {Thermalization of yang-mills theory in a ($3+1$)-dimensional small lattice system},\ }\href {https://doi.org/10.1103/PhysRevD.103.094502} {\bibfield  {journal} {\bibinfo  {journal} {Phys. Rev. D}\ }\textbf {\bibinfo {volume} {103}},\ \bibinfo {pages} {094502} (\bibinfo {year} {2021})}\BibitemShut {NoStop}%
\bibitem [{\citenamefont {White}(1992)}]{white1992density}%
  \BibitemOpen
  \bibfield  {author} {\bibinfo {author} {\bibfnamefont {S.~R.}\ \bibnamefont {White}},\ }\bibfield  {title} {\bibinfo {title} {Density matrix formulation for quantum renormalization groups},\ }\href {https://doi.org/10.1103/PhysRevLett.69.2863} {\bibfield  {journal} {\bibinfo  {journal} {Phys. Rev. Lett.}\ }\textbf {\bibinfo {volume} {69}},\ \bibinfo {pages} {2863} (\bibinfo {year} {1992})}\BibitemShut {NoStop}%
\bibitem [{\citenamefont {White}(1993)}]{white1993density}%
  \BibitemOpen
  \bibfield  {author} {\bibinfo {author} {\bibfnamefont {S.~R.}\ \bibnamefont {White}},\ }\bibfield  {title} {\bibinfo {title} {Density-matrix algorithms for quantum renormalization groups},\ }\href {https://doi.org/10.1103/PhysRevB.48.10345} {\bibfield  {journal} {\bibinfo  {journal} {Phys. Rev. B}\ }\textbf {\bibinfo {volume} {48}},\ \bibinfo {pages} {10345} (\bibinfo {year} {1993})}\BibitemShut {NoStop}%
\bibitem [{\citenamefont {Verstraete}\ \emph {et~al.}(2004)\citenamefont {Verstraete}, \citenamefont {Garc\'{\i}a-Ripoll},\ and\ \citenamefont {Cirac}}]{verstraete2004matrix}%
  \BibitemOpen
  \bibfield  {author} {\bibinfo {author} {\bibfnamefont {F.}~\bibnamefont {Verstraete}}, \bibinfo {author} {\bibfnamefont {J.~J.}\ \bibnamefont {Garc\'{\i}a-Ripoll}},\ and\ \bibinfo {author} {\bibfnamefont {J.~I.}\ \bibnamefont {Cirac}},\ }\bibfield  {title} {\bibinfo {title} {Matrix product density operators: Simulation of finite-temperature and dissipative systems},\ }\href {https://doi.org/10.1103/PhysRevLett.93.207204} {\bibfield  {journal} {\bibinfo  {journal} {Phys. Rev. Lett.}\ }\textbf {\bibinfo {volume} {93}},\ \bibinfo {pages} {207204} (\bibinfo {year} {2004})}\BibitemShut {NoStop}%
\bibitem [{\citenamefont {Haegeman}\ \emph {et~al.}(2011)\citenamefont {Haegeman}, \citenamefont {Cirac}, \citenamefont {Osborne}, \citenamefont {Pi\ifmmode~\check{z}\else \v{z}\fi{}orn}, \citenamefont {Verschelde},\ and\ \citenamefont {Verstraete}}]{haegeman2011time}%
  \BibitemOpen
  \bibfield  {author} {\bibinfo {author} {\bibfnamefont {J.}~\bibnamefont {Haegeman}}, \bibinfo {author} {\bibfnamefont {J.~I.}\ \bibnamefont {Cirac}}, \bibinfo {author} {\bibfnamefont {T.~J.}\ \bibnamefont {Osborne}}, \bibinfo {author} {\bibfnamefont {I.}~\bibnamefont {Pi\ifmmode~\check{z}\else \v{z}\fi{}orn}}, \bibinfo {author} {\bibfnamefont {H.}~\bibnamefont {Verschelde}},\ and\ \bibinfo {author} {\bibfnamefont {F.}~\bibnamefont {Verstraete}},\ }\bibfield  {title} {\bibinfo {title} {Time-dependent variational principle for quantum lattices},\ }\href {https://doi.org/10.1103/PhysRevLett.107.070601} {\bibfield  {journal} {\bibinfo  {journal} {Phys. Rev. Lett.}\ }\textbf {\bibinfo {volume} {107}},\ \bibinfo {pages} {070601} (\bibinfo {year} {2011})}\BibitemShut {NoStop}%
\bibitem [{\citenamefont {Haegeman}\ \emph {et~al.}(2016)\citenamefont {Haegeman}, \citenamefont {Lubich}, \citenamefont {Oseledets}, \citenamefont {Vandereycken},\ and\ \citenamefont {Verstraete}}]{haegeman2016unifying}%
  \BibitemOpen
  \bibfield  {author} {\bibinfo {author} {\bibfnamefont {J.}~\bibnamefont {Haegeman}}, \bibinfo {author} {\bibfnamefont {C.}~\bibnamefont {Lubich}}, \bibinfo {author} {\bibfnamefont {I.}~\bibnamefont {Oseledets}}, \bibinfo {author} {\bibfnamefont {B.}~\bibnamefont {Vandereycken}},\ and\ \bibinfo {author} {\bibfnamefont {F.}~\bibnamefont {Verstraete}},\ }\bibfield  {title} {\bibinfo {title} {Unifying time evolution and optimization with matrix product states},\ }\href {https://doi.org/10.1103/PhysRevB.94.165116} {\bibfield  {journal} {\bibinfo  {journal} {Phys. Rev. B}\ }\textbf {\bibinfo {volume} {94}},\ \bibinfo {pages} {165116} (\bibinfo {year} {2016})}\BibitemShut {NoStop}%
\bibitem [{\citenamefont {Pang}\ \emph {et~al.}(2020)\citenamefont {Pang}, \citenamefont {Hao}, \citenamefont {Dugad}, \citenamefont {Zhou},\ and\ \citenamefont {Solomonik}}]{pang2020efficient}%
  \BibitemOpen
  \bibfield  {author} {\bibinfo {author} {\bibfnamefont {Y.}~\bibnamefont {Pang}}, \bibinfo {author} {\bibfnamefont {T.}~\bibnamefont {Hao}}, \bibinfo {author} {\bibfnamefont {A.}~\bibnamefont {Dugad}}, \bibinfo {author} {\bibfnamefont {Y.}~\bibnamefont {Zhou}},\ and\ \bibinfo {author} {\bibfnamefont {E.}~\bibnamefont {Solomonik}},\ }\href {https://doi.org/10.48550/arXiv.2006.15234} {\bibinfo {title} {Efficient 2d tensor network simulation of quantum systems}} (\bibinfo {year} {2020}),\ \Eprint {https://arxiv.org/abs/2006.15234} {arXiv:2006.15234 [cs.DC]} \BibitemShut {NoStop}%
\bibitem [{\citenamefont {{NVIDIA cuQuantum team}}(2023)}]{cuQuantum}%
  \BibitemOpen
  \bibfield  {author} {\bibinfo {author} {\bibnamefont {{NVIDIA cuQuantum team}}},\ }\href {https://doi.org/10.5281/zenodo.10068206} {\bibinfo {title} {{NVIDIA/cuQuantum: cuQuantum v23.10}}} (\bibinfo {year} {2023})\BibitemShut {NoStop}%
\bibitem [{Note2()}]{Note2}%
  \BibitemOpen
  \bibinfo {note} {Note that this presentation only works as presented in the simplest topological sector of the allowed states. There are other states that allow for additional overall winding numbers, which will not be considered in this work. One can easily generalize the loop representation to also include winding loops.}\BibitemShut {Stop}%
\bibitem [{\citenamefont {Weingarten}(2008)}]{weingarten1978asymptotic}%
  \BibitemOpen
  \bibfield  {author} {\bibinfo {author} {\bibfnamefont {D.}~\bibnamefont {Weingarten}},\ }\bibfield  {title} {\bibinfo {title} {{Asymptotic behavior of group integrals in the limit of infinite rank}},\ }\href {https://doi.org/10.1063/1.523807} {\bibfield  {journal} {\bibinfo  {journal} {Journal of Mathematical Physics}\ }\textbf {\bibinfo {volume} {19}},\ \bibinfo {pages} {999} (\bibinfo {year} {2008})},\ \Eprint {https://arxiv.org/abs/https://pubs.aip.org/aip/jmp/article-pdf/19/5/999/11195862/999\_1\_online.pdf} {https://pubs.aip.org/aip/jmp/article-pdf/19/5/999/11195862/999\_1\_online.pdf} \BibitemShut {NoStop}%
\bibitem [{\citenamefont {Klco}\ and\ \citenamefont {Savage}(2020)}]{klco2020minimally}%
  \BibitemOpen
  \bibfield  {author} {\bibinfo {author} {\bibfnamefont {N.}~\bibnamefont {Klco}}\ and\ \bibinfo {author} {\bibfnamefont {M.~J.}\ \bibnamefont {Savage}},\ }\bibfield  {title} {\bibinfo {title} {Minimally entangled state preparation of localized wave functions on quantum computers},\ }\href {https://doi.org/10.1103/PhysRevA.102.012612} {\bibfield  {journal} {\bibinfo  {journal} {Phys. Rev. A}\ }\textbf {\bibinfo {volume} {102}},\ \bibinfo {pages} {012612} (\bibinfo {year} {2020})}\BibitemShut {NoStop}%
\end{thebibliography}%

\pagebreak

\onecolumngrid

\setcounter{equation}{0}
\setcounter{figure}{0}
\setcounter{table}{0}

\renewcommand{\theequation}{S\arabic{equation}}
\renewcommand{\thefigure}{S\arabic{figure}}

\begin{center}
\textbf{\large Quantum Simulation of SU(3) Lattice Yang Mills Theory at Leading Order in Large N: Supplemental Material}
\end{center}

\section{2x2 Plaquette Basis}
\FloatBarrier

As an explicit example, we will compare the different basis choices for a lattice consisting of $2\times2$ plaquettes with open boundary conditions and electric fields truncated at $p+q\leq 1$. A comparison of all different basis choices is summarized in Table~\ref{tab:basis choices}. Using the basis described in Ref~\cite{byrnes2006simulating} where no Gauss law constraints are utilized, the Hilbert space on a single link can be described by the basis states $\ket{\mathbf{1},0,0}$, $\ket{\mathbf{3},a,b}$, and $\ket{\mathbf{\Bar{3}},a,b}$ where $a$ and $b$ are labels of the components of the $\mathbf{3}$ and $\mathbf{\Bar{3}}$ representations. This results in a 19 dimensional Hilbert space per link and a total Hilbert space dimension of $19^{12}\approx2\times10^{15}$ for the full system. Encoding these states directly on a quantum computer as suggested in Ref~\cite{byrnes2006simulating} would require $5$ qubits per link and a total of $60$ qubits for the entire lattice.

\begin{figure}
    \centering
    \includegraphics[width=0.8\linewidth]{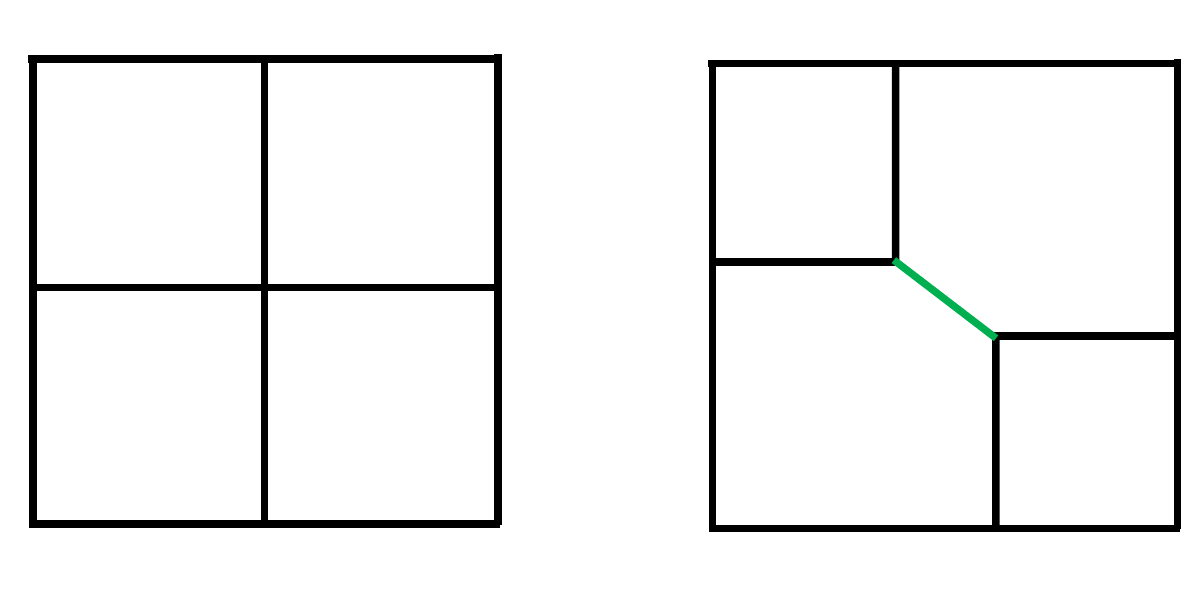}
    \caption{The left figure shows the original $2\times2$ lattice and the right figure shows the point split lattice. The green link is the virtual link inserted in the point splitting procedure.}
    \label{fig:LatticePointSplit}
\end{figure}

The multiplet basis introduced in Ref.~\cite{ciavarella2021trailhead} provides a more economical basis by taking advantage of the Gauss law constraints. However, it requires that the lattice consists of vertices where only three links meet. This can be accomplished through virtual point splitting as described in Ref.~\cite{raychowdhury2020loop}. The original and point split lattice are shown in Fig.~\ref{fig:LatticePointSplit}. At this truncation, the physical links in the lattice can be described by the basis states $\ket{\mathbf{1}}$, $\ket{\mathbf{3}}$, and $\ket{\mathbf{\Bar{3}}}$. The virtual link in the center of the lattice can potentially have any representation that results from $\mathbf{3}\otimes \mathbf{3}$, $\mathbf{\Bar{3}}\otimes \mathbf{\Bar{3}}$, or $\mathbf{3}\otimes \mathbf{\Bar{3}}$. Therefore, the virtual link can be described by the basis states  $\ket{\mathbf{1}}$, $\ket{\mathbf{3}}$, $\ket{\mathbf{\Bar{3}}}$, $\ket{\mathbf{8}}$, $\ket{\mathbf{6}}$, and $\ket{\mathbf{\Bar{6}}}$. The physical links can be represented using only two qubits per link and the virtual link can be represented with three qubits. This leads to a qubit count of $27$ for the entire lattice with this encoding. Note that not all states in this encoding are physical, as it contains states where the representations at a vertex do not add up to a singlet state.

The loop basis described previously consists of all oriented closed loops with at most one loop on each physical link and up to two loops on the virtual links. This basis only contains gauge invariant states, but is completely non-local and likely cannot be used for efficient encoding onto quantum computers. However, with the addition of the large $N_c$ approximation, the only remaining states are those where each plaquette has at most a single loop of electric flux flowing around it. As discussed in the main text, this results in a local Hamiltonian that only requires a single qubit per plaquette.

\begin{table}[]
    \centering
    \begin{tabular}{|c|c|c|c|}
    \hline
          & Byrnes-Yamammoto & Multiplet Basis & Large $N_c$ \\
          \hline
       \# Qubits  & 60 & 27 & 4\\ 
       \hline
       Gauss's Law Enforced & No & Partially & Partially \\
       \hline
    \end{tabular}
    \caption{Comparison of the different bases discussed in this work.}
    \label{tab:basis choices}
\end{table}

\FloatBarrier
\section{Plaquette Matrix Elements }
\label{app:plaq_elements}
The plaquette matrix elements (introduced in~\cite{ciavarella2021trailhead}) can be conveniently represented on a point-split vertex where it will be expanded as a sum over generators of gauge-invariant rotations in the electric multiplet basis. While the generic expression depends on ${\rm SU}(N)$ $9J$ symbols formed by contractions of Clebsch-Gordan coefficients~\cite{klco20202,ciavarella2021trailhead}, analytic expressions for matrix elements of low lying representations can be determined for arbitrary ${\rm SU}(N)$.

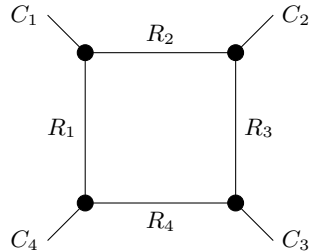
\begin{figure}
    \centering
    \begin{tikzpicture}
        \filldraw (0,0) circle (3pt);
        \filldraw (0,2) circle (3pt);
        \filldraw (2,0) circle (3pt);
        \filldraw (2,2) circle (3pt);

        \draw (0,0) -- node[anchor=east] {$R_1$} ++(0,2);
        \draw (0,2) -- node[above] {$R_2$} ++(2,0);
        \draw (2,0) -- node[anchor=west] {$R_3$} ++(0,2);
        \draw (2,0) -- node[below] {$R_4$} ++(-2,0);

        \draw (-0.5,-0.5) node[anchor=east]{$C_4$} -- (0,0);
        \draw (2.5,-0.5) node[anchor=west]{$C_3$} -- (2,0);
        \draw (-0.5,2.5) node[anchor=east]{$C_1$} -- (0,2);
        \draw (2.5,2.5) node[anchor=west]{$C_2$} -- (2,2);
    \end{tikzpicture}
    \caption{A single plaquette with only a single external link at each vertex.}
    \label{fig:plaquette}
\end{figure}

In the electric basis, the matrix elements of the link operator take the form
\begin{equation}
    \bra{K,P,p} \hat{U}_{\alpha,\beta} \ket{J,M,m} = \sqrt{\frac{\text{dim}(J)}{\text{dim}(K)}} C^{K,P}_{N,\alpha;JM} C^{\Bar{K},\Bar{p}}_{\Bar{N},\Bar{\beta};\Bar{J}\Bar{m}} \,,
    \label{eq:link_elements}
\end{equation}
where $C^{A,a}_{B,b;C,c}$ is the Clebsch-Gordan coefficient describing the combination of representations $B$ and $C$ into representation $A$. Gauge invariant states in the theory will be formed from singlets at each vertex whose wavefunctions will be denoted by
\begin{equation}
    \ket{\phi(A,B,C)} = \sum_{a,b,c} \frac{C^{\Bar{C},\Bar{c}}_{A,a;B,b}}{\sqrt{\text{dim}(C)}} \ket{A,a} \ket{B,b} \ket{C,c} \,.
\end{equation}
Using these singlet states, the gauge-invariant wavefunction for the plaquette in Fig.~\ref{fig:plaquette} can be written as
\begin{equation}
    \ket{\psi \begin{pmatrix}
        C_1 & R_2 & C_2 \\
        R_1 &     & R_3 \\
        C_4 & R_4 & C_3
    \end{pmatrix}} = \ket{\phi(R_1,\Bar{R_2},C_1)} \ket{\phi(R_2,\Bar{R_3},C_2)} \ket{\phi(R_2,\Bar{R_3},C_3)} \ket{\phi(R_4,\Bar{R_1},C_4)} \,.
    \label{eq:plaq_wvfn}
\end{equation}

Using Eq.~\eqref{eq:link_elements} and Eq.~\eqref{eq:plaq_wvfn}, the matrix elements of the plaquette operator can be written as
\begin{align}
    \bra{\psi \begin{pmatrix}
        C_1 & R_{2f} & C_2 \\
        R_{1f} &     & R_{3f} \\
        C_4 & R_{4f} & C_3
    \end{pmatrix}} & \hat{\Box} \ket{\psi \begin{pmatrix}
        C_1 & R_{2i} & C_2 \\
        R_{1i} &     & R_{3i} \\
        C_4 & R_{4i} & C_3
    \end{pmatrix}}  = \\
    & \sqrt{\frac{\text{dim}(R_{1i}) \text{dim}(R_{2i}) \text{dim}(R_{3i}) \text{dim}(R_{4i})}{\text{dim}(R_{1f}) \text{dim}(R_{2f}) \text{dim}(R_{3f}) \text{dim}(R_{4f})}} \nonumber \\
    & V(C_1,R_{1f},R_{1i}, R_{2f},R_{2i}) V(C_2,R_{2f},R_{2i}, R_{3f},R_{3i}) \nonumber \\
    & V(C_3,R_{3f},R_{3i}, R_{4f},R_{4i}) V(C_4,R_{4f},R_{4i}, R_{1f},R_{1i}) \,,
    \label{eq:plaq_el}
\end{align}
where the $V$'s are vertex factors defined by
\begin{align}
    & V(C,A_f,A_i,B_f,B_i)  = \nonumber \\
    & \sum_{\alpha_i,\alpha_f,\beta_i,\beta_f,\gamma} \bra{\phi(A_f,\Bar{B}_f,C)} 
C^{A_f,\alpha_f}_{N,\gamma; A_i \alpha_i} C^{\Bar{B}_f,\beta_f}_{\Bar{N},\Bar{\gamma}; \Bar{B}_i \beta_i} \ket{A_f,\alpha_f} \bra{A_i,\alpha_i} \otimes \ket{\Bar{B}_f,\beta_f} \bra{\Bar{B}_i,\beta_i} \ \ket{\phi(A_i,\Bar{B}_i,C)} \nonumber \\
& = \frac{1}{\text{dim}(C)} \sum_{\alpha_i,\alpha_f,\beta_i,\beta_f,\gamma,c} C^{\Bar{C},\Bar{c} \ *}_{A_f,\alpha_f;\Bar{B}_f,\beta_f} C^{A_f,\alpha_f}_{N,\gamma; A_i \alpha_i} C^{\Bar{B}_f,\beta_f}_{\Bar{N},\Bar{\gamma}; \Bar{B}_i \beta_i} C^{\Bar{C},\Bar{c}}_{A_i,\alpha_i;\Bar{B}_i,\beta_i} \,.
\end{align}
When the lattice is a 1D chain of plaquettes, this is sufficient to describe all plaquette matrix elements between gauge invariant states. In higher spatial dimensions, this result can still be used provided the theory is placed on a point-split lattice.

To evaluate vertex factors with arbitrary \SU{N}, a few general identities for the Clebsch-Gordan coefficients will be needed. If one of the lower entries of the Clebsh-Gordan coefficients is the singlet, then the coefficient is given by
\begin{equation}
    C^{R,a}_{R,b;1,0} = \delta_{a,b} \,.
\end{equation}
Also, if two representations are combining to form a singlet the only non-zero Clebsch-Gordan coefficients are given by
\begin{equation}
    C^{1,0}_{R,a;\Bar{R},\Bar{a}} = \frac{1}{\sqrt{\text{dim}(R)}} \,.
\end{equation}
Adding representations gives orthonormal states which gives the identity
\begin{equation}
    \sum_{a,b} C^{C,c}_{A,a;B,b} C^{C,c'}_{A,a;B,b} = \delta_{c,c'}\,.
\end{equation}
Another useful identity follows from multiple choices of decompositions of the singlet into three representations, i.e.
\begin{equation}
    \ket{1,0} = \sum_{a,b,c} \frac{C^{\Bar{A},\Bar{a}}_{B,b;C,c}}{\sqrt{\text{dim}(A)}} \ket{A,a} \ket{B,b} \ket{C,c} = \sum_{a,b,c} \frac{C^{\Bar{B},\Bar{b}}_{A,a;C,c}}{\sqrt{\text{dim}(B)}} \ket{A,a} \ket{B,b} \ket{C,c}  \,,
\end{equation}
implies 
\begin{equation}
   C^{\Bar{A},\Bar{a}}_{B,b;C,c} = \sqrt{\frac{\text{dim}(A)}{\text{dim}(B)}} C^{\Bar{B},\Bar{b}}_{A,a;C,c} \,.
\end{equation}
Using these identities, we find
\begin{equation}
    V(\mathbf{1},B,A,B,A) = V(A,B,A,N,\mathbf{1}) = \sqrt{\frac{\text{dim}(B)}{\text{dim}(A)}}  \,.
    \label{eq:vertexF}
\end{equation}

For the truncations considered in this work, each plaquette is constrained to have a single loop of flux running around it. Generically, the matrix elements of plaquette operators only depends on the links on the plaquette and the electric flux flowing into each vertex of the plaquette. By considering Eq.~\eqref{eq:plaq_el} on a point-split lattice, it can be seen that at this truncation in the plaquette basis, the matrix elements of a plaquette operator only depend on the plaquette being acted on and the plaquettes that share a link with it. Using Eq.~\eqref{eq:plaq_el} and Eq.~\eqref{eq:vertexF}, it can be seen that when the neighboring plaquettes are unexcited, the matrix elements for allowed transitions is always $1$. Explicitly for ${\rm SU} (N)$, we have
\begin{equation}
    \bra{\circlearrowright} \hat{\Box} + \hat{\Box}^\dagger \ket{0} = \bra{\circlearrowleft} \hat{\Box} + \hat{\Box}^\dagger \ket{0} = 1 \,,
\end{equation}
and specific to the case of ${\rm SU} (3)$
\begin{equation}
    \bra{\circlearrowright} \hat{\Box} + \hat{\Box}^\dagger \ket{\circlearrowleft} = 1 \,.
\end{equation}
When neighboring plaquettes are allowed to have a shared link in the representation given by anti-symmetric combination of fundamental representations, more matrix elements need to be considered. These can be evaluated using the above vertex factors and the fact that plaquette operators commute. For example, consider a $1D$ chain of three plaquettes. Using Eq.~\eqref{eq:plaq_el} and Eq.~\eqref{eq:vertexF}, we find
\begin{equation}
    \bra{\circlearrowright}_1 \bra{\circlearrowleft}_2 \bra{0}_3 \hat{\Box}_2 \ket{\circlearrowright}_1 \ket{0}_2 \ket{0}_3 = \sqrt{\frac{1}{2} \left(1 - \frac{1}{N}\right)} \,.
\end{equation}
The matrix element $\bra{\circlearrowright}_1 \bra{\circlearrowleft}_2 \bra{\circlearrowright}_3 \hat{\Box}_2 \ket{\circlearrowright}_1 \ket{0}_2 \ket{\circlearrowright}_3$ can be evaluated in the same manner or by using the previous matrix element and the commutativity of plaquette operators. Explicitly, we have
\begin{align}
    \bra{\circlearrowright}_1 \bra{\circlearrowleft}_2 \bra{\circlearrowright}_3 \hat{\Box}_2 \ket{\circlearrowright}_1 \ket{0}_2 \ket{\circlearrowright}_3 & = \bra{\circlearrowright}_1 \bra{\circlearrowleft}_2 \bra{\circlearrowright}_3 \hat{\Box}_2  \hat{\Box}^\dagger_3 \ket{\circlearrowright}_1 \ket{0}_2 \ket{0}_3 \nonumber \\
    & = \bra{\circlearrowright}_1 \bra{\circlearrowleft}_2 \bra{\circlearrowright}_3 \hat{\Box}^\dagger_3 \hat{\Box}_2 \ket{\circlearrowright}_1 \ket{0}_2 \ket{0}_3 \nonumber \\
    & = \sqrt{\frac{1}{2} \left(1 - \frac{1}{N}\right)} \bra{\circlearrowright}_1 \bra{\circlearrowleft}_2 \bra{\circlearrowright}_3 \hat{\Box}^\dagger_3 \ket{\circlearrowright}_1 \ket{\circlearrowleft}_2 \ket{0}_3 \nonumber \\
    & = \frac{1}{2} \left(1 - \frac{1}{N}\right) \,.
\end{align}
The same style of argument can be used on a $2D$ square lattice to find
\begin{equation}
    \bra{ \begin{matrix}
            & C_1 &  \\
        C_4 & \circlearrowleft  & C_2 \\
            & C_3 & 
        \end{matrix}}
        \hat{\Box}
    \ket{ \begin{matrix}
            & C_1 &  \\
        C_4 & 0  & C_2 \\
            & C_3 & 
        \end{matrix}} = \left(\frac{1}{2} \left(1 - \frac{1}{N}\right)\right)^{n_e/2} \,,
\end{equation}
where $n_e$ is the number of $C_i$ in the state $\ket{\circlearrowright}$. Note that this method can also be used to derive the non large $N$ suppressed plaquette matrix elements needed to include the $\mathbf{8}$ and $\mathbf{6}$ representations.

\FloatBarrier

\section{Monte Carlo}
For the truncated Hamiltonians studied in this work, static properties at finite temperature can be determined using Monte Carlo techniques. In general, the Boltzmann distribution can be approximated by
\begin{equation}
    e^{-\beta \sum_i \hat{H}_i} \approx \prod_i e^{-\beta \hat{H}_i} \,.
\end{equation}
By inserting a complete set of states, one can obtain a sum over states that can be sampled with the Metropolis algorithm. If all matrix elements of each of the $e^{-\beta \hat{H}_i}$ are positive, then this sampling can be done without a sign problem.

Explicitly for the single loop Hamiltonian in the main text, the Boltzmann distribution can be approximated by
\begin{equation}
    e^{-\beta \hat{H}} \approx \prod_{i=1}^{L_t} e^{-\frac{\beta}{L_t} \hat{H}_{\text{even}}} e^{-\frac{\beta}{L_t} \hat{H}_{\text{odd}}} \,,
\end{equation}
where $\hat{H}_{\text{even}}$ is the sum over all terms in the Hamiltonian that act on plaquettes on the even sublattice and $\hat{H}_{\text{odd}}$ is the sum over all terms in the Hamiltonian that act on plaquettes on the odd sublattice. Inserting a sum over electric basis states yields
\begin{equation}
    \Tr{\prod_{L_t} e^{-\frac{\beta}{L_t} \hat{H}_{\text{even}}} e^{-\frac{\beta}{L_t} \hat{H}_{\text{odd}}}} = \sum_{s_{x,y,t} = 0}^1 e^{-S(s_{x,y,t})} \,,
\end{equation}
where the action $S(s_{x,y,t})$ is given by
\begin{align}
    S(s_{x,y,t}) & =  \sum_{x=0}^{L} \sum_{y=0}^{L} \sum_{t=0}^{2L_t} \frac{\beta}{L_t} \left(\frac{8}{3}g^2 - \frac{1}{2g^2} \right) s_{x,y,t} \nonumber \\
    & - \log(\cosh\left(\frac{\beta}{L_t g^2 \sqrt{2}}\right)) \left( s_{x,y,t} s_{x,y,t+1} + (1-s_{x,y,t}) (1-s_{x,y,t+1}) \right) \nonumber \\
    & - \log(\sinh\left(\frac{\beta}{L_t g^2 \sqrt{2}}\right))  \left( s_{x,y,t} (1-s_{x,y,t+1}) + (1-s_{x,y,t}) s_{x,y,t+1}  \right) \,,
\end{align}
and is subject to the constraints that if $s_{x,y,t} = 1$ all of its neighbors in the spatial directions are $0$ and $s_{x,y,t} = s_{x,y,t+1}$ for $x+y = t \mod{2}$. $e^{-S(s_{x,y,t})}$ is positive definite and therefore the sum over field configurations can be sampled with the Metropolis algorithm without a sign problem. Note that the derivation of this action is analogous to the relationship between the quantum Ising model and the classical Ising model in one higher dimension~\cite{suzuki2013transverse}. A similar decomposition can be done for the Hamiltonian including vertices with three incoming $\mathbf{3}$ representations.

\begin{figure}
    \centering
    \includegraphics[width=8.7cm]{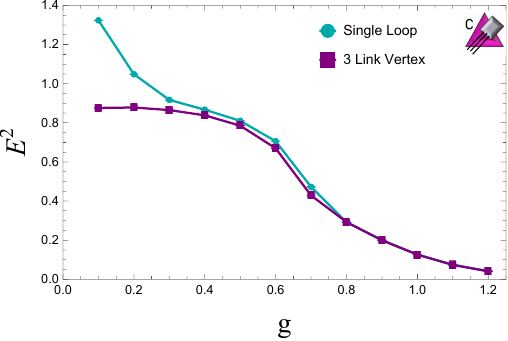}
    \caption{Monte Carlo calculation of $\hat{E}^2$ at $\beta=4$ on a $8\times8$ lattice with periodic boundary conditions. The blue points were computed using the single loop Hamiltonian in the main text and the purple points were computed using the Hamiltonian obtained by allowing three incoming or outgoing $\mathbf{3}$ representations at a vertex. }
    \label{fig:MC}
\end{figure}

As an example of a calculation that could be performed, a $8\times8$ spatial lattice with periodic boundary conditions was simulated at $\beta = 4$ with $L_t=40$ imaginary time steps for various values of $g$. For each coupling $g$, $8$ samples of $8100$ field configurations were generated with the Metropolis algorithm. The first $100$ field configurations of each sample were discarded and the remaining were used to compute observables. Error bars were computed through bootstrap resampling of the $8$ samples. Fig.~\ref{fig:MC} shows the expectation of $\hat{E}^2$ as a function of $g$ for the truncated Hamiltonians studied in this work. As this figure shows, the truncated Hamiltonians are in agreement at large $g$ and discrepancies grow as $g$ is lowered. 

\section{Quantum Simulation Implementation}
\subsection{Interaction Picture Trotterization}
The standard strategy for simulating time evolution in a lattice gauge theory on a quantum computer has been to utilize a Trotterized time evolution operator where the time evolution generated by $\hat{H} = \hat{H}_E + \hat{H}_B$ is approximated by
\begin{equation}
    e^{-i \hat{H} \Delta t} = e^{-i \hat{H}_E \Delta t} e^{-i \hat{H}_B \Delta t} + \mathcal{O}(\Delta t^2) \,.
    \label{eq:EBTrotter}
\end{equation}
The errors induced by Trotterizing into electric and magnetic terms will be $\mathcal{O}(\Delta t^2)$.
In an electric basis, the electric Hamiltonian, $\hat{H}_E$, is diagonal and $e^{-i \hat{H}_E \Delta t}$ can be implemented with single and controlled $\hat{Z}$ rotations. 
The magnetic Hamiltonian, $\hat{H}_B$, will generate rotations between different electric basis states and its implementation will generally require further Trotterization.  
The errors from this further Trotterization of the magnetic Hamiltonian will be $\mathcal{O}\left(\Delta t^2/g^4\right)$. 
In the limit of strong coupling, the dominant source of error in the Trotterized time evolution will therefore come from the splitting into electric and magnetic terms. 

This source of error can be reduced by instead performing the simulation in an interaction picture where the electric energy is taken to be the ``free'' term in the Hamiltonian. 
Explicitly, if we define $\hat{H}_{B,I}(t) = e^{i \hat{H}_E t} \hat{H}_{B} e^{-i \hat{H}_E t}$, then the time evolution operator is given by
\begin{equation}
    e^{-i \hat{H} t} = e^{-i \hat{H}_E t} \  \mathcal{T} e^{-i \int_0^t ds \hat{H}_{B,I}(s)} \,.
    \label{eq:int_pic_ev}
\end{equation}
By breaking up the time ordered matrix exponential into small time steps, it can be approximated by
\begin{equation}
    \mathcal{T} e^{-i \int_0^t ds \hat{H}_{B,I}(s)} = \prod_n \left(e^{-i \int_{\Delta t n}^{\Delta t (n+1)} ds \hat{H}_{B,I}(s)} +\mathcal{O}\left(\frac{\Delta t^2}{g^4}\right) \right) \,.
\end{equation}
The size of the error in this approximation was estimated from the next to leading order term in the Magnus expansion. 
Therefore, by performing time evolution in the simulation in the interaction picture and Trotterizing $e^{-i \int_{\Delta t n}^{\Delta t (n+1)} ds \hat{H}_{B,I}(s)}$ into an implementable circuit, the error in a single time step will be $\mathcal{O}\left(\Delta t^2/g^4\right)$, as desired.
This is a factor of $1/g^4$ difference from the standard Trotterization in Eq.~\eqref{eq:EBTrotter} and will be helpful when performing simulations at large $g$.

The quantum simulations of the single loop Hamiltonian presented in the main text were performed using this interaction picture Trotterization.
Note that the magnetic term in this Hamiltonian contains a diagonal piece that will be absorbed into the electric Hamiltonian when going to the interaction picture. 
The interaction picture magnetic Hamiltonian is given by
\begin{equation}
    \hat{H}_{B,I}(t) = -\frac{1}{g^2 \sqrt{2}}\sum_p \left(\cos(\varepsilon t) \hat{X}_p + \sin(\varepsilon t) \hat{Y}_p\right) \prod_{q \in \partial p} \hat{P}_{0,q} 
    \,,
\end{equation}
where $p$ labels a plaquette on the lattice, $\partial p$ denotes the set of plaquettes neighboring $p$ and $\varepsilon = \frac{8}{3}g^2 - \frac{1}{2g^2}$. 

To Trotterize time evolution generated by this operator, the sum over plaquettes will be split into two separate sum over sublattices. 
Each plaquette on a $2D$ lattice can be labeled by $(x,y)$ coordinates placed at the center of the plaquette. 
The lattice will be separated into an even sublattice, $E$, and odd sublattice, $O$, where the even sublattice contains all plaquettes with coordinates $(x,y)$ such that $(x+y)\mod{2} = 0$ and the odd sublattice contains all remaining plaquettes. 
This separation is chosen such that all plaquettes in each sublattice commute with each other and when mapped to circuits can be implemented in parallel. 
With this decomposition, the time evolution operator can be Trotterized as
\begin{equation}
    e^{-i \int_0^{\Delta t} ds \hat{H}_{B,I}(s)} = e^{-i \int_0^{\Delta t} ds \hat{H}_{B,E}(s)} e^{-i \int_0^{\Delta t} ds \hat{H}_{B,O}(s)} + \mathcal{O}\left(\frac{\Delta t^2}{g^4}\right) \ \ \ ,
    \label{eq:checker_trotter}
\end{equation}
where $\hat{H}_{B,E}(s)$ is the same as $\hat{H}_{B,I}(s)$ except the sum has been restricted to plaquettes in the even sublattice and $\hat{H}_{B,O}(s)$ is restricted to the odd sublattice. 

\subsection{Time Evolution Circuits}
\label{app:trotter_circuit}
\begin{figure}
    \centering
    \includegraphics[width=0.6\textwidth]{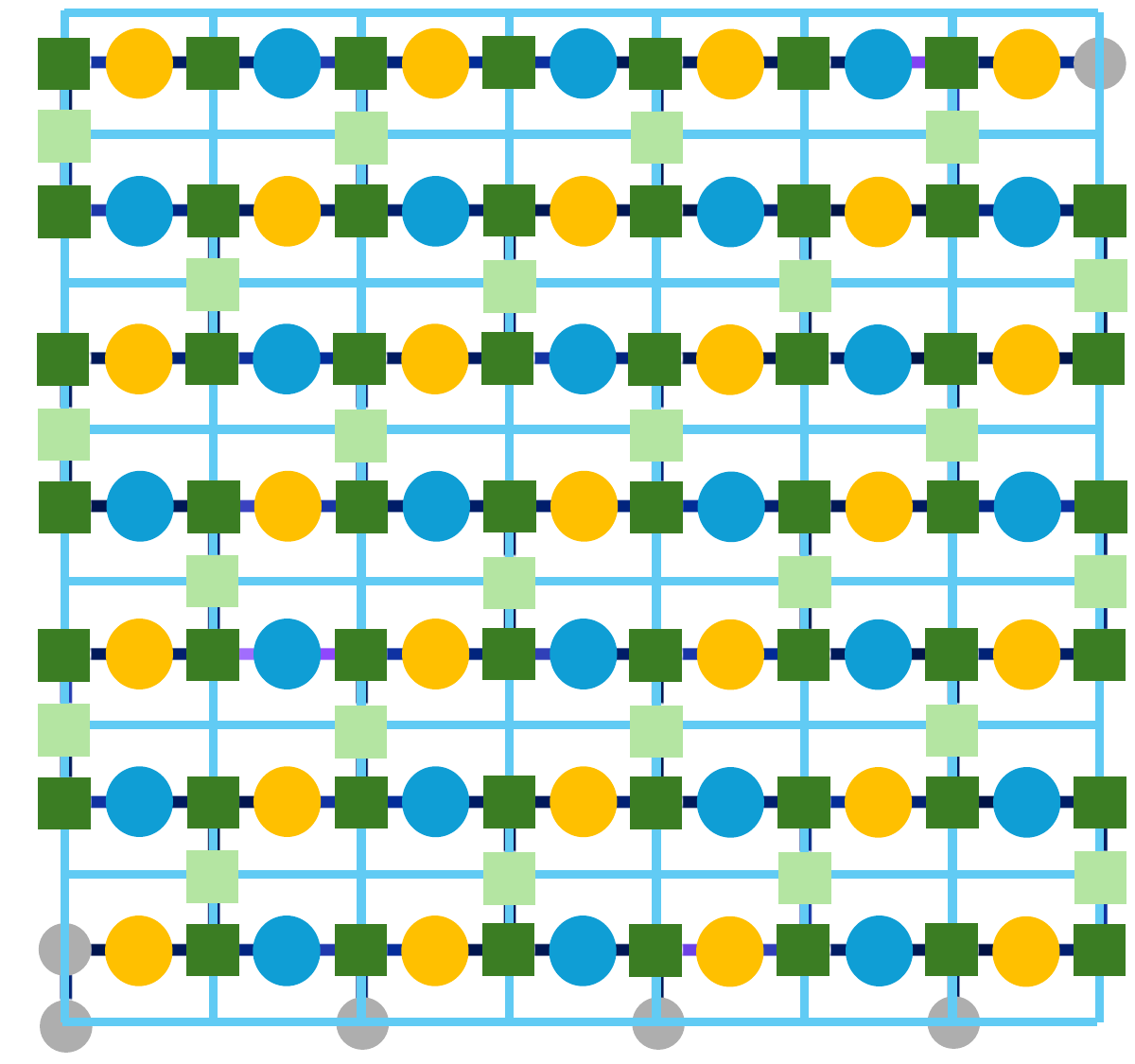}
    \caption{Qubit assignment onto {\tt ibm\_torino}. Light blue lines indicate the lattice being simulated and dark lines indicate the connectivity between qubits. Circular qubits are used to represent the state of the system and square qubits are used as ancillas to help implement the time evolution operator. The yellow qubits represent the states of plaquettes on the even sublattice and the blue qubits represent plaquettes on the odd sublattice. The dark green qubits are ancilla qubits used to directly implement controls on the physical qubits (corresponding to $A_1$ or $A_3$ in Fig.~\ref{fig:ancilla_prep}) and the light green qubits represent ancilla qubits corresponding to $A_2$ in Fig.~\ref{fig:ancilla_prep}. Gray qubits denote qubits on the chip that were not used in the simulation.}
    \label{fig:IBM_torino_circ}
\end{figure}

Implementation of the Trotterized time evolution described in the previous section requires a mapping onto the qubits of a quantum computer and circuits that respects the computer's connectivity. 
The Trotterized time evolution operator in Eq~\eqref{eq:checker_trotter}, can be decomposed into 
\begin{equation}
     e^{-i \int_0^{\Delta t} \!{\rm d}s \,\hat{H}_{B,E}(s)} e^{-i \int_0^{\Delta t}  \!{\rm d}s \, \hat{H}_{B,O}(s)} = \left[ e^{i \phi \sum_p \hat{Z}_p}\right] \, \left[e^{i \theta \sum_{p \in E}  \hat{X}_p \prod_{q \in \partial p} \hat{P}_{0,q} }\right] \, \left[e^{i \theta \sum_{p \in O}  \hat{X}_p \prod_{q \in \partial p} \hat{P}_{0,q} }\right] \, \left[e^{-i \phi \sum_p \hat{Z}_p}\right] \,,
\end{equation}
for appropriately chosen $\theta$ and $\phi$. 
A complication arises when implementing a circuit for $e^{i \theta \sum_{p \in E}  \hat{X}_p \prod_{q \in \partial p} \hat{P}_{0,q} }$ or $e^{i \theta \sum_{p \in E}  \hat{X}_p \prod_{q \in \partial p} \hat{P}_{0,q} }$ on IBM's superconducting quantum processors as they do not have a square geometry~\cite{ibmTorino}. 
However, a square lattice can be embedded on their heavy hex connectivity as shown in Fig.~\ref{fig:IBM_torino_circ}, where the blue lines denote the links on the 2 dimensional lattice.
The circular qubits at the center of plaquettes are used to represent the state of the plaquette at that position.
The remaining qubits on the chip will be used as ancillas to help implement these unitaries.
To explain this better, we have color coded the qubits in~\ref{fig:IBM_torino_circ} according to their role in the circuit. Circular qubits are used to represent the state of the system and square qubits are ancillas.
Yellow qubits (blue) correspond to physical qubits on even (odd) lattice sites. Dark green square qubits denote the ancilla qubits on the left and right of the physical qubits storing the information about the four neighboring qubits for implementing controlled operations. 
Finally, light green qubits are used to communicate information between the dark green ancillas.

\begin{figure}
    \centering
    \subfigure[Qubit Layout (odd lattices left, even lattices right)]{
\includegraphics[width=0.7\textwidth]{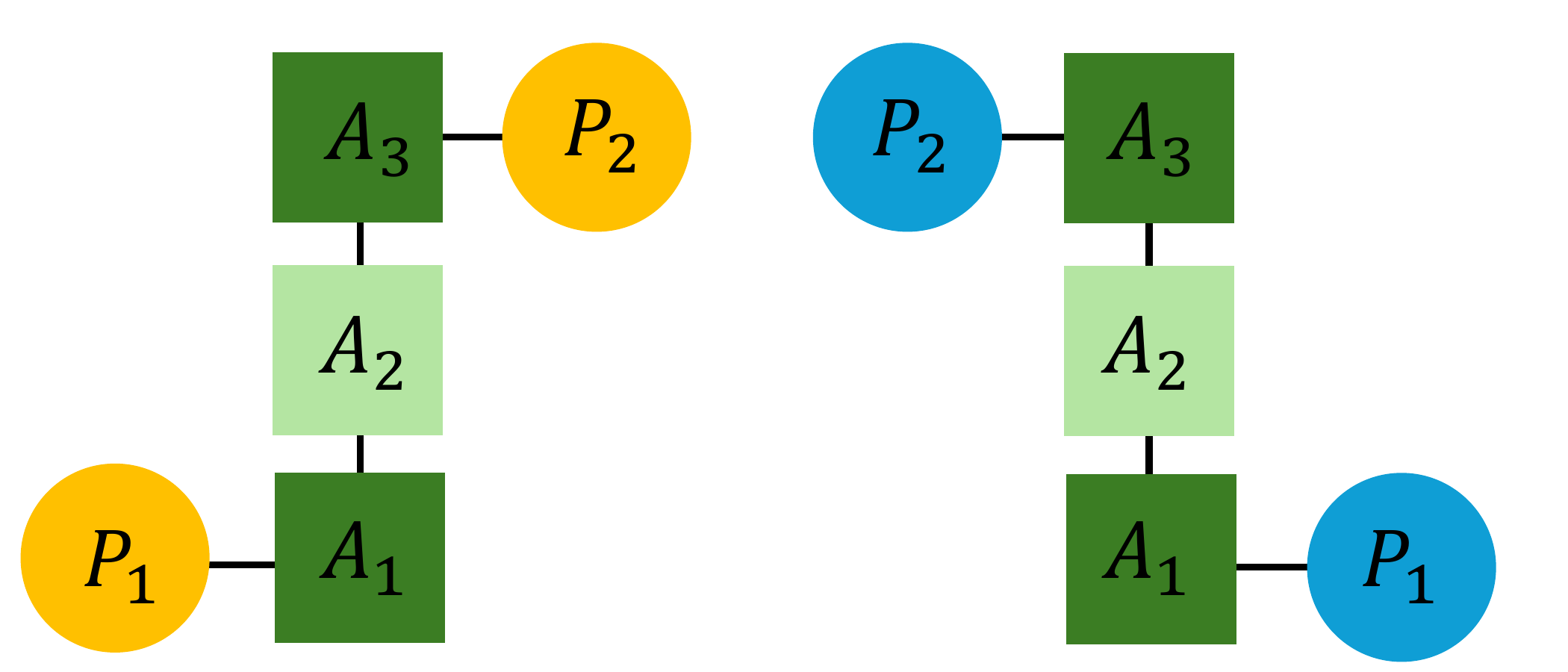}}
    
    \subfigure[Ancilla Preparation Circuit]{

\Qcircuit @C=1em @R=1em {
\lstick{P_1} &\ctrl{1} &\qw                              & \qw       & \qw                       & \qw      & \qw                               & \qw       & \qw                      & \qw     &\ctrl{1}     &\qw&\qw &\qw\\
\lstick{A_1} &\targ    &\qw                              & \qw       & \qw                       & \ctrl{1}       & \qw                               & \qw       & \qw            & \ctrl{1}    &\targ&\targ&\qw &\qw\\
\lstick{A_2} &\gate{H} &\gate{e^{-i\frac{\pi}{8}\hat{Z}}}&\targ&\gate{e^{-i\frac{\pi}{8}\hat{Z}}}&\targ& \gate{e^{-i\frac{\pi}{8}\hat{Z}}}& \targ     & \gate{e^{-i\frac{\pi}{8}\hat{Z}}} & \targ&\gate{H}&\ctrl{-1}&\ctrl{1} &\qw \\
\lstick{A_3} &\targ    &\qw                              & \ctrl{-1} & \qw                       & \qw      & \qw                               & \ctrl{-1} & \qw                     & \qw      &\targ     &\qw&\targ &\qw \\
\lstick{P_2} &\ctrl{-1}&\qw                              & \qw       & \qw                       & \qw & \qw                               & \qw       & \qw                               & \qw &\ctrl{-1}   &\qw&\qw &\qw
}
    
    }
    \caption{Circuit that sends all ancillas ($A_1$, $A_2$, and $A_3$ beginning in the state $\ket{0}$) to the state $\ket{1}$ if the physical qubits ($P_1$ and $P_2$) are in the state $\ket{0}$, and leaves them alone otherwise. The upper diagram shows the connectivity of the qubits on the hardware and the lower shows the circuit applied to these qubits.}
    \label{fig:ancilla_prep}
\end{figure}

To implement the circuit, we recall that the magnetic component of the Hamiltonian given in equation (10) of the main text requires applying an $\hat X$ gate to a plaquette if the neighboring plaquettes are in the zero state. 
The strategy used to implement these unitaries on a sublattice will be to store the information about neighboring plaquettes in the two ancilla qubits located adjacent to the physical qubit being evolved. 
To be explicit, the ancilla qubits $A_1$ -- $A_3$ will be put in the $\ket{1}$ state if both of the neighboring plaquettes $P_1$ and $P_2$ it represents are in the $\ket{0}$ state, and left in the $\ket{0}$ state otherwise.
This is essentially an implementation of the standard technique to implement multi-control gates using ancilla qubits on the geometry of {\tt ibm\_torino} \cite{nielsen2001quantum}. 
The explicit circuit to prepare the ancilla qubits is shown in Fig.~\ref{fig:ancilla_prep}. 

Using this circuit, $e^{i \theta \hat{X}_p \prod_{q \in \partial p} \hat{P}_{0,q} }$ can be implemented by applying the ancilla preparation circuit to the neighboring ancillas, applying $e^{i \theta \hat{X}_p \hat{P}_{1,a_1} \hat{P}_{1,a_2} }$ (where $a_1$ and $a_2$ are the ancillas neighboring qubit $p$), and then undoing the ancilla preparation circuit. 
Note that each qubit in the even (or odd) sublattice will require the same set of ancillas to be prepared. 
Therefore, one can apply the ancilla preparation to all ancillas on the lattice, evolve the entire sublattice in parallel and then undo the ancilla preparation circuits. 
To be more explicit, a classification of the qubits on {\tt ibm\_torino} is shown in Fig.~\ref{fig:IBM_torino_circ}. 
Qubits on the even sublattice are shown in yellow and qubits on the odd sublattice are shown in blue. 
To evolve the qubits on the even sublattice, the ancilla preparation circuit would be applied to each line of ancilla qubits between the blue qubits. 
Then, the yellow qubits are evolved with controls from the green qubits. The ancilla circuits are then undone before evolving the other sublattice. 
By also alternating the order in which the even and odd lattices are evolved, i.e. in one step evolve even then odd and in the next step evolve odd then even, half of the ancilla circuits can be cancelled. 
This results in a CNOT depth of 45 per Trotter step (and a depth of 23 for the first Trotter step by cancelling CNOT gates against the initial state).

\FloatBarrier
\subsection{Error Mitigation}
\label{app:error_mit}
Current quantum hardware suffers from noise and gate errors that limit the accuracy of simulations. 
These errors can be reduced using error suppression and error mitigation techniques. 
In this work, dynamical decoupling with an $XX$ sequence was used to reduce the coherent errors in the calculation~\cite{viola1999dynamical}. 
Pauli twirling was applied to all CNOT gates in the circuit to convert all coherent noise into Pauli error channels~\cite{urbanek2021mitigating,Rahman:2022rlg,rahman2022real}. 
These error suppression techniques reduce the size of errors in the calculation, but do not correct them. 
Measurement errors in the calculation were mitigated using Twirled Readout EXtincation (T-REX) mitigation~\cite{trexmit}. 
In this technique, a Pauli operator is applied at the end of the circuit and the change in sign of the measured $\hat{Z}$ operator is corrected in post processing. 
If the incoherent noise in the circuit can be modelled by a Pauli error channel following the correct circuit, the errors in Pauli operators can be corrected using Operator Decoherence Renormalization (ODR)~\cite{urbanek2021mitigating,Rahman:2022rlg,rahman2022real,farrell2023scalable}. 
In ODR, two circuits are run on the hardware, a ``physics'' and a ``mitigation'' circuit. The mitigation circuit is chosen such that it can be simulated classically.
For the simulation in this work, the physics circuit is the Trotterized time evolution circuit described above, and the mitigation circuit is the same circuit except with $\Delta t = 0$. 
Under the assumption that the incoherent noise in the circuit can be modelled by a Pauli error channel following the correct circuit, the measured Pauli operator is proportional to the value that would be measured in the absence of noise. 
The constant of proportionality can be determined using the mitigation circuit, because the noiseless value can be computed classically, and used to correct the results of the physics circuit. 
This can significantly reduce the effect of noise but does not correct all errors because not all noise can be modelled this way. 
The dominant source of error in the computation comes from the CNOT gates. 
In the absence of noise, replacing a CNOT gate with a sequence of three CNOT gates would not change the circuit. 
In reality, this enhances the amount of noise in the circuit. By probabilistically inserting pairs of CNOT gates, the noise in the circuit can be amplified by a continuous parameter $r$. 
This can then be used to extrapolate to the zero noise limit~\cite{urbanek2021mitigating,he2020zero,pascuzzi2022computationally}. 
In this work, CNOT gates were replaced by three CNOT gates with probability $p=0.25$ which corresponds to $r=1.5$. 
Both the $r=1$ and $r=1.5$ circuits were mitigated using ODR and T-REX before performing a linear extrapolation to the zero noise limit.

\begin{figure}
    \centering
    \includegraphics[width=8.7cm]{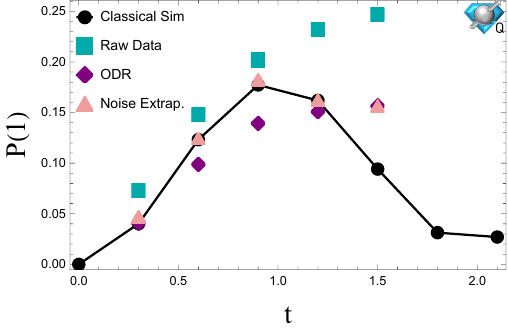}
    \caption{Average probability of a plaquette being excited from the electric vacuum as a function of time on a $4\times4$ plaquette lattice with open boundary conditions and $g=1$. The black line shows a classical simulation of the Trotterized time evolution operator that was implemented. The blue points show the unmitigated data from simulations on {\tt ibm\_torino}. The purple points show the data from {\tt ibm\_torino} mitigated using ODR and T-REX. The pink points show the results of a linear noise extrapolation applied to the data mitigated with ODR and T-REX. All error bars in this plot were computed by bootstrap resampling circuits.}
    \label{fig:4x4error_mit}
\end{figure}

To test the effectiveness of these error mitigation techniques for the simulations in this work, the single loop Hamiltonian in the main text was simulated on a lattice with $4\times4$ plaquettes with open boundary condititions and $g=1$. 
The time evolution circuit described in the previous section was used to evolve the electric vacuum and was implemented using $\Delta t = 0.3$. 
For a lattice of this size, 16 qubits are needed to represent the state of the system, and on {\tt ibm\_torino} a 39 qubit block was used due to the connectivity constraints of the hardware. 
For each time slice, $75$ different Pauli twirls were used for physics and mitigation circuits for both $r=1$ and $r=1.5$. 
In each separate $r=1.5$ twirl, CNOT gates were inserted probabilistically. 
Each individual circuit was evaluated with $5,000$ shots. 
The probability of a plaquette being excited averaged over the lattice as a function of time is shown in Fig.~\ref{fig:4x4error_mit}. 
In these simulations, the probability of each qubit being excited was individually mitigated before being averaged over the lattice. 
This figure shows that the error mitigation procedure used in this work can be used to reliably implement up to $4$ Trotter steps.

\begin{figure}
    \centering
    \includegraphics[width=8.7cm]{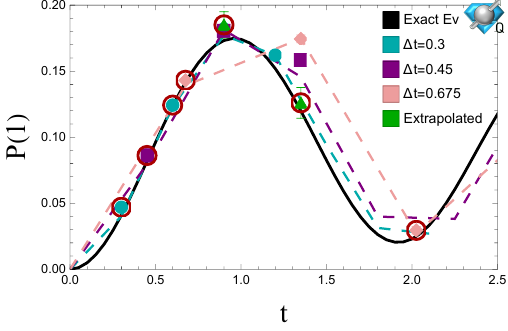}
    \caption{Average probability of a plaquette being excited from the electric vacuum as a function of time on a $4\times4$ plaquette lattice with open boundary conditions and $g=1$. The black line is the exact time evolution computed classically. The colored dashed lines show the results of classical simulation of the Trotterized circuit implemented on hardware. The blue, purple, and pink points are the error mitigated results from {\tt ibm\_torino}. The green points are the linear extrapolation to $\Delta t =0$ for time slices that were sampled with multiple $\Delta t$. The red circles indicate the points displayed in the main text.
    }
    \label{fig:torino_sim_4x4_ex}
\end{figure}

\begin{figure}
    \centering
    \includegraphics[width=8.7cm]{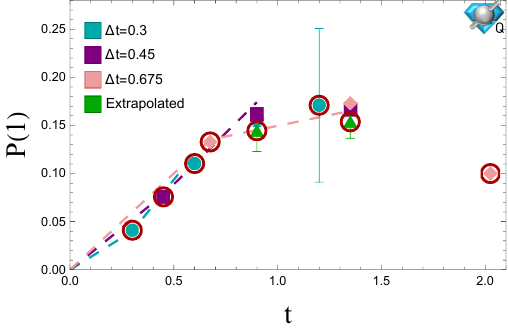}
    \caption{Average probability of a plaquette being excited from the electric vacuum as a function of time on a $7\times7$ plaquette lattice with open boundary conditions and $g=1$. The black line is the exact time evolution computed classically. The colored dashed lines show the results of tensor network simulation of the Trotterized circuit implemented on hardware. The blue, purple, and pink points are the error mitigated results from {\tt ibm\_torino}. The green points are the linear extrapolation to $\Delta t =0$ for time slices that were sampled with multiple $\Delta t$. The red circles indicate the points displayed in the main text.
    }
    \label{fig:torino_sim_7x7_ex}
\end{figure}

In addition to the errors from imperfect hardware, the quantum simulation performed on {\tt ibm\_torino} has errors from the Trotterization of the time evolution operator. 
These errors can be mitigated by performing simulations with different time steps, $\Delta t$, that sample the same points in time. 
For each point in time that is sampled by multiple $\Delta t$, a linear extrapolation to $\Delta t =0$ can be performed. 
Fig.~\ref{fig:torino_sim_4x4_ex} shows the values of $\Delta t$ that were sampled in the $4\times4$ plaquette lattice simulation on {\tt ibm\_torino}. The green points indicate the results of the extrapolation in $\Delta t$.
The circled points at each $t$ indicate the point with the smallest theoretical error at that $t$. 
For times that were sampled by multiple $\Delta t$, this would be the value obtained by the $\Delta t$ extrapolation and for points sampled only once that would be the result of the error mitigation described in the previous paragraph. 
Fig.~\ref{fig:torino_sim_7x7_ex} shows the same data for the $7\times7$ plaquette lattice simulation. 
Note that because statevector simulation of systems of this size are beyond what can be done with classical computers, the classical simulations in this figure were performed with tensor networks. 
The tensor network simulations were performed using {\tt CuQuantum}~\cite{cuQuantum} on a single NVIDIA A100 GPU.

\end{document}